\documentclass[useAMS,usenatbib]{mn2e}
\usepackage[totalwidth=515pt,totalheight=680pt,left=1.4cm,right=1.4cm]{geometry}
\usepackage{graphicx,amssymb}
\input psfig

\title[Full-Lifetime Two-Planet Simulations]
{Simulations of two-planet systems through all phases of stellar evolution: implications for the instability
boundary and white dwarf pollution}
\author[Veras, Mustill, Bonsor \& Wyatt]{Dimitri Veras$^{1}$\thanks{E-mail: veras@ast.cam.ac.uk}, Alexander J. Mustill$^{2}$\thanks{E-mail: alex.mustill@uam.es}, Amy Bonsor$^{3}$\thanks{E-mail: amy.bonsor@obs.ujf-grenoble.fr} and Mark C. Wyatt$^{1}$\thanks{E-mail: wyatt@ast.cam.ac.uk} \\
$^{1}$Institute of Astronomy, University of Cambridge, Madingley Road, Cambridge CB3 0HA
\\
$^{2}$Universidad Aut\'{o}noma de Madrid, Departamento de F\'{i}sica Te\'{o}rica C-XI, 28049 Madrid, Spain
\\
$^{3}$UJF-Grenoble 1 / CNRS-INSU, Institut de Plan\'{e}tologie et d'Astrophysique de Grenoble (IPAG), UMR 5274, BP 53, F-38041
\\ 
Grenoble cedex 9, France}

\begin{document}

\date{Accepted 2013 February 13.  Received 2013 January 15; in original form 2012 November 20}

\pagerange{\pageref{firstpage}--\pageref{lastpage}} \pubyear{2013} 

\maketitle

\label{firstpage}

\begin{abstract}
Exoplanets have been observed at many stages of their host
star's life, including the main sequence (MS), subgiant and red giant 
branch stages.  Also, polluted white dwarfs (WDs) likely represent 
dynamically active systems at late times.  Here, we perform 
3-body simulations which include realistic post-MS stellar mass loss and span 
the entire lifetime of exosystems with two 
massive planets, from the endpoint of formation to several Gyr into
the WD phase of the host star.  We find that both MS and WD
systems experience ejections and star-planet collisions (Lagrange instability) 
even if the planet-planet separation well-exceeds 
the analytical orbit-crossing (Hill instability) boundary.  Consequently, {\it MS-stable 
planets do not need to be closely-packed to experience instability during 
the WD phase}.  This instability may pollute the WD directly through collisions,
or, more likely, indirectly through increased scattering of smaller bodies such as 
asteroids or comets.  Our simulations show that this instability occurs predominately 
between tens of Myr to a few Gyrs of WD cooling. 
\end{abstract}

\begin{keywords}
planet-star interactions, planets and satellites: dynamical evolution and stability, 
stars: evolution, stars: AGB and post-AGB, stars: white dwarfs
\end{keywords}

\section{Introduction}

A planet's life may be split into four distinct stages:
1) formation and concurrent dynamical excitation, 
2) main sequence (MS) evolution, 3) evolution during post-MS 
stellar phase changes, and 4) white dwarf (WD) evolution. 
The first stage generally lasts no longer than $0.1\%$ of the 
entire MS lifetime.  The second stage is relatively dynamically
quiescent, with only occasional 
but often important scattering interactions.  In the third stage,
the planet is subject to dynamical changes due to the
star's violent actions as it becomes a giant.  In the final stage, 
the star has become a WD, and the planet again enters and 
remains in a phase of relative 
dynamical quiescence occasionally punctuated by scattering interactions
or external forcing. This general picture, which does not include 
possibilities such as the capture of free-floating planets, 
planetary destruction due to supernovae, or multiple host stars, describes 
the life cycle of the vast majority of known exoplanets.

The volume of planetary literature investigating the first two stages dwarfs 
the literature describing the final two stages, despite the fact that the Universe
is already over 13.5 Gyr old \citep{jaretal2011} and that the Milky Way
contains about $10^9$ WDs (\citealt*[][Pgs. 2-3]{bintre2008} 
and \citealt*{holetal2008}).  Further, 
these final two stages are becoming increasingly relevant given
the suggestions or discoveries of exoplanets in post-MS systems 
\citep{wolfra1992,wolszczan1994,sigetal2003,siletal2007,muletal2008,geietal2009,leeetal2009,muletal2009,setetal2010,wicetal2010,chaetal2011,adaetal2012,faretal2012a,leeetal2012a,leeetal2012b,satetal2012a}.

Explorations of exosystem evolution in the third stage include
one-planet studies \citep{villiv2007,villiv2009,veretal2011,kraper2012,musvil2012,norspi2012,spimad2012,vertou2012,adaetal2013},  
just a few dedicated multiple-planet studies \citep{debsig2002,por2012,voyetal2013}, and studies
focusing on the evolution of comets \citep{alcetal1986,paralc1998}.
Further, \cite{bonwya2010} consider the effect of 
post-MS evolution on debris discs. Motivated by observations of 
metal-polluted WDs, \cite{bonetal2011,bonetal2012} 
and \cite{debetal2012} model the interplay between a planet and a 
belt of smaller material amidst stellar mass loss.  

Here, we self-consistently simulate the second, third and fourth
stages together.  We combine stellar evolution with planetary gravitational
scattering amongst multiple massive planets, and extend the work of \cite{debsig2002} 
by considering full-lifetime simulations with realistic mass
loss prescriptions at each post-MS phase. Following the evolution 
over the whole stellar lifetime means that we can be sure that 
systems whose stability is investigated on the giant and WD stages 
will have survived the long MS evolution. Through these integrations, we can  
determine what types of planetary architectures might be expected in exoplanet-hosting
WD systems, and could allow us to extrapolate backwards in time
from observed WD systems.  We restrict our explorations to
two-planet systems in this initial study given the vast phase space 
to explore; three-planet simulations will be presented in a 
follow-up paper.  First, we briefly summarize our knowledge of
planetary instability for one- and two-planet systems
during the MS (Subsection 1.1) and
post-MS (Subsection 1.2).

\subsection{Instability in Main Sequence Planetary Systems}

Dynamical instability in planetary systems is often said to occur
when a planet suffers a close encounter with the star 
or another planet, or is ejected from the system.  Occasionally, investigators 
use stricter definitions of instability, such as when the semimajor axis or
eccentricity variation of a planet exceeds a certain per cent of
its nominal value.  Additionally, a wide body of literature has arisen
characterizing chaotic orbits as a precursor to instability; 
\cite{daretal2012} and references therein summarize many of these techniques.

\subsubsection{One-Planet Instability}

One planet orbiting a MS star will typically remain stable 
throughout the star's MS lifetime in the absence of external forces.
Exceptions may include planets which are close enough to their parent stars
to be tidally disrupted \citep[e.g.][]{guetal2003} and possibly evaporated 
\citep[e.g.][]{guietal1996}.  In the opposite extreme, a planet which is far 
enough away from its parent star may be ejected due to external forces
such as passing stars \citep{zaktre2004,vermoe2012} or achieve a high enough
eccentricity through Galactic tides to cause a collision with the star
\citep{vereva2012a,vereva2012b}.

\subsubsection{Two-Planet Instability}

In addition to tidal interactions and external forces, the mutual 
perturbations between two planets may also create instability.  Partially motivated by 
tractable analytical solutions to the general three-body problem,
the source of this instability has been studied extensively.  If the 
orbits of two planets are guaranteed to never overlap (precluding
a collision between both planets), then they are said to be ``Hill stable''.  \cite{gladman1993}
pioneered the analytic use of Hill stability for planetary systems in specific
cases and has motivated many subsequent analyses, as recently summarized by, e.g., 
\cite{donnison2009,donnison2010a,donnison2010b,donnison2011}.
Hill stability does not guarantee that the outer planet
remains bound to the system, nor does it prevent the inner planet from
colliding with the star.  If both planets remain bound and
retain their ordering, and no collision with the star occurs,
then the system is ``Lagrange stable''\footnote{This type of stability
has also been referred to as ``Laplace stability'' \citep[e.g.][]{kubetal1993}
and featured but remained unnamed in many papers published before the discovery of
exoplanets and high-speed computing.}.  Unlike Hill stability, Lagrange stability 
does not benefit from a known analytical formulation, but rather empirical estimates
based on numerical simulations. 

The analytical Hill stability boundary is conservative.  Two planets
whose initial separation is less than the Hill stable distance
may in fact remain stable.  If the initial separation is greater than the  
Hill stable distance, then the planets are guaranteed to retain their
ordering.  In simulations of the HD 12661 and 47 Uma systems, 
\cite{bargre2006} found that pairs of planets 
close to the Hill stability boundary are not Lagrange stable,
and hence are not generally stable.  They tentatively suggest that the 
Lagrange stability boundary exceeds the Hill stable boundary by at least
21\% as measured by the semimajor axis ratio.  Subsequent work \citep{bargre2007} 
revealed how mean motion commensurabilities can broaden the divide between
the Hill and Lagrange stable boundaries; \cite{kopbar2010} demonstrated how the
boundary between stable and unstable systems is not sharp. Hence, numerical 
validation of analytical stability estimates is crucial.

\subsection{Instability in Post-Main Sequence Planetary Systems}

Mass loss from a dying star can trigger planetary instability
in different ways, which are outlined below.  A common assumption 
amongst the studies
which have considered instabilities in post-MS systems is isotropic
stellar mass loss.  We also adopt this assumption here, as
modelling non-isotropic mass loss would significantly complicate
both numerical and analytical modelling and is best left to
separate, dedicated studies.  One such dedicated post-MS 
study \citep{paralc1998} importantly observes that the speed of
(effectively massless) comets near the boundary of a planetary 
system may be comparable to the recoil velocity of the parent star due
to asymmetric mass loss.  That study suggests that 
anisotropic mass loss will affect the details of planets being 
ejected after scattering but is unlikely to have a significant 
effect on the prior dynamics.  Other studies modelling planetary
dynamics due to non-isotropic mass loss instead focus on jet
accelerations present at the birth sites of planets 
\citep{namouni2005,namouni2007,namouni2012}.

Further, in all cases we assume the planets are orbiting
a single star.  Extensions to the multiple-star case 
\citep{kraper2012,por2012,vertou2012} are likely to be nontrivial.

\subsubsection{One-Planet Instability}

For decades, binary star investigations revealed that stellar mass
loss causes orbital semimajor axis expansion.  Less 
well-known is that when the mass loss is rapid enough, the eccentricity
of the companion's orbit can change as well 
\citep{omarov1962,hadjidemetriou1963,hadjidemetriou1966,veretal2011}.
If the eccentricity is great enough, a planetary companion may escape
from the system.  

In the more commonly-used adiabatic limit,
$de/dt = 0$ and $da/dt = -\left(a/\mu\right)\left(d\mu/dt\right)$.  
Here, $\mu \equiv G \left(M_{\star} + M_p\right)$, 
$M_{\star}$ and $M_p$ are the masses of the star and planet,
$a$ is the planet's semimajor axis, and $e$ is the planet's eccentricity.
\noindent{This} limit holds when the mass loss timescale is
much longer than the orbital timescale.  Adiabaticity will be broken,
even briefly, if at any point a sudden burst of mass loss
causes the timescales to become comparable 
\citep{vertou2012,verwya2012}.  Hence, characterizing whether
or not planetary evolution is adiabatic amidst mass loss
will be important for any post-MS scattering study.

Another source of instability for one-planet systems
could come from tidal orbital decay and potentially
direct engulfment by the rapidly expanding 
stellar envelope.  \cite{villiv2007,villiv2009} and \cite{villaver2011}
treat this effect in detail with additional physics
such as frictional drag, planet accretion and planet
evaporation.  \cite{musvil2012} model individual
thermal pulses and demonstrate how they affect
planetary stability.  
In this study, we only consider planets that are too distant 
to be affected by the stellar envelope
expansion and so are not affected by tides, accretion or
evaporation (see Section \ref{sec:radius}).

\subsubsection{Two-Planet Instability}

\cite{debsig2002} considered adiabatic evolution of 
two-planet systems while exposed to 
a $1 M_{\odot}$ star losing half of its mass
over 1000 planetary orbits.  This foundational study
considered circular and coplanar equal mass planets, 
with planet/star mass ratios ranging from $10^{-3}$ 
to $10^{-7}$.  They discovered importantly that 
although adiabatic evolution causes both planets
to move outward and maintain their initial semimajor
axis ratio, their critical Hill separation changes.

The rate of change of the separation measured in 
units of Hill's radii is equal 
to $\mu^{-2/3} d\mu/dt$.  This dependence causes 
previously Hill stable planetary
systems to become unstable, and incite gravitational
scattering which could not occur on the MS.
Their simulation results suggest that scattering
instabilities may be more widespread during post-MS
evolution than during MS evolution.  Here, we investigate
this claim in significant detail.  For discussion
on the high mass-loss non-adibatic multi-planet case 
recently presented by \cite{voyetal2013}, please 
see Section 6.4.

\subsection{Paper Outline}

We begin in Section 2 with a description of the challenges of using
N-body numerical simulations for gravitational scattering amidst mass loss.
In Section 3 we determine the regimes where engulfment and tides from stellar
envelope expansion can be neglected for this study.
Section 4 presents a general formulation of the Hill stability limit and 
shows how it changes due to stellar mass loss.  We use the results of Sections 2-4
to motivate the setup for our numerical scattering simulations.  In Section 5, we 
perform these simulations, and report the results.  We discuss the
consequences in Section 6 and conclude in Section 7.

\section{Numerical Convergence}

\begin{figure}
\centerline{
\psfig{figure=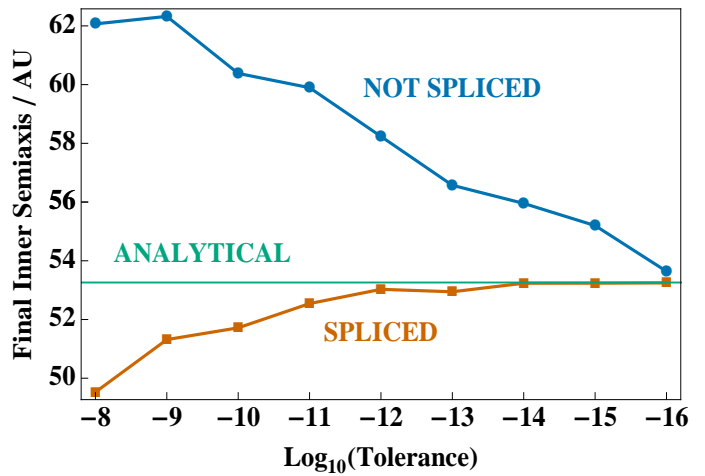,height=6.2cm,width=9.0cm} 
}
\caption{
Difference between interpolating {\it SSE}-outputted mass at 
every {\it Mercury} timestep (blue circles) versus
interpolating this mass within {\it Mercury} timesteps 
(orange squares). ``Spliced'' indicates the latter and
``Not Spliced'' indicates the former.  Shown are the 
final values of the inner planet's semimajor
axis for a pair of $0.001 M_{\odot}$-planets with initial 
semimajor axes of $10$ AU and $30$ AU, and initial 
eccentricities of $0.0$ and 
$0.5$, respectively.  The parent Solar-metallicity star 
was modeled to lose $\approx 6.22 M_{\odot}$
of its initial $\approx 7.66 M_{\odot}$ at the start of an 
asymptotic giant branch phase lasting
almost $5 \times 10^5$ yr.  The ``Analytical'' line 
refers to the final semimajor axis of the inner planet 
predicted by adibatic mass loss.
The convergence properties of the spliced BS integrator for 
systems with both gravitational scattering and mass loss 
is a significant improvement.}
\label{splicetest}
\end{figure}

In this section, we highlight the difficulty in achieving accurate N-body simulations 
that model both central star mass loss and gravitational scattering amongst multiple
massive planets, and implement a solution. 

\subsection{Stellar Evolution Code}

We utilize the {\it SSE} stellar evolution code \citep{huretal2000}, which adopts 
empirically-derived algebraic formulations in order to quickly generate a stellar
evolutionary track solely from a given progenitor mass and metallicity, 
and stellar model parameters such as the Reimers mass-loss 
coefficient.  We use the same mass loss prescriptions as described
in Section 7.1 of \cite{huretal2000} with their Reimers mass-loss coefficient 
default value of 0.5.
Their choice is observationally motivated by the Horizontal Branch morphology
in Galactic globular clusters \citep{iberen1983}, and lies in the center of the
range recently considered by \cite{verwya2012}, who discuss this choice in light
of an updated version of the Reimers law \citep{schcun2005}.
The {\it SSE} code allows us to sample many different evolutionary tracks 
easily, and outputs the important parameters, $M_{\star}(t)$ and $R_{\star}(t)$,
where $R_{\star}$ is the radius of the star. 

\subsection{Planetary Evolution Code}

We also use the {\it Mercury} integration package \citep{chambers1999}, which 
specializes in modeling planetary dynamical evolution.  In order to accurately 
model close encounters between planets and the parent star -- a 
necessity for this study -- we use the Bulirsch-Stoer (BS) integrator.  This 
integrator features an adaptive timestep, which is determined by a tolerance 
parameter given at the start of the simulation.  A tolerance of $10^{-12}$ is 
considered to be highly accurate \citep{jurtre2008}.  Smaller tolerance values 
should roughly converge to the same result; Fig. 7b of \cite{smilis2009} demonstrates 
that in crowded 5-planet systems separated by several Hill radii, tolerances of 
$10^{-12}$, $10^{-13}$, $10^{-14}$, $10^{-15}$ and $10^{-16}$ will yield instability 
timescales which are all within the same order of magnitude.  Tolerances below 
$10^{-16}$ generally cannot be achieved because in those cases the accuracy 
requested is greater than machine precision.

\subsection{Merging Both Codes}

\cite{veretal2011} found that linearly interpolating {\it SSE} stellar mass output at 
each {\it Mercury} timestep adequately models the dynamical evolution of a single 
planet amidst stellar mass loss because the numerical simulations reproduced the analytical 
results.  For multi-planet systems, this technique alone is inadequate.  The 
interaction between both planets coupled with stellar mass loss causes a 
failure of convergence of orbital parameters as the tolerance is decreased.

In order to improve the accuracy, we have performed an additional interpolation of 
the {\it SSE} stellar mass output in between each {\it Mercury} timestep at each BS substep.  
The resulting finer gradation makes a crucial difference, as demonstrated by 
Fig. \ref{splicetest}.  The figure plots the final semimajor axis 
values for the inner $10^{-3} M_{\odot}$ planet and the outer $10^{-3} M_{\odot}$ 
planet in a system with initial semimajor axes of $10$ AU and $30$ AU, and 
initial eccentricities of $0.0$, $0.5$,  respectively.  All initial orbital angles 
were set to $0^{\circ}$.  The simulations were run for the entire evolution of the 
Thermally Pulsing Asymptotic Giant Branch (TPAGB) phase of a 
Z=Z$_{\odot} = 0.02$ (Solar metallicity) star.
We chose a progenitor mass of $8 M_{\odot}$ to model particularly violent mass
loss.  Our simulations ran during the TPAGB phase only, when $M_{\star}$
was reduced from $7.659 M_{\odot}$ to $1.438 M_{\odot}$ in 
about 492,744 yr.


The plot contains two curves from the simulation output, representing final 
values of the semimajor axis due to {\it SSE}-outputted mass interpolation at 
each {\it Mercury} timestep 
(blue circles; ``non-spliced''), and with an additional interpolation in-between 
timesteps (orange squares; ``spliced'').  The third, green, curve is the analytic prediction
for the final semimajor axis of the inner planet.  This value can be 
determined because the mass loss is adiabatic and the planets are not 
near a strong mean motion commensurability (hence their semimajor axes remain
secularly unaffected).

Without the additional interpolation, the results do not appear to converge until perhaps at the machine precision limit for the BS tolerance\footnote{In the one-planet case, when the outer planet is removed from these particular simulations, then all three curves are visually indistinguisable from one another on this plot.  This result reinforces the finding of \cite{veretal2011} that splicing within timesteps is generally not necessary in one-planet simulations.}.  Further, the extent of the variance in the non-spliced curves may fundamentally change the endstate of the system if any more close encounters occur.  Therefore, we use the spliced BS integrator throughout the rest of this work.  Convergence with the spliced integrator is achieved in this case at an accuracy of $\sim10^{-12}$; we are conservative and adopt the value of $10^{-13}$ for our integrations. 

Another consideration is the ejecta-crossing lag time.
Stellar ejecta will cause the inner planet's orbit to shift before the outer planet's orbit.
In some cases, this ``lag time'' between orbital shifts may produce a noticeable change in 
the dynamics that is missed by assuming both planets simultaneously change their orbits.
The weakness of this assumption is accentuated for widely-spaced orbits and for systems which are not 
in the adiabatic regime.  For the (adiabatic) systems studied here, however, this assumption likely produces
a negligible effect\footnote{By using the observed mass ejecta speed in the post-MS system 
R Sculptoris of $\approx 14.3$ km/s \citep{maeetal2012}, one can estimate that the ejecta will 
take 121 days to travel 1 AU.  Thus, for an inner planet at 10 AU and an outer planet at 12 or 13 AU,
the travel time is less than a year, a small fraction of the inner planet's orbital period.}, 
and hence is neglected for the remainder of this study.



\subsection{Further Adaptations}

All orbital elements in this work are reported in Jacobi coordinates. 
Therefore, as {\it Mercury} receives input in astrocentric coordinates,
we performed the conversion.  Further, we had to modify the default
version of {\it Mercury} to account for a changing stellar mass in the 
output file {\tt xv.out} so that the conversion from Cartesian
output to Jacobi elements was performed correctly.  Consequently,
the size of {\tt xv.out} nearly triples in size.  Although
this increase might be prohibitive for high-resolution studies of individual
systems, here we are interested primarily in the final
stability state of each system.  Therefore, in our case, outputs 
at a Myr resolution are all that is required.  Independent of 
the paucity of outputs, {\it Mercury} does record the times
of collisions to within a timestep.

\section{Treating the Stellar Radius} \label{sec:radius}

Additionally, we modified {\it Mercury} to incorporate the stellar 
radius evolution profiles from {\it SSE}.  Over its lifetime, a star's 
radius evolves nonlinearly and nonmonotonically.  Because 
these variations are modest and all occur within $0.05$ AU during the MS,
most previous planet scattering studies treated the radius as static and/or negligible.
However, during the post-MS, the radius variations can be violent 
and extend beyond several AU.

\subsection{Expansion}

Because of the potential for planetary collisions, evaporation and/or envelopment due
to the expanding stellar envelope, the variations in stellar radius must
be taken into account during post-MS scattering simulations.  An important
question is whether or not a planet, expanding its orbit due to mass loss,
can outrun an expanding stellar envelope.  The answer is complicated by
the fact that the timescales for and amplitudes of mass loss and 
radial increase are not completely in sync, although they often are similar.
Additionally, a star's radius may decrease. WD radii are even more
compact than MS radii.

In order to characterize these variations, we generated Figs. 
\ref{envexp}-\ref{SSERadius}, with SSE data.  The figures 
characterize the maximum stellar radius 
for different metallicities and stellar phases, respectively.
In particular, Fig. \ref{envexp} suggests that the maximum
stellar radius is largely independent of metallicity,
and that roughly the number of AU at the maximum radius
is equal to the number of initial $M_{\odot}$.  Figure
\ref{SSERadius} illustrates that the stellar radius
generally increases during post-MS phases, although 
for the $1 M_{\odot}$ case, there is an order-of-magnitude
decrease after the RGB stage.  This decrease becomes progressively
smaller as the progenitor stellar mass is increased until vanishing
at about $3 M_{\odot}$.  

These trends are model-dependent.  Other stellar evolution codes 
\citep[e.g.][]{vaswoo1993} demonstrate that as a result of thermal pulses on 
the AGB the expanding stellar radius can expand up to 1 AU more than the rule-of-thumb maximum
radius from the last paragraph.  Further, the maximum radius at each phase 
might differ depending on the model used; see, for example, \cite{villiv2009}.

\subsection{Tides}

The maximum stellar radius is just a physical boundary; stellar tides
can extend beyond the reach of the star.  Because modeling tides is both beyond 
the scope of this study and remains the subject of debate,
we choose initial conditions for our numerical simulations where 
we can safely neglect tidal effects.  Planets are not necessarily 
destroyed by tides nor by being engulfed in the stellar envelope.
The remarkable sub-10 hour periods of the two planets in the hot B subdwarf 
star KIC 05807616 system \citep{chaetal2011} 
provide strong evidence that 
planets can survive deep immersion into the stellar envelope.  

\begin{figure}
\centerline{
\psfig{figure=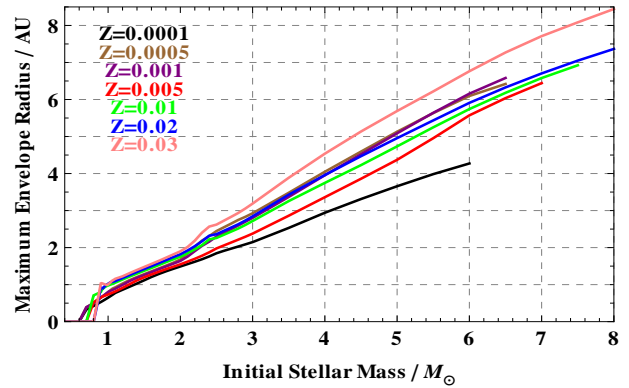,width=8.5cm,height=5.5cm} 
}
\caption{
Maximum stellar radius as a function of progenitor mass
and metallicity.  Colours denote different metallicities; the blue curve 
(Z=0.02) represents Solar metallicity.  Curves end
when supernovae occur.  This plot demonstrates a rule of thumb:
generally, the number of AU at the maximum radius
is approximately equal to the number of
initial $M_{\odot}$.
}
\label{envexp}
\end{figure}

\begin{figure}
\centerline{
\psfig{figure=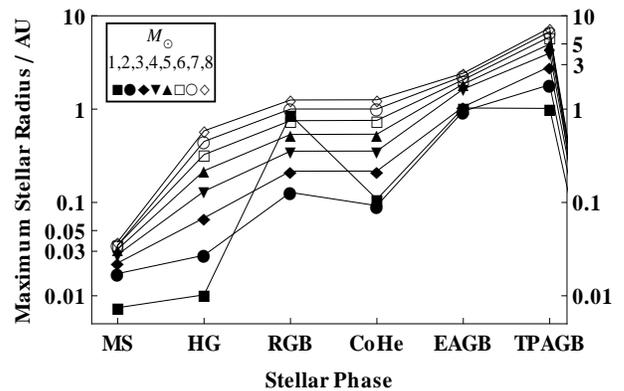,width=8.5cm,height=5.5cm} 
}
\caption{
Maximum stellar radius as a function of progenitor mass
and stellar phase.  The y-axis is logarithmic, and the x-axis
is monotonically increasing in time.  Symbols denote different 
progenitor masses, and the stellar phases are MS = Main Sequence,
HG = Hertzprung Gap, RGB = Red Giant Branch, CoHe = Core Helium
Burning, EAGB = Early Asymptotic Giant Branch, and 
TPAGB = Thermally Pulsing Asymptotic Giant Branch.  For Solar
metallicity, stellar radii generally increase monotonically throughout 
post-MS phases except for progenitor masses approximately 
equal to $1 M_{\odot}$.
}
\label{SSERadius}
\end{figure}

Both \cite{villiv2009} and \cite{kunetal2011} have analyzed planet engulfment by 
Red Giant Branch (RGB) stars by including a number of physical factors, and 
use more detailed stellar evolution models than {\it SSE}.  Their results indicate 
that, for lower-mass stars, tides can significantly affect planets 
with semimajor axes that are about 
2.5 times as high as the maximum stellar RGB radius, and that engulfment is 
sensitively dependent on progenitor stellar mass. However, the RGB radius 
does not greatly exceed 1 AU.  Although AGB radii are significantly larger 
(Figs. \ref{envexp} and \ref{SSERadius}), \cite{musvil2012} found that for 
higher-mass stars, the most distant Jovian planet that becomes engulfed is initially 
at approximately the maximum stellar radius; tides can slightly shrink 
the orbits of planets for about an AU beyond this maximum. Hence, we 
adopt 10 AU as the initial orbit of the innermost planet in our simulations.  
This planet will certainly be safe from engulfment in the envelope, and will 
experience no significant tidal decay except possibly from the most massive stars.

\subsection{Radius-based Code Modifications}

Our simulation initial conditions are chosen such that
stable planetary systems are well beyond the influence of 
tidal effects.  However, instabilities which arise
during the simulations may cause planets to approach or
collide with the expanding stellar envelope.  If a collision
occurs, the system is flagged as unstable and is stopped.
In reality, if a star engulfs a planet, the star's mass
would increase slightly and as a result its radius might
change as well.  Remaining planets in the system would
then be affected because angular momentum must be conserved.

Additionally, {\it Mercury's} collision detection algorithm {\tt mce\_cent} 
checks to see if the pericenter of a planet's
orbit lies within the star's radius.  If so, a collision
is flagged.  However, if the star's reflex velocity is sufficiently
high, then a collision might not occur.  Therefore, we modified 
{\tt mce\_cent} to account for the stellar reflex velocity.  For planetary
mass companions, however, the reflex velocities are low,
and are not likely to factor into collision statistics.

\begin{figure}
\centerline{
\psfig{figure=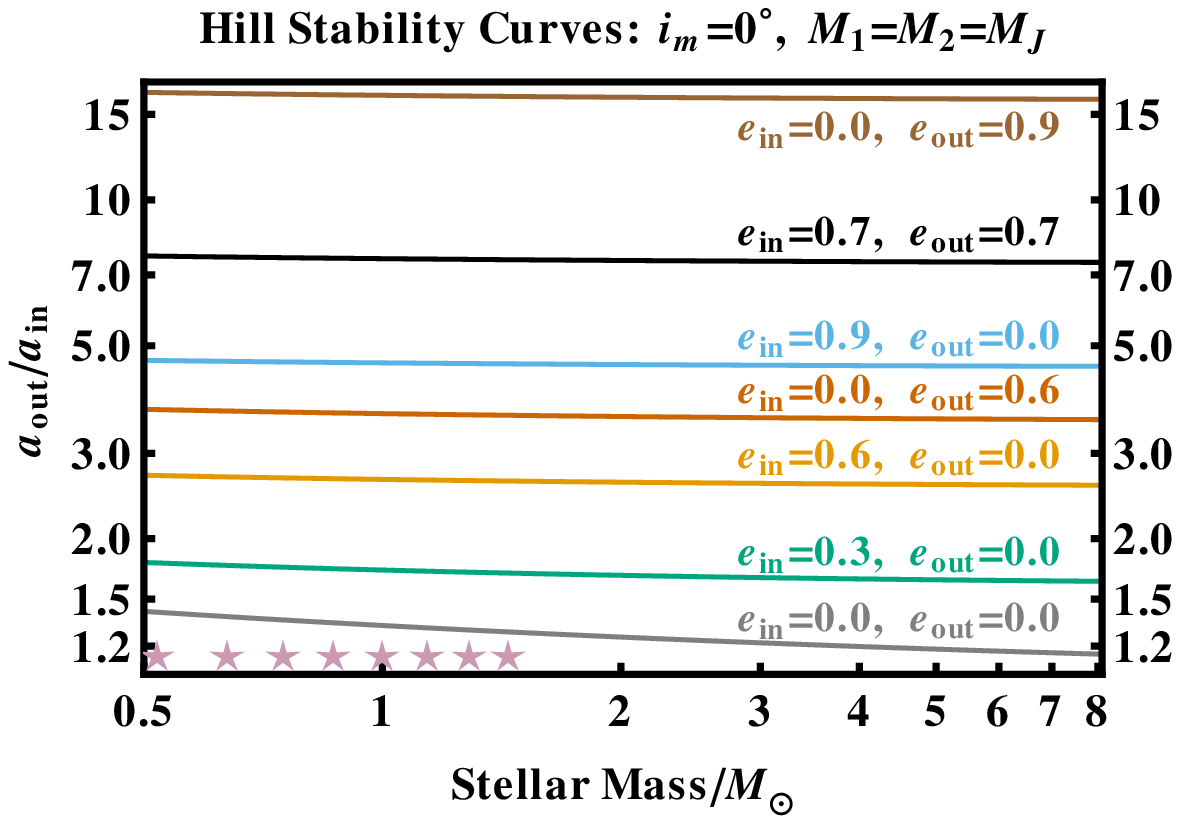,width=8.5cm,height=5.5cm}
}
\centerline{ }
\centerline{
\psfig{figure=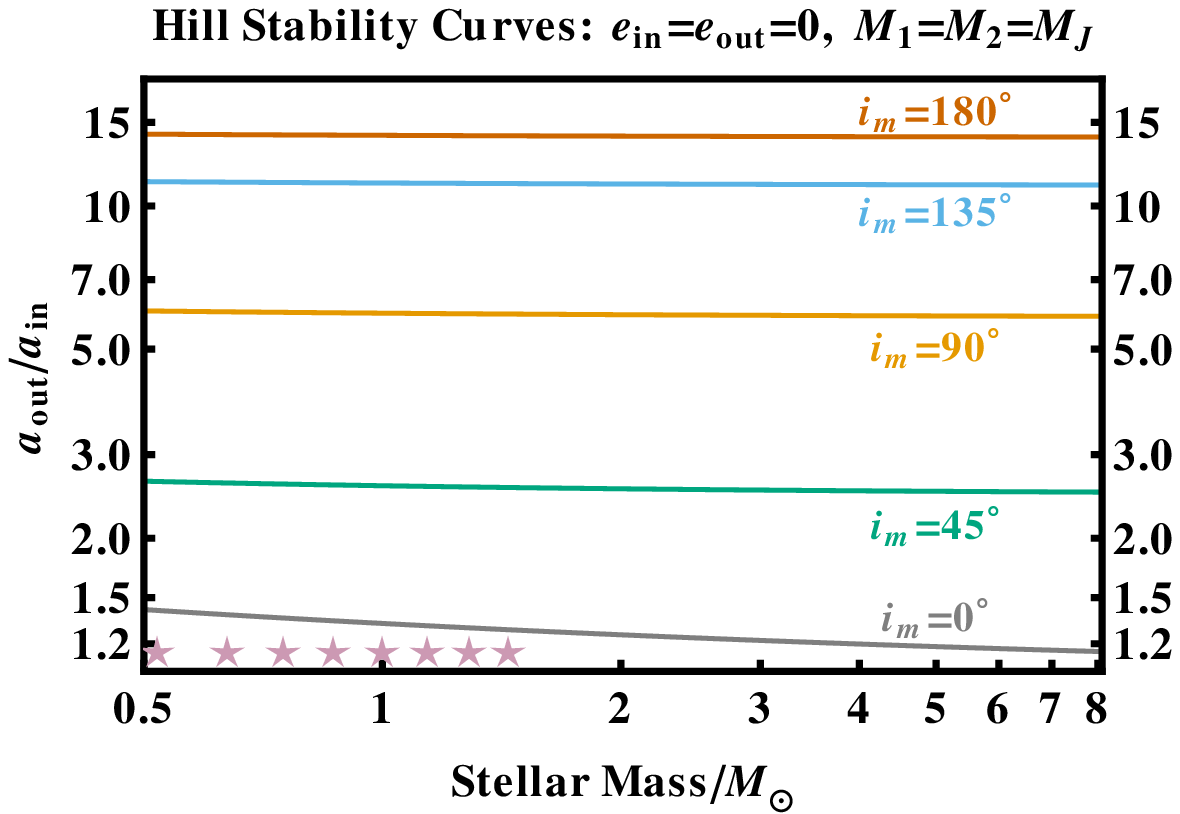,width=8.5cm,height=5.5cm} 
}
\centerline{ }
\centerline{
\psfig{figure=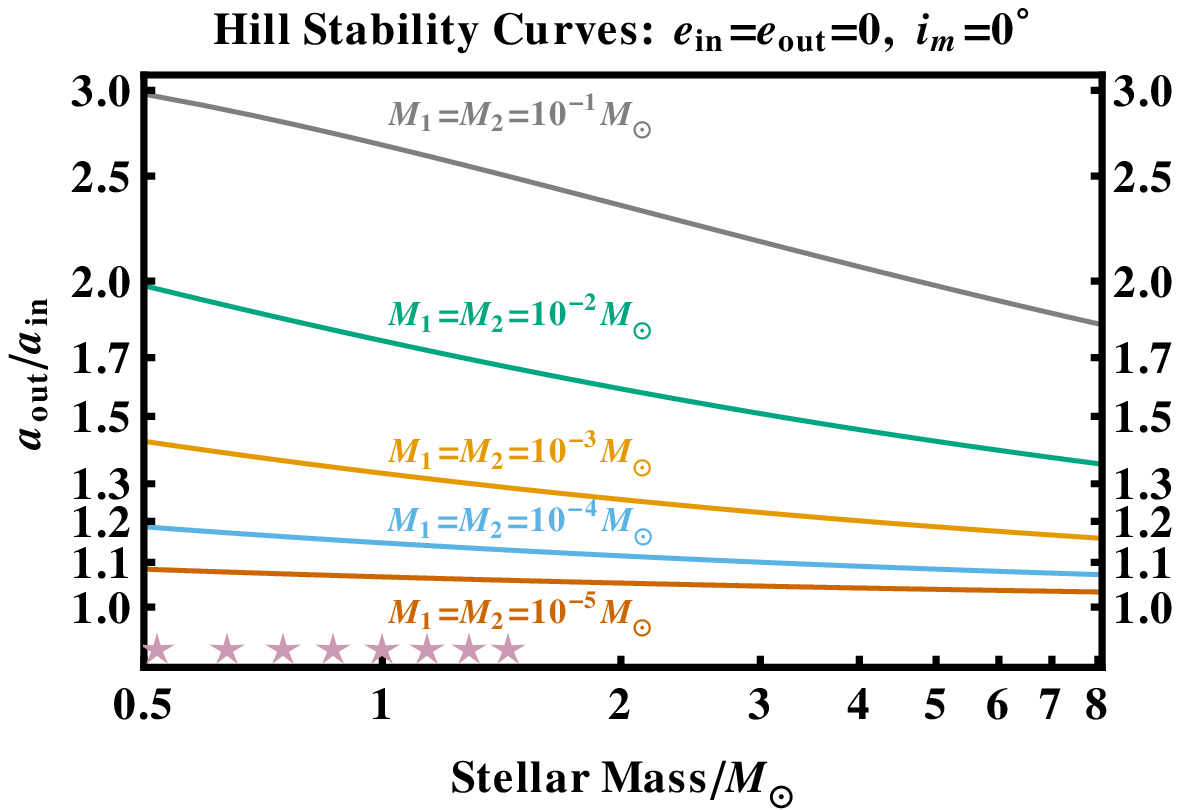,width=8.5cm,height=5.5cm} 
}
\caption{
Critical Hill semimajor axis ratios as a function of 
stellar mass for different eccentricities ({\it upper panel}),
different mutual inclinations ({\it middle panel}) and
different planetary masses ({\it bottom panel}).  The purple
stars on the bottom of each plot, from right to left, 
represent the eventual WD mass [in brackets in the following]
for MS progenitor masses of
$8 M_{\odot} [1.44 M_{\odot}]$, 
$7 M_{\odot} [1.29 M_{\odot}]$, 
$6 M_{\odot} [1.14 M_{\odot}]$, 
$5 M_{\odot} [1.00 M_{\odot}]$, 
$4 M_{\odot} [0.87 M_{\odot}]$, 
$3 M_{\odot} [0.75 M_{\odot}]$, 
$2 M_{\odot} [0.64 M_{\odot}]$ and 
$1 M_{\odot} [0.52 M_{\odot}]$.  The figure illustrates
that the Hill radius is sensitively dependent on
$e_{\rm out}$, $e_{\rm in}$, $i_m$, $M_1$, and $M_2$,
but weakly dependent on $M_{\star}$.  For Jovian and
terrestrial-mass planets, stellar mass
loss can change the Hill stability limit by at most
a few tenths in $a_{\rm out}/a_{\rm in}$.  
}
\label{hille1}
\end{figure}

\section{Hill Stability}

We now make some analytical stability estimates to identify the systems 
of interest for our N-body runs. We particularly seek systems that are
likely to be stable on the MS and subsequently destabilised during the 
giant or WD stages. \cite{donnison2011} provides a formulation of Hill 
stability in Jacobi coordinates which allows for arbitrarily inclined and eccentric orbits.  
His treatment is fully general with one exception: the expression for the 
system energy is the two-body approximation.  This approximation is necessary
in order to obtain an analytically tractable (but not strictly correct) solution;
the intractable terms appear, for example, in Eq. (2.27) of \cite{veras2007}, which
provides the complete expression for the energy of a three-body system in terms of 
Jacobi orbital elements.  See Subsection \ref{sharpness} of this paper for further discussion
on this point.

In the following, the subscripts ``1'' and ``2'' refer to the inner and outer planets, 
the subscript ``in'' refers to the star/innermost-planet binary, and the subscript ``out'' refers to 
the outer planet properties measured with respect to the inner binary.  Let $i_{\rm m}$ 
represent the mutual inclination of the inner and outer binaries.   Then the Hill
stability curve is given by \citep{donnison2011}:

\begin{eqnarray}
&&  \left(1 + y^2\right) \left(y^2 \beta^2 + 2 y \beta \cos{i_{\rm m}} + 1 \right) 
=
\nonumber
\\
&&
- \frac{2 S_{\rm cr} \left(M_{\star} + M_1 + M_2 \right)y^2}{M_{2}^3 \left(M_{\star} + M_1\right)^3 
\left(1 - e_{\rm out}^2 \right)}
\label{first}
\end{eqnarray}

\noindent{where}

\begin{eqnarray}
y &\equiv& 
\sqrt{\frac{a_{\rm in} M_2 \left(M_{\star} + M_1\right)}
{a_{\rm out} M_{\star} M_1 }}
\label{second}
\\
\beta &\equiv&
\left( \frac{M_{\star} M_1}{M_2}  \right)^{3/2}
\sqrt{ \frac{M_{\star} + M_1 + M_2}
{\left(M_{\star} + M_1 \right)^4}
\frac{\left(1-e_{\rm in}^2\right)}{\left(1-e_{\rm out}^2\right)}
     }
\label{second2}
\end{eqnarray}

\noindent{and} with \citep{donnison2006}:

\begin{eqnarray}
S_{\rm cr} &\equiv& \frac{1}{2 \left(M_{\star} + M_1 + M_2 \right)}
\left(M_{\star}M_1 + \frac{M_{\star} M_2}{1+x_{\rm crit}} + \frac{M_1M_2}{x_{\rm crit}}  \right)^2
\nonumber
\\
&&\times \left(M_{\star}M_1 + M_{\star}M_2 \left(1 + x_{\rm crit}\right)^2 + M_1 M_2 x_{\rm crit}^2\right)
\end{eqnarray}

\noindent{such} that $x = x_{\rm crit}$ is the unique real solution to the following quintic equation:

\begin{eqnarray}
&&\left(M_{\star} + M_1\right) x^5 + \left(3 M_{\star} + 2 M_{1} \right) x^4 + \left(3 M_{\star} + M_1 \right) x^3
\nonumber
\\
&&- \left(3M_2 + M_1 \right)x^2 - \left(3M_2 + 2 M_1\right)x = \left(M_2 + M_1\right)
\label{last}
\end{eqnarray}

\noindent{}Care should be taken when choosing the root of the quartic equation in
Eq. (\ref{first}) when solving for $y$.

The literature is replete with special-case solutions to Eqs. (\ref{first})-(\ref{last})
[see \citealt*{georgakarakos2008} for a review]\footnote{We have discovered
two typographic errors in the previous literature:  The last quantity of the LHS of Eq. (2.13)
in \cite{verarm2004} should not be squared, and the sign in front of $A$ in Eq. (9)
of \cite{donnison2011} is incorrect.} 
but typically treat the masses as static 
and define a ``separation'' as a modulated
{\it ratio} of the planetary semimajor axes.  Equation (\ref{second}) demonstrates why.
In order to model how the Hill stability curves 
change with mass loss, one need only to evaluate Eqs. (\ref{first})-(\ref{last})
at different times during a star's evolution.


In the circular, coplanar, equal-planetary mass 
case presented in \cite{debsig2002}, the critical Hill
separation is shown to vary by a few tenths due to significant mass loss.  
We have undertaken a wider parameter exploration of phase space, and
have discovered that this result generally holds true for 
orbits of any eccentricity, inclination and stellar mass
as long as the planetary masses are at most about one Jupiter-mass each.  
Our results are presented
in Fig. \ref{hille1}, where each panel showcases a different 
parameter dependency.

The variation in the Hill stability limit due to stellar mass loss
is often equivalent to several AU for planets which reside
beyond about 10 AU.  Consequently, if planetary packing \citep[e.g.][]{rayetal2009} 
produces planets near the Hill stability limit during the MS, 
then post-MS evolution may trigger instability.

Despite these considerations, one must remember that
the Hill stability criterion is a {\it sufficient but not necessary}
condition for the planets to remain ordered.  Hill stable
planets may be Lagrange unstable, and planets failing to satisfy the 
Hill stability condition may nevertheless be stable. Regarding the 
latter case, one outstanding feature of 
Fig. \ref{hille1} is that moderately
eccentric or inclined orbits yield critical semimajor axes
ratios that are high -- much higher, for example, than the
semimajor axis ratios of any adjacent pair of planets in the Solar System.

Large critical semimajor axis ratios may strongly influence 
the location where two planets become Lagrange stable.
\cite{verarm2004} show that as the mutual inclination of the
circular orbits of two equal mass planets increases,
the critical Hill stability limit becomes a progressively
worse indicator of the separations at which planets
may actually become Lagrange stable.  Their Fig. 5 illustrates
that for $i_m = 35^{\circ}$, instability occurs effectively
for values under half of the critical separation.  However,
their numerical simulations were run for just 2 Myr, almost certainly
missing instances of longer-term instability.

Therefore, determining how mass loss affects the stability
prospects of the orbits of two planets is perhaps more
complex than just considering the analytic effect on the Hill stability
limit.  Hence, we now turn to N-body simulations.

\section{N-body Simulations}

Ideally, we could sample the entire two-planet/single-star phase space with
numerical simulations.  Realistically, we must take measures
to restrict our studies to computationally feasible and insightful
simulations.   To better understand how to restrict the phase space,
we consider typical MS lifetimes, the mass lost at each
evolutionary phase, and the planetary period enhancement during the
WD phase, in Subsections 5.1-5.3.  We motivate and state 
our initial conditions in Subsection 5.4.  Subsections 5.5 and 5.6 present 
the simulation results.

\subsection{Main Sequence Timescales}

\begin{figure}
\centerline{
\psfig{figure=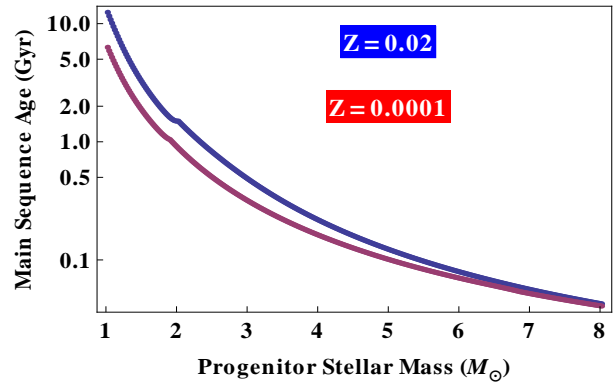,width=8.5cm,height=5.5cm} 
}
\caption{
The MS age of stars with Solar metallicity (blue, top curve)
and with a very low (Z=0.0001) metallicity (red, bottom curve).
Higher mass progenitors significantly reduce the CPU time needed
to integrate planets over a star's entire MS lifetime.
}
\label{MSage}
\end{figure}

\begin{figure}
\centerline{
\psfig{figure=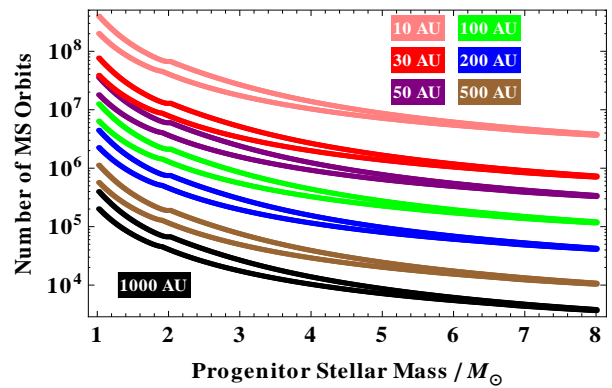,width=8.5cm,height=5.5cm} 
}
\caption{
The number of orbits taken by an isolated planet around a
star throughout its MS phase.  In each pair of 
equivalently-coloured curves, the top curve
is for Z$=0.02$ and the bottom curves is for Z$=0.0001$.
The number of orbits decrease with higher stellar mass
because the decreased MS timescale
dominates the shortened orbital period.
These values are important both physically -- to determine
instability -- and computationally -- to assess the feasibility
of integrating over the entire main sequence phase.
}
\label{MSorbits}
\end{figure}

MS ages are given in Fig. \ref{MSage}.  These ages
may vary by Gyrs depending on stellar metallicity, and are at least 1 Gyr
long for any progenitor mass less than $2 M_{\odot}$.  With 
a MS lifetime of over $10$ Gyr, a Solar-mass, Solar-metallicity
star with orbiting planets is particularly prohibitive
to integrate.  This long timescale helps explain the uncertainties in 
long-term evolution of the Solar System planets
\citep{khokuz2007,lasetal2011}.  Figure \ref{MSorbits} provides
estimates for the number of planetary orbits that would
be achieved during the MS as a function of
stellar progenitor mass, for a variety of planetary semimajor
axes.  The curves result from the competition between
the decreased orbital timescale with increased stellar mass, and 
the decreased MS lifetime with increased stellar mass; the 
latter overwhelmingly wins. 
Note the number of orbits tails off significantly as $M_{\star}(0)$
is increased, for all semimajor axes.  A one order-of-magnitude
change in semimajor axis corresponds to 1.5-order-of-magnitude 
change in the number of MS orbits.

\subsection{Post-Main Sequence Phase Properties}

\begin{figure}
\centerline{
\psfig{figure=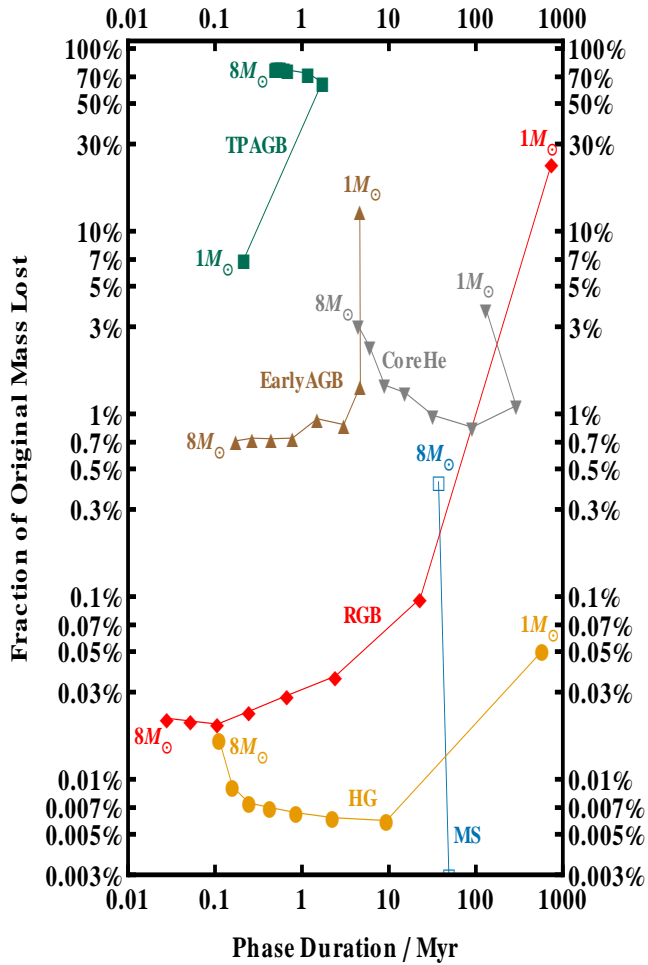,width=9cm,height=13.5cm} 
}
\caption{
Correlating mass loss fractions, phase durations and progenitor
masses for stars of Solar metallicity and a Reimers mass loss
coefficient of 0.5.  Each curve contains 8 symbols representing
stellar progenitor masses of $8 M_{\odot}$, $7 M_{\odot}$, 
$6 M_{\odot}$, $5 M_{\odot}$, $4 M_{\odot}$, $3 M_{\odot}$, 
$2 M_{\odot}$ and $1 M_{\odot}$, ordered monotonically.
Green squares represent the
Thermally Pulsing Asymptotic Giant Branch (TPAGB),
brown upward-pointing triangles the Early Asymptotic 
Giant Branch (EAGB), gray downward-pointing triangles
the Core Helium Burning phase (CoreHe), red diamonds
the Red Giant Branch (RGB), yellow circles the Hertzprung
Gap (HG), and the blue open squares the MS.
Most MS mass loss is too small for this plot.
Except for the $1 M_{\odot}$ case,
the most important mass loss phases are the TPAGB, which
all last on the order of $1$ Myr.
}
\label{SSEMass}
\end{figure}

The evolution timescales of the intermediate post-MS, 
pre-WD phases are short compared to the MS timescale.
Figure \ref{SSEMass} provides timescales for each stellar 
phase, and relates the phase to the percentage of the 
star's {\it original} mass lost, for Solar-metallicity 
stars.  The plot demonstrates that except for
the $1 M_{\odot}$ case, most mass is lost during the TPAGB
and negligible percentages of mass are lost in the other phases.

\begin{figure}
\centerline{
\psfig{figure=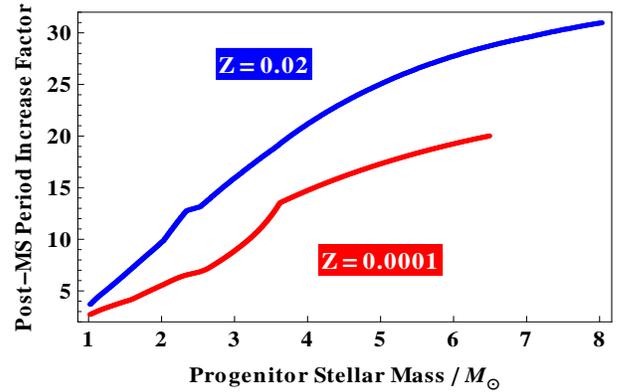,width=8.5cm,height=5.5cm} 
}
\caption{
The planetary period enhancement factor due to post-MS
evolution, supposing that the planet has evolved entirely adiabatically
and is isolated from any other perturbations.
This enhancement factor is independent of semimajor axis.
The top curve is for Z$=0.02$ and the bottom curve is for Z$=0.0001$.
}
\label{postMSenhance}
\end{figure}

\subsection{White Dwarf Period Enhancement}

After the star has become a WD, the star stops losing
mass and gradually cools down.  Compared to its MS mass, the WD 
mass is greatly reduced, and cannot exceed the Chandrasekhar
Limit of $\approx 1.4 M_{\odot}$.   The result is that the planetary
period may be drastically increased.  Assume that the planet's
evolution is entirely adiabatic.  A reduction of the star's mass by a 
factor of $k$ will cause the planet's semimajor axis to be increased
by a factor of $k$.  Hence, the planetary period around the WD is $k^2$ 
times the period around the MS star.
Figure \ref{postMSenhance} plots this enhancement factor as a function
of progenitor stellar mass, and demonstrates both that
WD planets perform fewer orbits than MS counterparts in the
same amount of time (with possible implications for scattering) 
and that WD numerical simulations may proceed much more 
quickly than MS simulations.

\subsection{Initial Conditions}

\subsubsection{Fiducial Choices} \label{subsec:fiducial}

The above considerations lead us to choose
an integration duration of 5 Gyr and
Solar-metallicity progenitor masses of $8 M_{\odot}$,
$7 M_{\odot}$, $6 M_{\odot}$, $5 M_{\odot}$, $4 M_{\odot}$
and $3 M_{\odot}$.  This mass range extends
down to the upper mass-end of the observed 
range of exoplanet host stars \citep{satetal2012b}.
This combination allows us to sample an ensemble of multi-planet
systems over every phase of evolution, including a substantial
sampling (over 4.5 Gyr) of evolution in the WD phase.
Simulation output occurs at a frequency of 1 Myr.
Performing a statistically significant simulation ensemble 
for $1 M_{\odot}$ stars is well-beyond our available resources;
for more details on the planetary consequences of the
possible evolutionary tracks of $1 M_{\odot}$ stars, see \cite{verwya2012}.

We adopt one Jupiter mass for the mass of each planet
($M_1 = M_2 = M_J$), assume the planets are on coplanar
orbits ($i_m = 0^{\circ}$), 
and assume they have small but
non-negligible MS eccentricities ($e_1(t=0) = e_2(t=0) = 0.1$).
These eccentricities are low compared to the observed MS values
for massive exoplanets beyond the tidal circularlization limit,
but higher than the near-circular orbits predicted by
core accretion theory.  The inner planet is initially set 
at $a_1 = 10$ AU to avoid
tides with the expanding stellar envelope; over 
25 known planets are thought to harbor $a > 10$ AU
\footnote{See the Extrasolar Planet Encyclopedia at  
http://exoplanet.eu/}.  Also, this semimajor axis 
guarantees that orbital evolution due to mass loss
will be adiabatic unless in the presence of an event 
such as a supernova, which is not modelled here.

We perform 632 simulations per ensemble, where each
ensemble features a different progenitor mass.  
In each individual simulation, the orbital angles 
(mean anomalies and longitudes of pericenter) of both 
planets are selected from a uniform random distribution.  
We adopt 79 values of $a_2$ so that we sample 8 different 
sets of orbital angles for each $a_2$ value. 
The range of $a_2$ values sampled encompasses both
the ``chaos limit'' and the Hill stability limit
in order for us to sample many different types of dynamical behaviour.
 
The chaos limit refers to a maximum semimajor axis 
ratio separation at which mean motion resonances do not 
necessarily overlap.  This limit is smaller than the Hill 
and Lagrange stability limits, and represents a fuzzy boundary within 
which instability occurs readily and quickly.  
\cite{wisdom1980} found the chaos limit to be

\begin{equation}
\frac{a_2 - a_1}{a_1} \approx 1.3 \left(\frac{M_p}{M_{\star}}\right)^{2/7}
\label{wisdom}
\end{equation}

\noindent{for} two equal mass circular-orbit planets.  Recently,
\cite{quifab2006} and \cite{muswya2012} expanded on this result
by considering bodies' eccentricity. \cite{muswya2012} 
discovered that for $e_m > 0.21 (M_p/M_{\star})^{3/7}$, where $e_m$
is the eccentricity of the least massive planet, the
chaos limit is (for $M_p = M_1 = M_2$):

\begin{equation}
\frac{a_2 - a_1}{a_1} = 1.8 e_{\mathrm{m}}^{1/5} \left(\frac{M_p}{M_{\star}}\right)^{1/5}
.
\end{equation}

We set our minimum value of $a_2$ to be less than
the limit from the more conservative definition (Eq. \ref{wisdom}) 
for each progenitor mass to help ensure that we have a tail of unstable simulations.  
Similarly, we wish to have a tail of stable simulations
for large separations.  Hence, we select a maximum value of $a_2$ 
that exceeds the MS Hill stability limit in each case.  


\subsubsection{Additional Simulations}
 
Motivated by the outcome of our fiducial simulations, we
performed two additional ensembles of simulations 
(632 simulations per ensemble).  In both, we adopted
a stellar progenitor mass of $5 M_{\odot}$. The first case 
assumed different planetary masses; we adopted 1 Earth-mass
for each planet ($M_1 = M_2 = M_{\oplus}$).  In the second,
we adopted $e_2 = 0.5$ to model a moderately
eccentric outer planet.  Doing so yields a much wider Hill
stability separation (see Fig. \ref{hille1}) than in the fiducial case.

\subsection{Simulation Results: Fiducial Cases}

\subsubsection{Overview}

Our goal is to identify instability and when it occurs.
We define instability as Lagrange instability:
if the planets at any point are found to achieve
a hyperbolic orbit, cross orbits, or collide with each other
or the star.  Hill instability includes just a few
of these possibilities: planet-planet collisions and crossing orbits.
Therefore, Hill stable systems may eventually be Lagrange unstable.
Those that do will feature ejection of the outer planet
and/or collision of the inner planet with the star.

We plot instability time vs. initial semimajor
axis ratio in Fig. \ref{Main}, which represents our main result.  
Dots indicate unstable systems. No dot appears
for a system that has remained stable over the entire 5 Gyr 
integration. If all 8 simulations for a given semimajor axis
ratio are stable over 5 Gyr, then we place a blue star at $10^{10}$
yr in the appropriate horizontal position, even though the 
vertical position of the star has no physical meaning and is
selected for visual impact.

The figure includes the six
ensembles of simulations with different progenitor
stellar masses.  Each post-MS phase change occurs
at different times on each plot.
See \cite{huretal2000} for detailed physical 
descriptions of each phase.
Although the horizontal lines are close together in
Fig. \ref{Main}, they are clearly distinguishable
on the zoomed-in Fig. \ref{Zoom} plots.  Different progenitor
masses also cause differences in the location
of the Hill stability limit.
On each plot is a black vertical dashed line, representing the 
Hill stability limit computed from the star's MS mass.  The 
black dotted lines represent the Hill stability limits computed 
with the star's WD mass (see the Fig. \ref{hille1} caption for these values).

\subsubsection{Physical Description of Figure \ref{Main}}

The Hill Stability limits and the onset of post-MS evolution 
provide boundaries within which one can understand the different
regions in the Fig. \ref{Main} plots:

1) During the MS, represented by the region
under all the coloured horizontal lines, dots appear predominately
inside of the MS Hill stability limit and predominately at times
under $10^7$ yr.  Hence, the limit is useful for identifying
short-term instability.

2) Some dots appear on the MS but outside of the MS Hill stability limit 
in the plots with
progenitor stellar masses of $5 M_{\odot}$, $4 M_{\odot}$ and $3 M_{\odot}$.
All these dots indicate long-term instability (occurring after $\sim 10^7$ yr).
The long-term MS unstable simulations
with initial separations exceeding the Hill stability limit
must be (and indeed are) Lagrange unstable: where the outer planet is ejected
and/or the inner planet collides with the star.  As the 
progenitor mass is decreased, the number of unstable MS systems beyond 
the MS Hill stability boundary increases.  
One possible reason is because
the longer MS lifetimes (see Fig. \ref{MSage}) translate into more 
orbits for the planets (see Fig. \ref{MSorbits}), and hence a greater 
opportunity for instability to occur.  Another potential
reason is that smaller planet-star mass ratios broaden the boundaries
between Hill and Lagrange instability.

3) MS instability beyond the MS Hill stability limit appears
to extend only as far as the WD Hill stability limit.  However,
this must be coincidence -- due to our choice of initial conditions -- as 
the planetary system has no knowledge of the post-MS mass loss that will occur.

4) The WD Hill stability limits ensure
that any WD instability occurring beyond this limit must be Lagrange
instability.  Our simulations corroborate this statement.

5) Each plot in Fig. \ref{Main} demonstrates that all systems become
Lagrange stable for $a_2/a_1 \gtrsim 1.55$, just inside the 2:1 commensurability.  
Reinforcing this estimate
are blue stars which were excluded from the plot (for visual clarity) that
extend all the way out in an unbroken chain to 
$a_2/a_1 \approx [1.779,1.775,1.770,1.765,1.757,1.747]$ for
$M_{\odot}(0) = [8M_{\odot},7M_{\odot},6M_{\odot},5M_{\odot},4M_{\odot},3M_{\odot}]$.
This sampled range helps to establish the robustness of
the Lagrange stability boundary for our chosen 5 Gyr integration duration.  
This boundary lies at a 
distance corresponding to approximately $[178\%,176\%,176\%,172\%,167\%,163\%]$ 
of the MS Hill stability limit and $[138\%,134\%,133\%,130\%,126\%,123\%]$
of the WD Hill stability limit.

In all our cases, planets with initial 
separations that exceed the $2$:$1$ commensurability are stable throughout
the 5 Gyr integration\footnote{The strong, first-order $2$:$1$ mean 
motion commensurability perhaps plays a role in establishing the 
Lagrange stability boundary (see, e.g., \citealt*{bargre2007}) for our fiducial
cases.  }.  
However, the Lagrange stability limit may be higher than the values
reported here for progenitor masses lower than those sampled here.  
Especially for $1 M_{\odot}$ stars, the outcome
is uncertain, given the long MS lifetime and strong mass loss over both
the RGB and AGB (see Fig. \ref{SSEMass}).  The distribution of unstable
systems in Fig. \ref{Main} shows a wide variation of instability times,
and suggests that longer simulations, perhaps out to the age of the Universe,
could yield additional instability.

\subsubsection{Description of Figures \ref{Zoom}-\ref{FracIns}}

Post-MS pre-WD phase changes can prompt instability, which can
be seen more clearly in Fig. \ref{Zoom}.  The phases
which cause the greatest mass loss (see Fig. \ref{SSEMass}) also
tend to be the phases which are most likely to trigger instability.
This tendency is indicated by the number of dots between the
TPAGB and WD lines versus the number of dots appearing below
the TPAGB line.  Note however, that the relatively long
length of the core helium burning phase helps to increase 
the number of unstable systems during that phase.
The abundance of dots just above the WD line indicates that the
TPAGB can unsettle stable systems enough to cause slightly delayed
instability.  In Fig. \ref{Zoom}, the WD Hill stability limit
appears to provide an effective boundary beyond which post-MS pre-WD
instability does not occur.  However, this apparent boundary again must be 
coincidence because the systems are unaware of forthcoming post-MS
mass loss.

\begin{figure*}
\centerline{
\psfig{figure=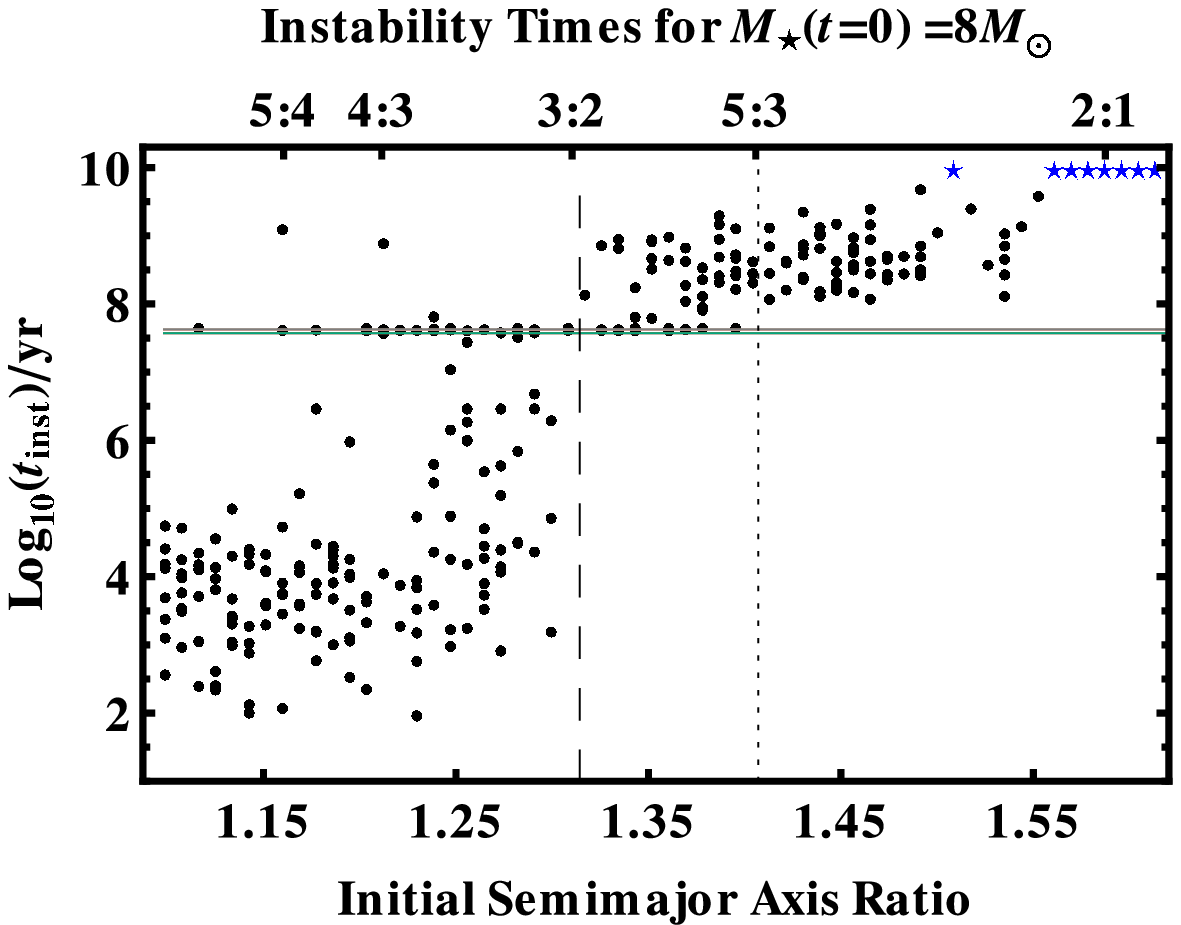,width=8.5cm,height=6.3cm} 
\psfig{figure=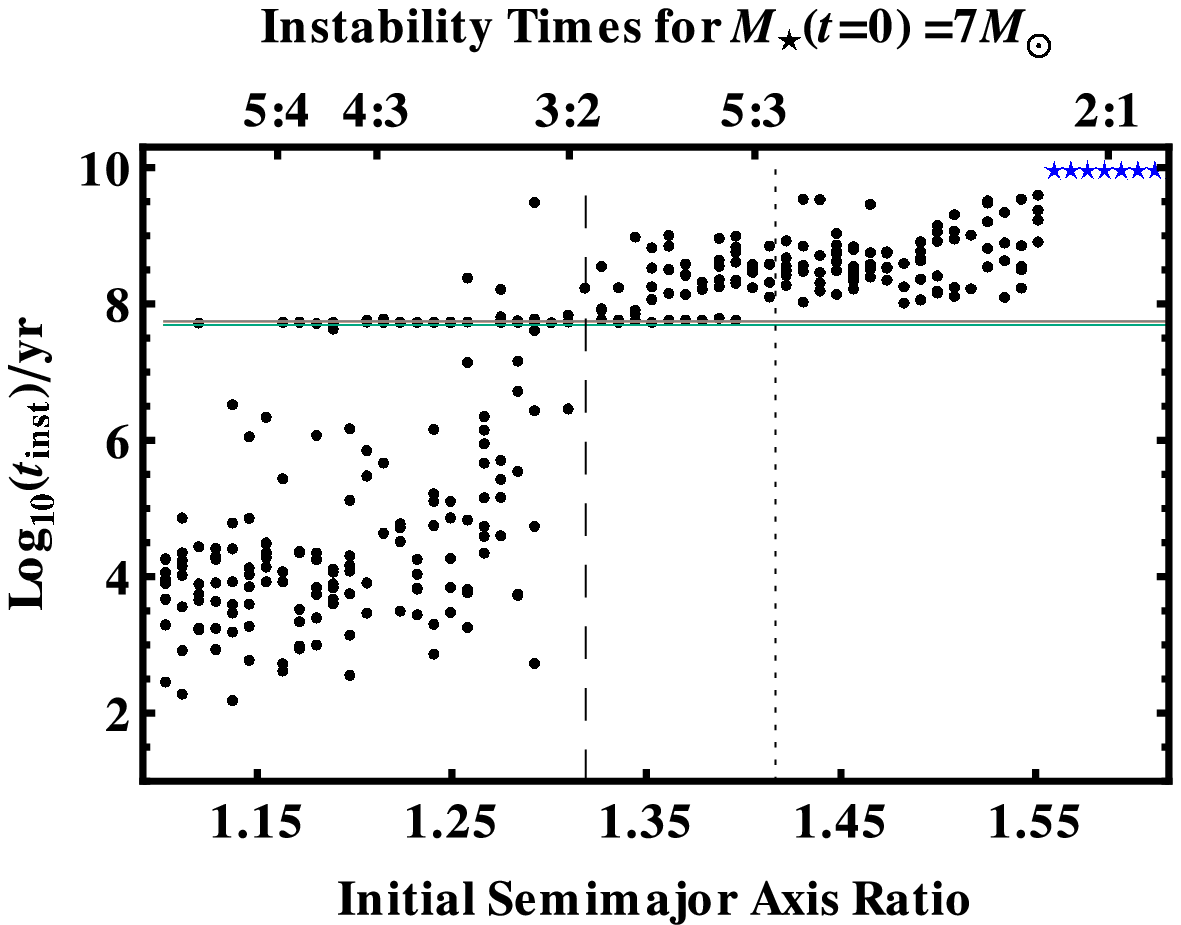,width=8.5cm,height=6.3cm}
}
\centerline{}
\centerline{
\psfig{figure=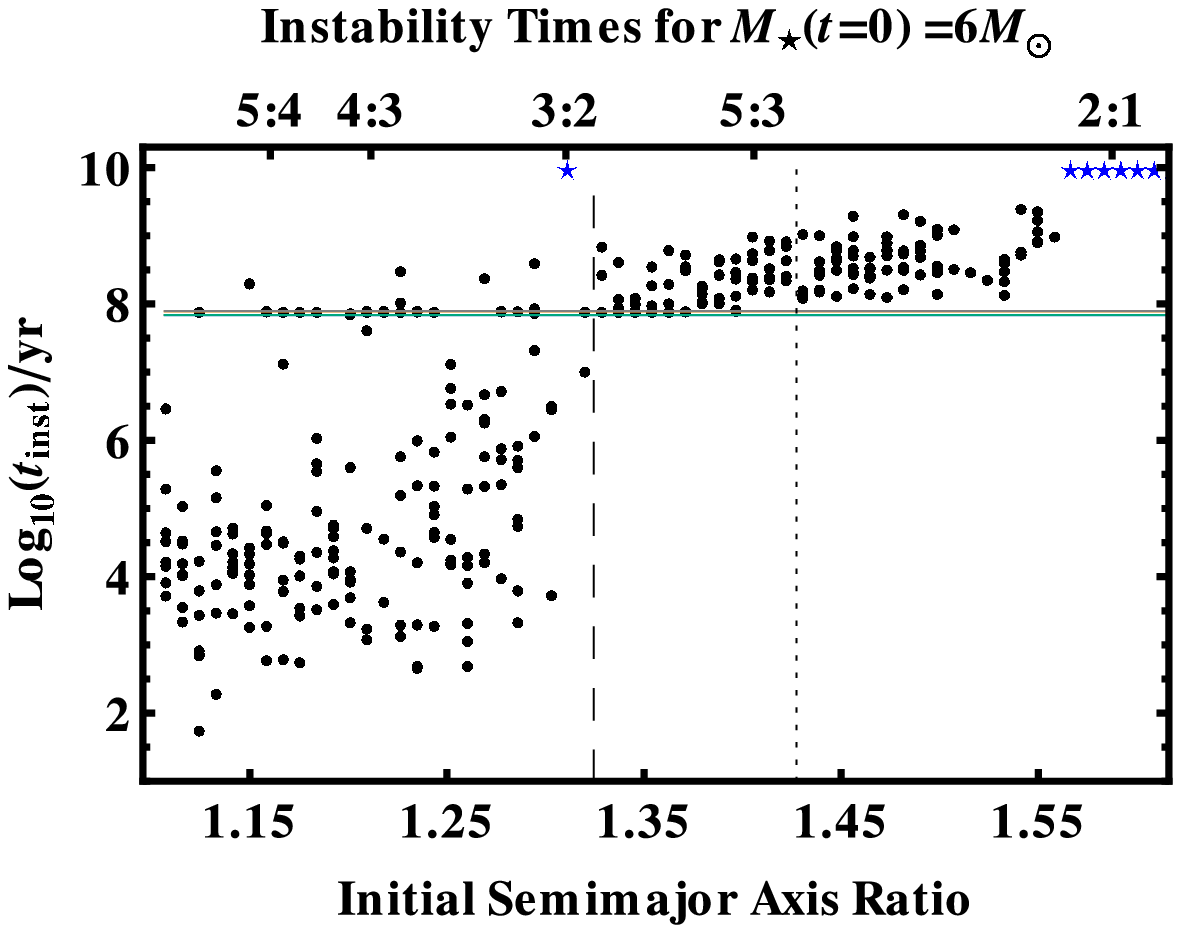,width=8.5cm,height=6.3cm} 
\psfig{figure=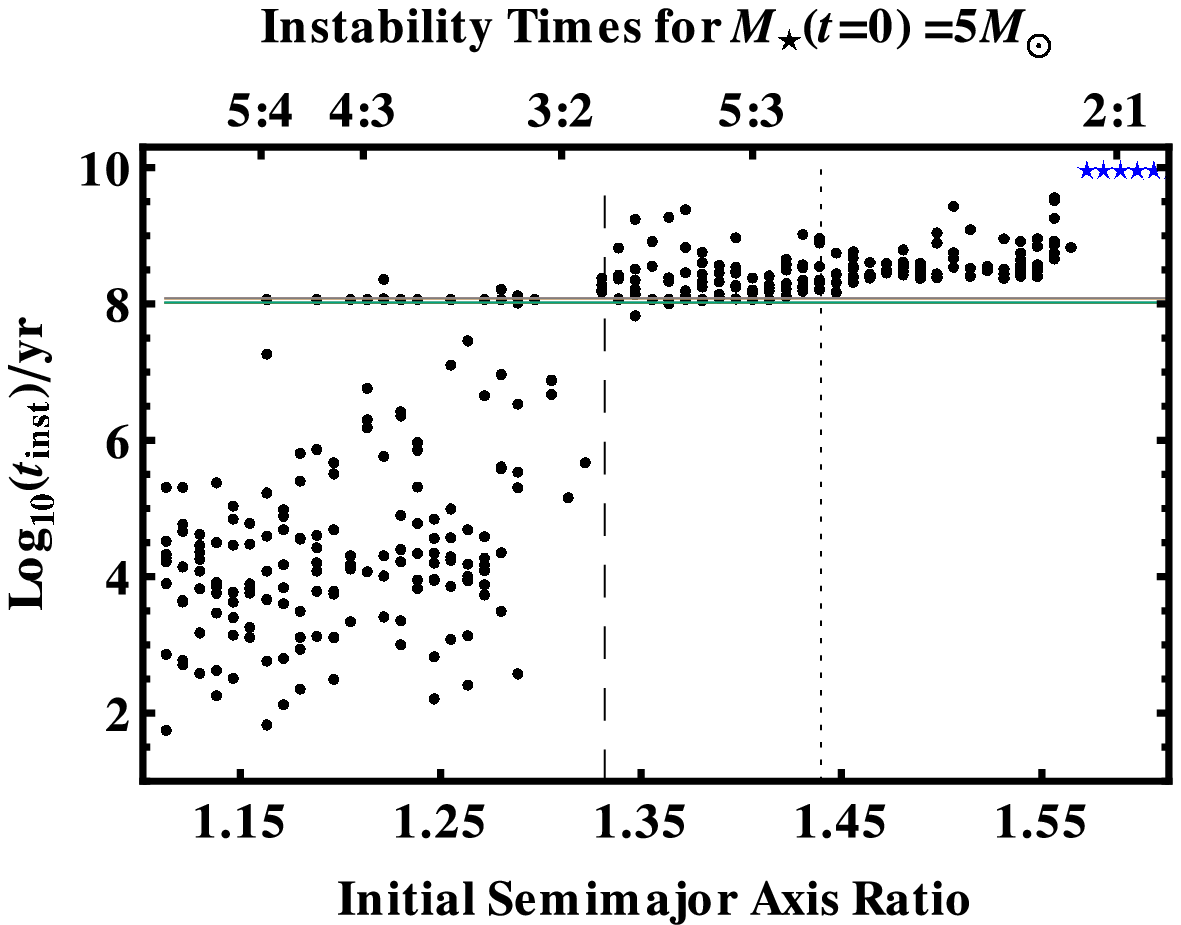,width=8.5cm,height=6.3cm}
}
\centerline{}
\centerline{
\psfig{figure=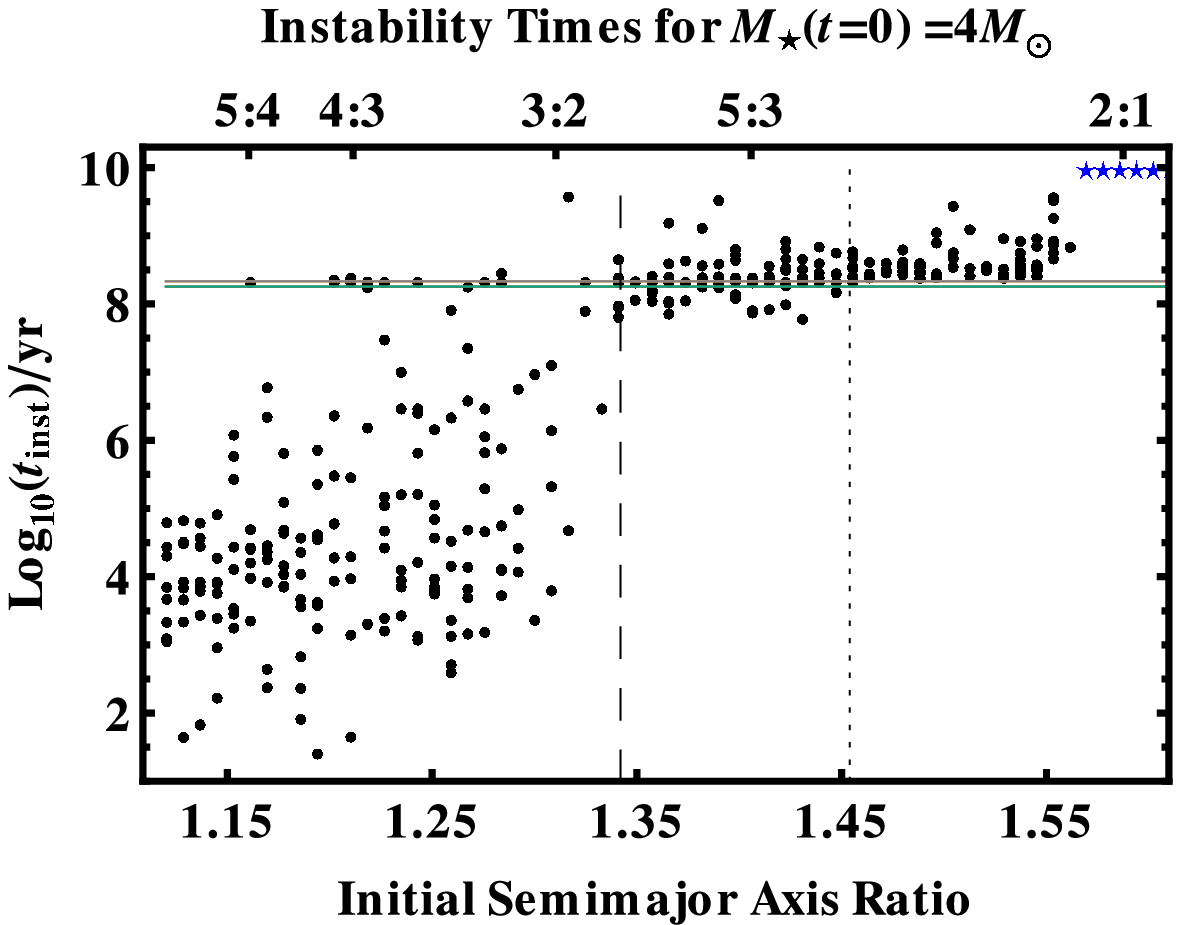,width=8.5cm,height=6.3cm} 
\psfig{figure=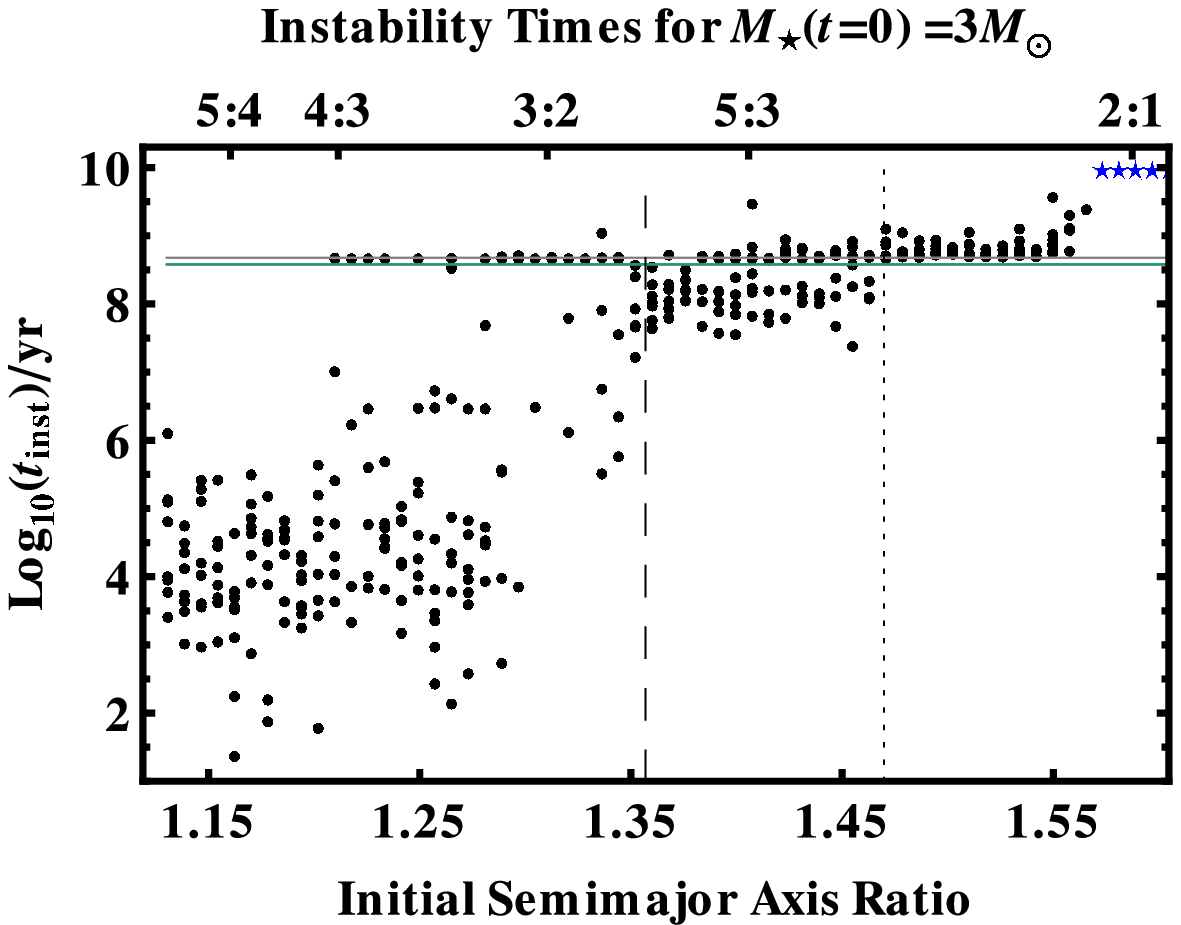,width=8.5cm,height=6.3cm}
}
\caption{
Instability times vs. initial semimajor axes ratios
for full-lifetime 2-planet simulations.
Dots indicate individual unstable systems, and blue stars indicate
all 8 systems at that separation ratio were stable over 5 Gyr.
Blue stars not shown extend out to at least $a_2/a_1 = 1.747$ 
in an unbroken string in each plot.  
Each coloured horizontal line represents a stellar phase change,
and is at a different position on each plot.
The two vertical lines represent the Hill
stability limit for the MS (dashed)
and WD phase (dotted) for each progenitor mass.  
Any unstable systems on the MS
beyond the MS Hill stability limit (such as for the
$5M_{\odot}$, $4M_{\odot}$ and $3M_{\odot}$ cases)
or during the WD phase beyond the WD Hill stability limit
are Lagrange unstable.  The plot demonstrates that instability 
during the WD phase can be achieved at separations that well-exceed both 
the MS and WD Hill stability limits.
}
\label{Main}
\end{figure*}

\begin{figure*}
\centerline{
\psfig{figure=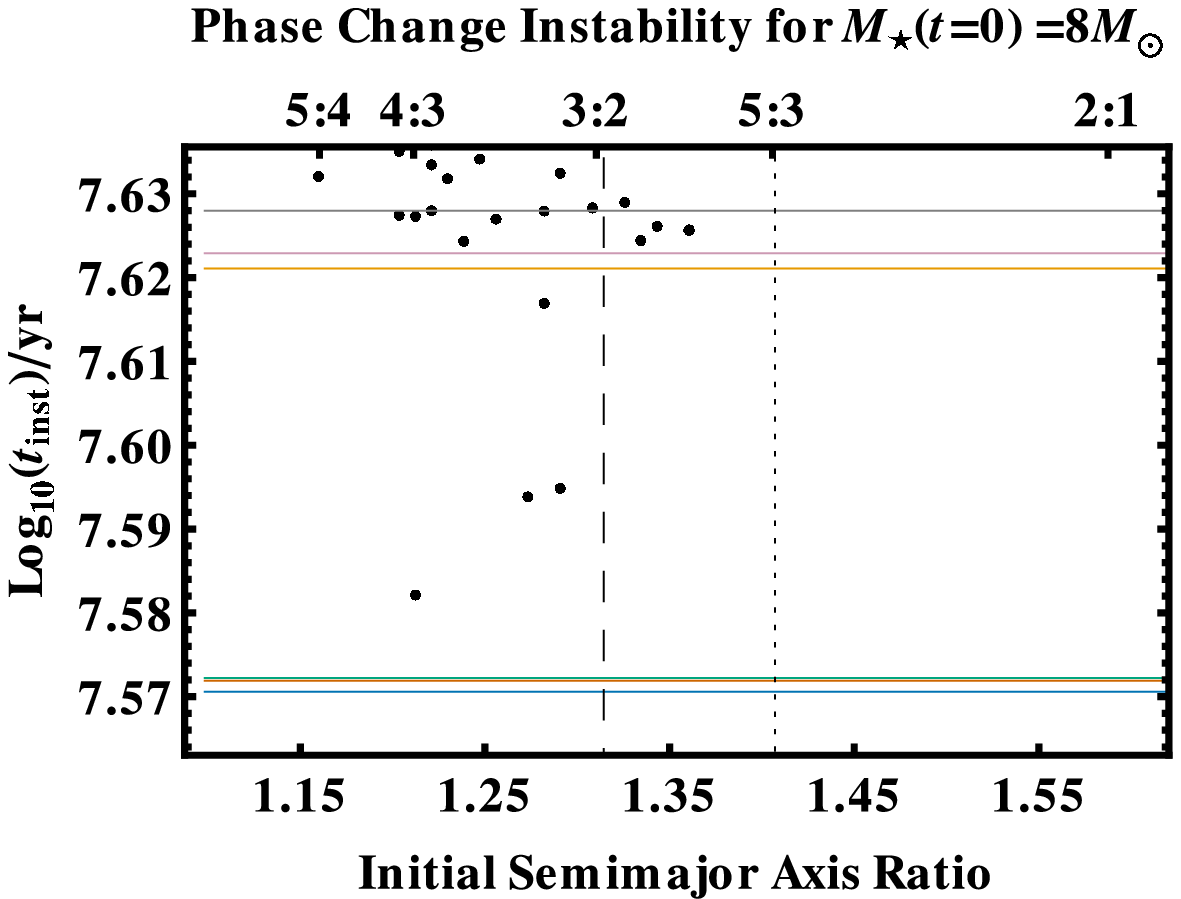,width=8.5cm,height=6.3cm} 
\psfig{figure=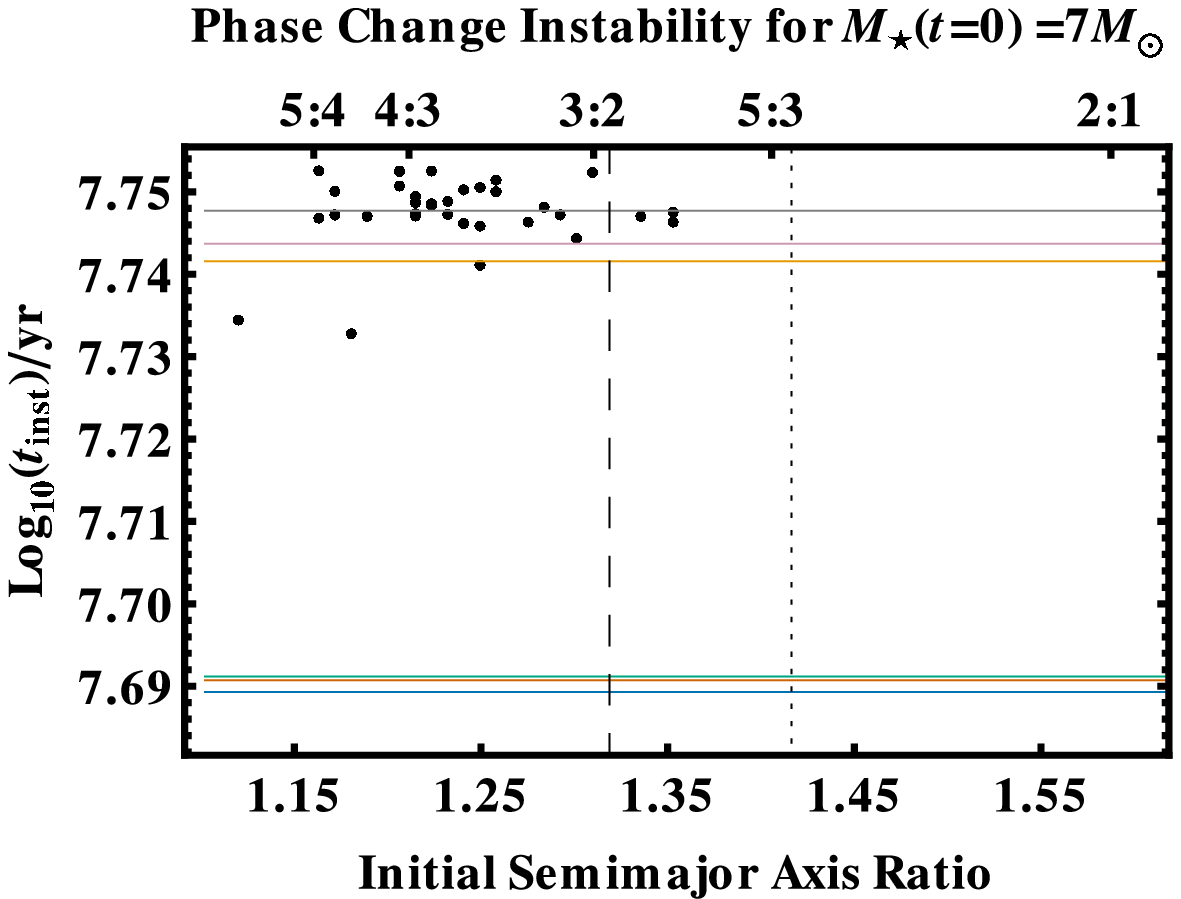,width=8.5cm,height=6.3cm}
}
\centerline{}
\centerline{
\psfig{figure=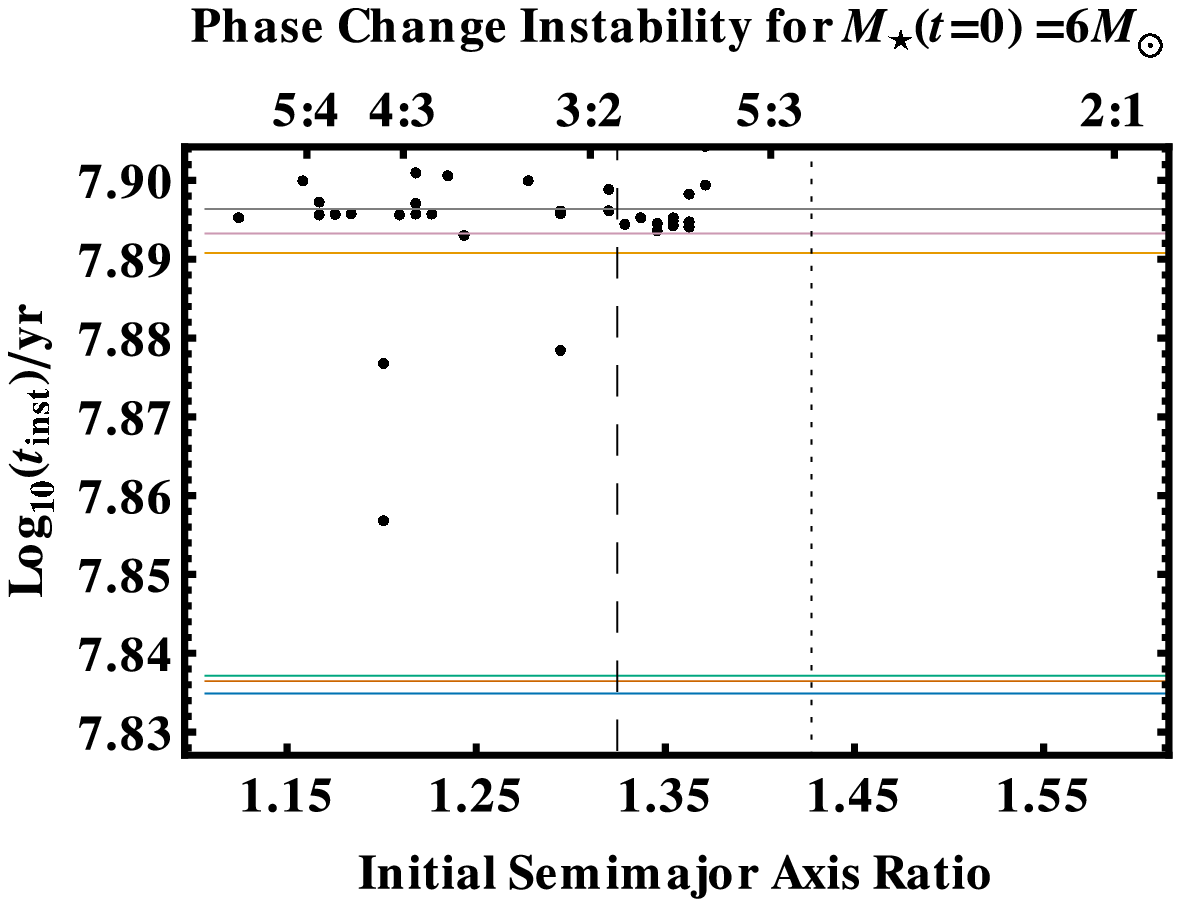,width=8.5cm,height=6.3cm} 
\psfig{figure=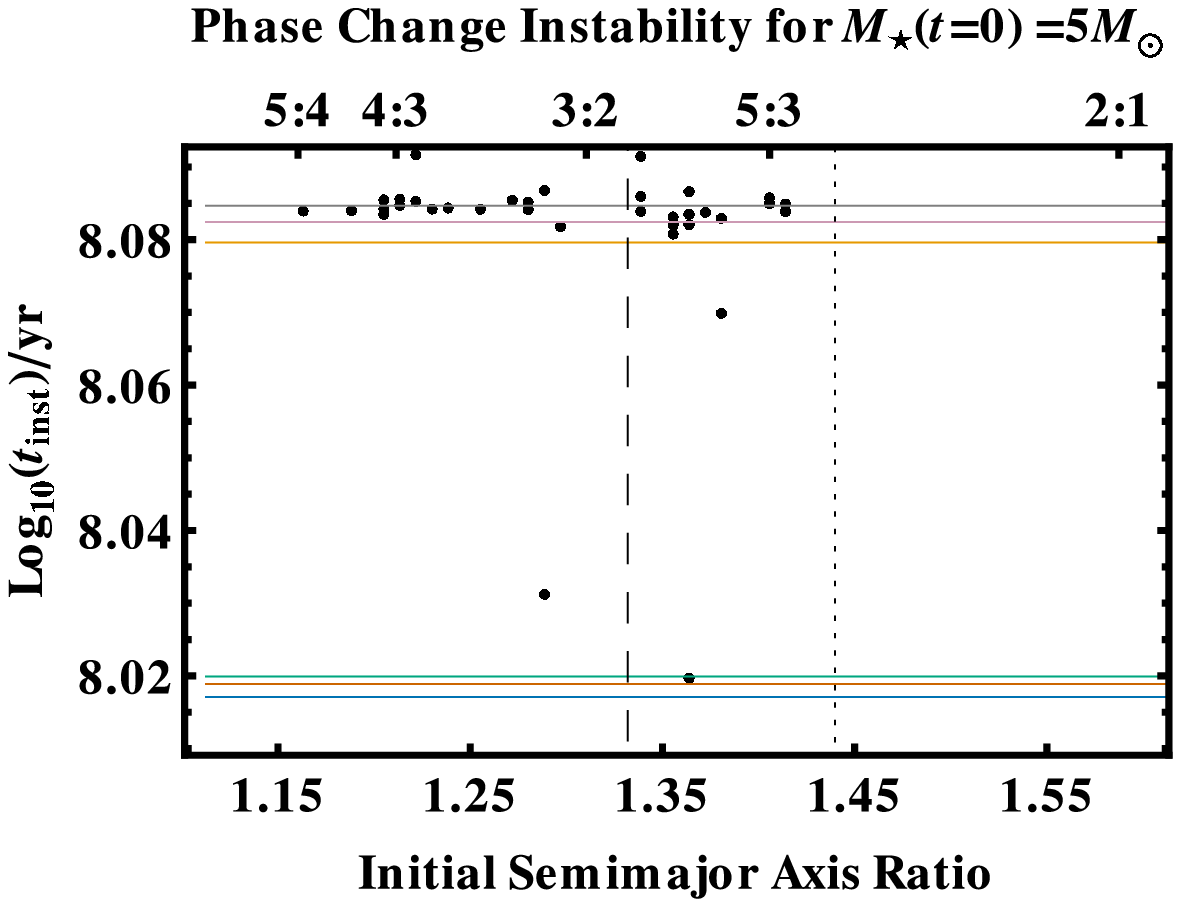,width=8.5cm,height=6.3cm}
}
\centerline{}
\centerline{
\psfig{figure=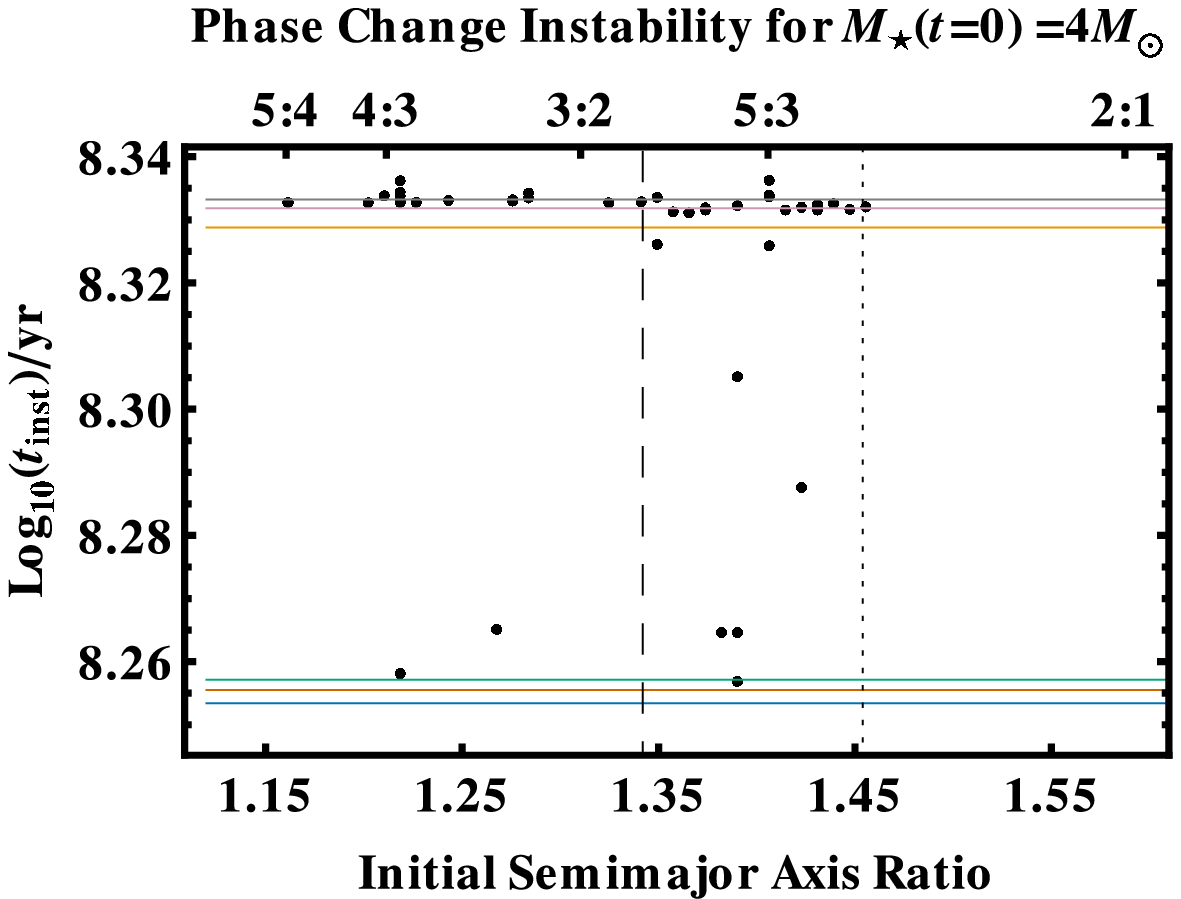,width=8.5cm,height=6.3cm} 
\psfig{figure=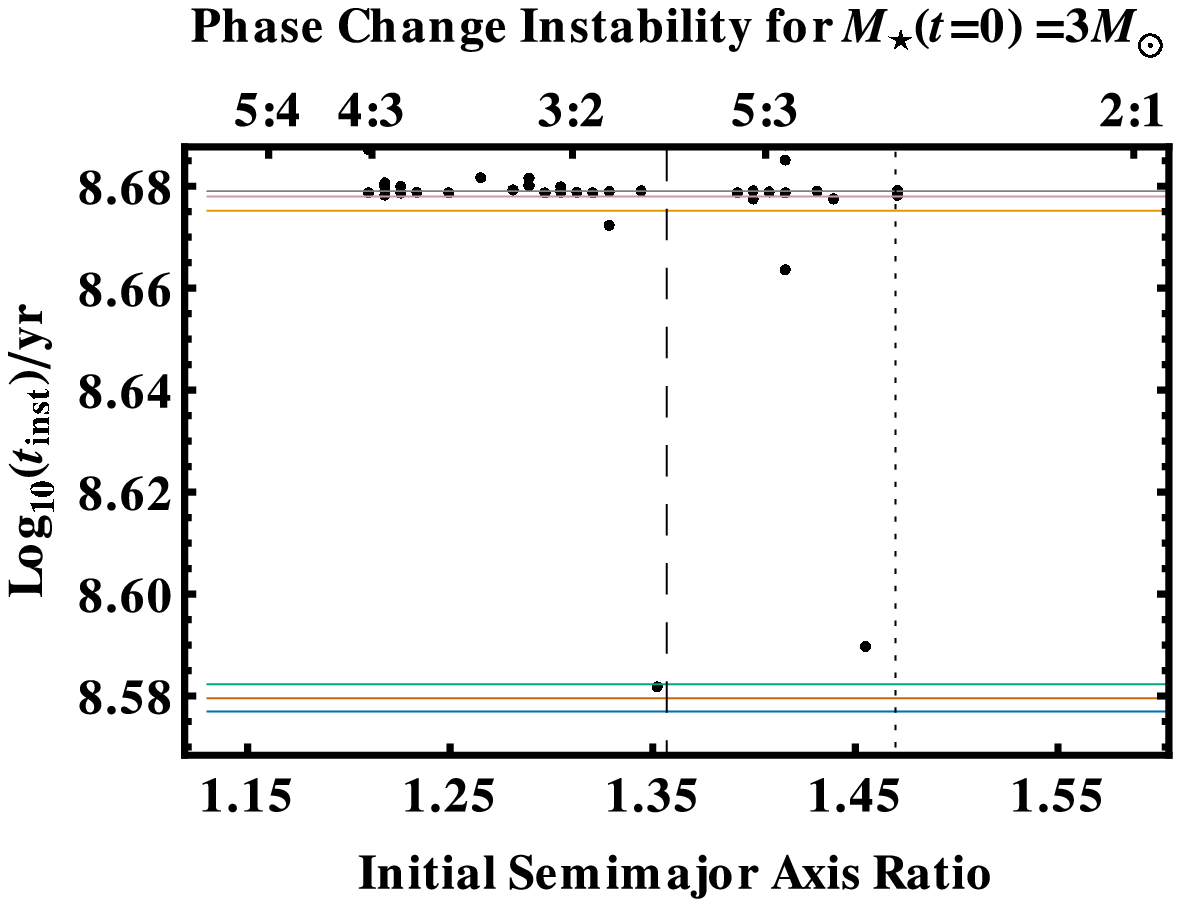,width=8.5cm,height=6.3cm}
}
\caption{
Zoomed-in versions of Fig. \ref{Main} to show
detail at times of stellar phase changes.
In ascending vertical order, the phases are ``Hertzprung Gap'' = blue; 
``Red Giant Branch'' = red; ``Core Helium Burning'' = green; 
``Early AGB'' = orange; ``TPAGB'' = purple; ``WD'' = gray.
The longest pre-WD post-MS phase is the 
Red Giant Branch; the most violent
phase (with the greatest mass loss, and causing
the greatest amount of instability) is the
TPAGB. The WD Hill Stability limit acts as an effective
empirical boundary for pre-WD post-MS instability.
}
\label{Zoom}
\end{figure*}

\begin{figure*}
\centerline{
\psfig{figure=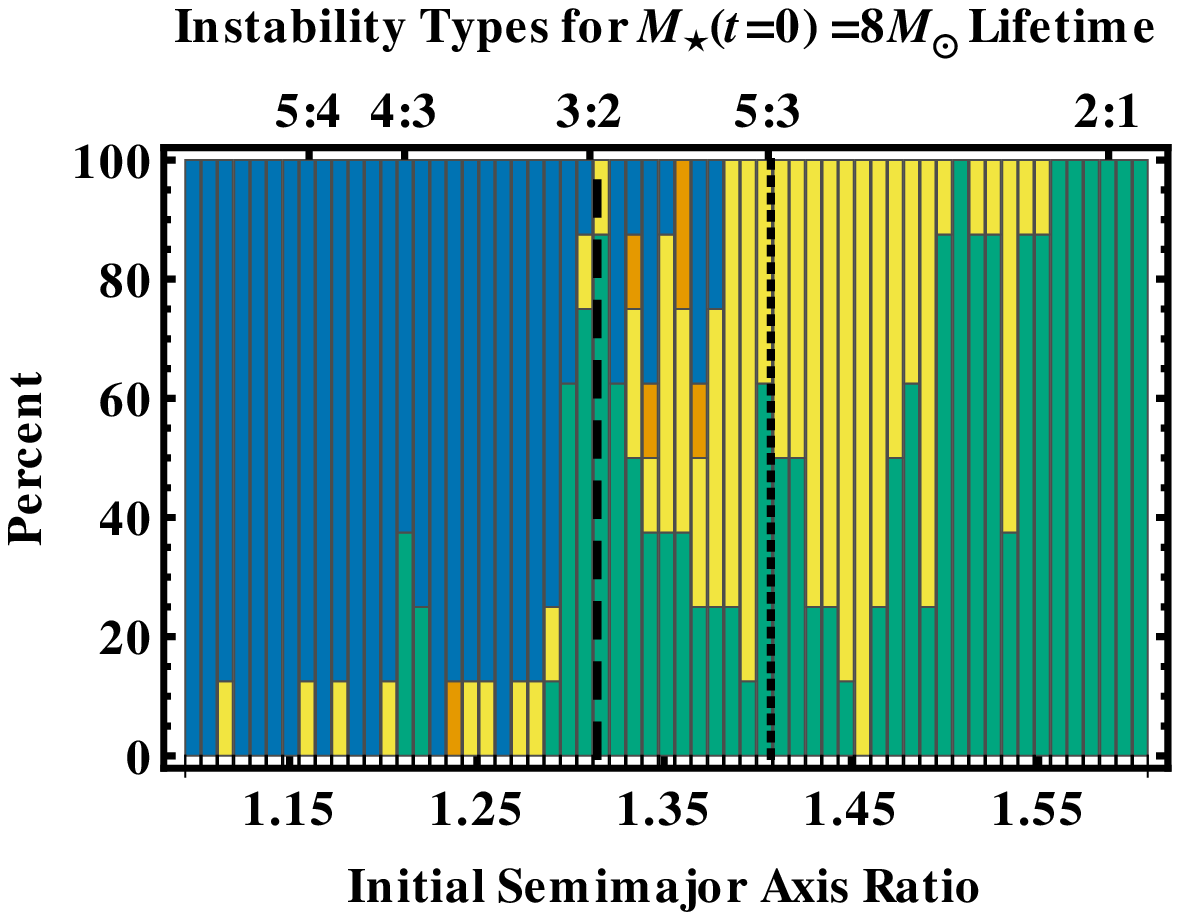,width=8.5cm,height=6.3cm} 
\psfig{figure=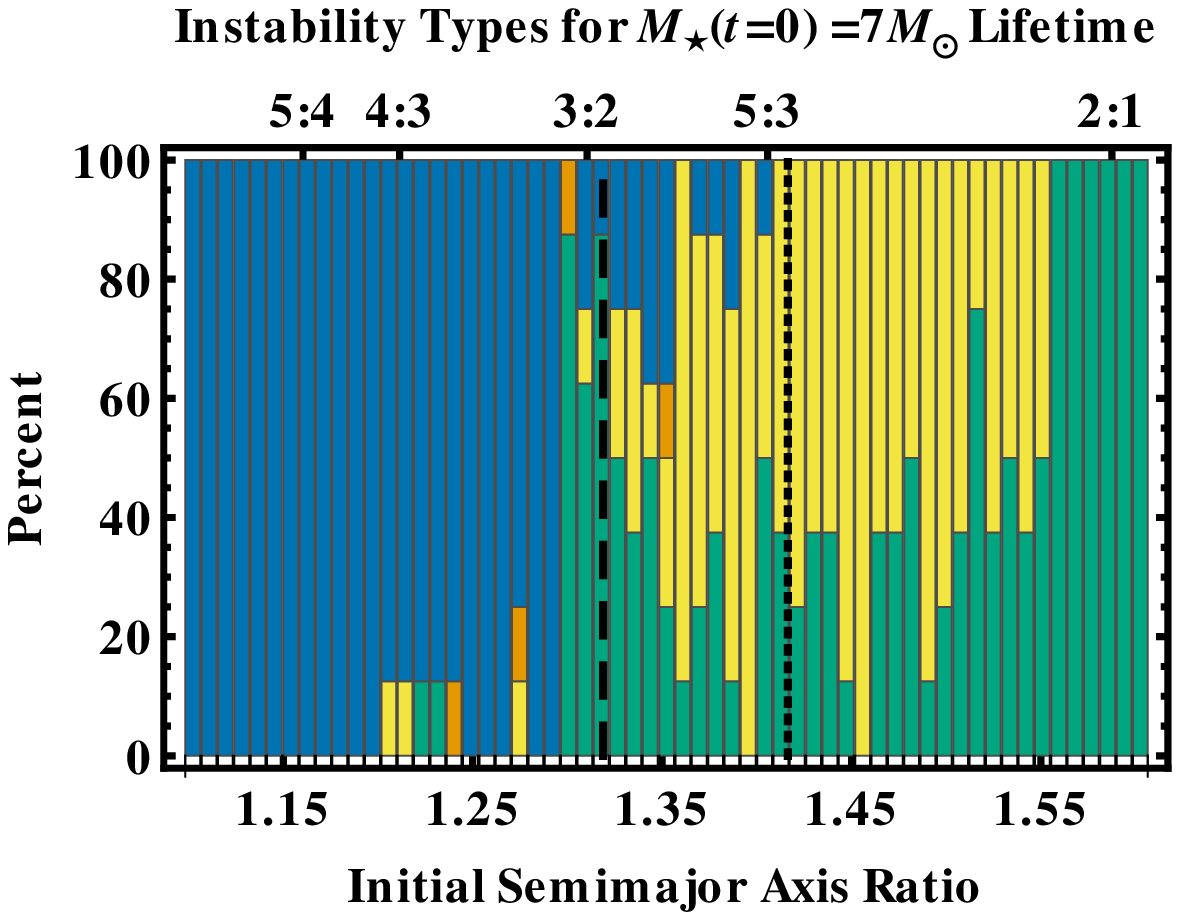,width=8.5cm,height=6.3cm}
}
\centerline{}
\centerline{
\psfig{figure=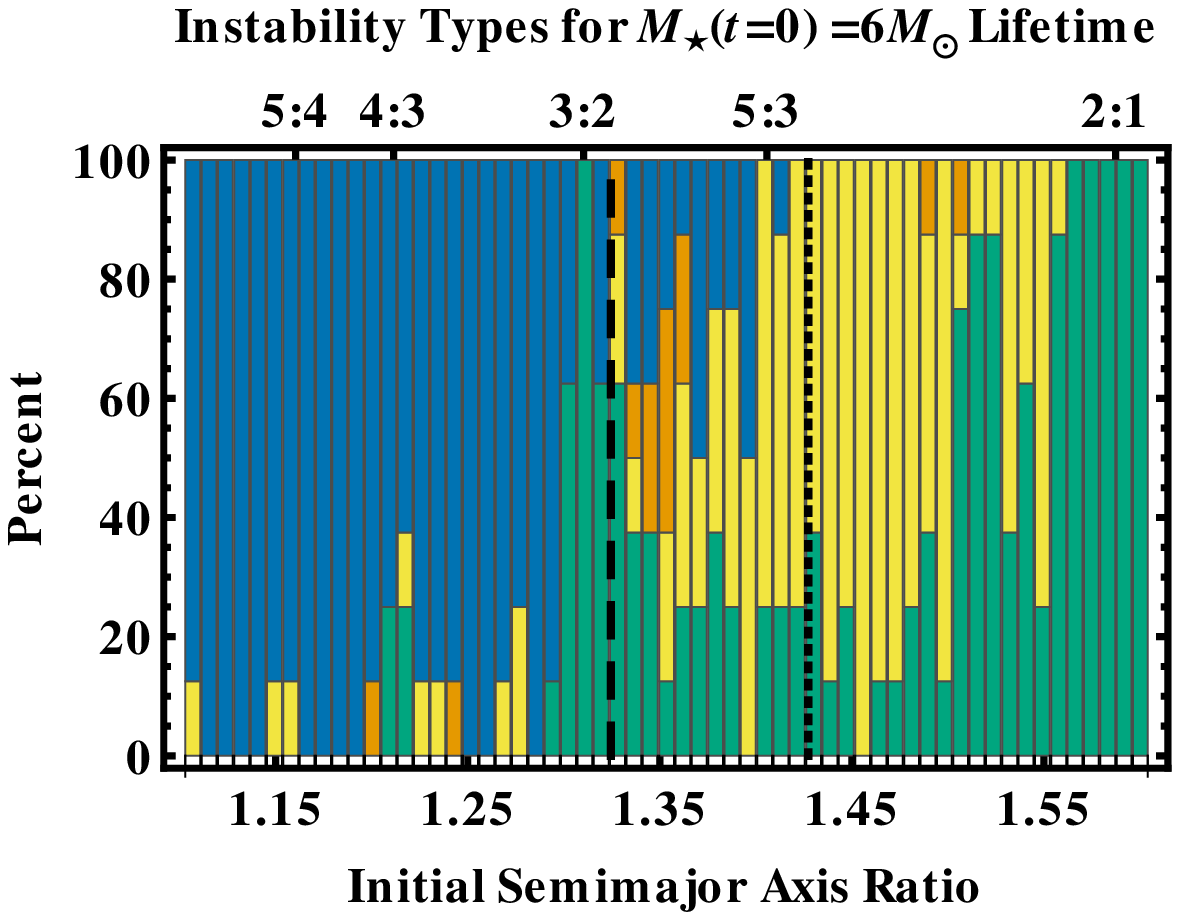,width=8.5cm,height=6.3cm} 
\psfig{figure=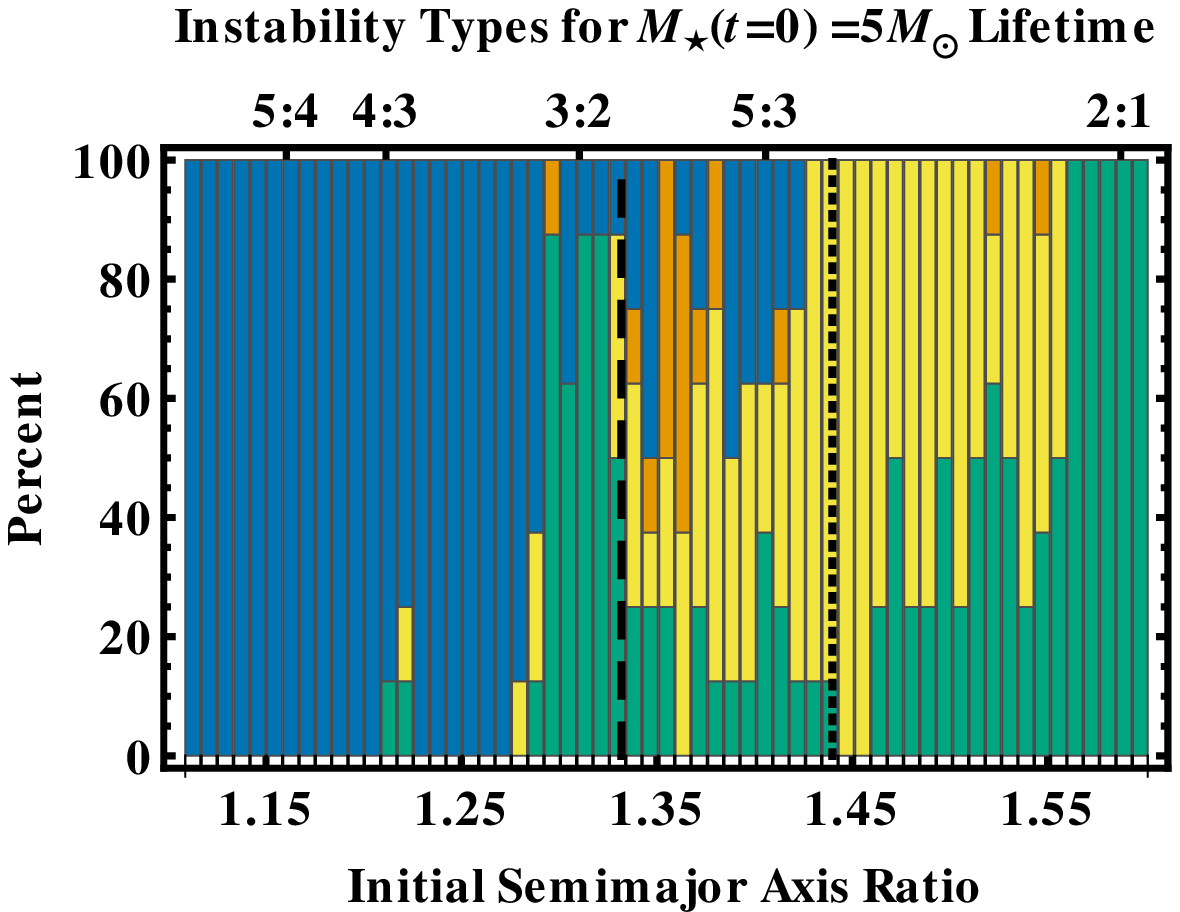,width=8.5cm,height=6.3cm}
}
\centerline{}
\centerline{
\psfig{figure=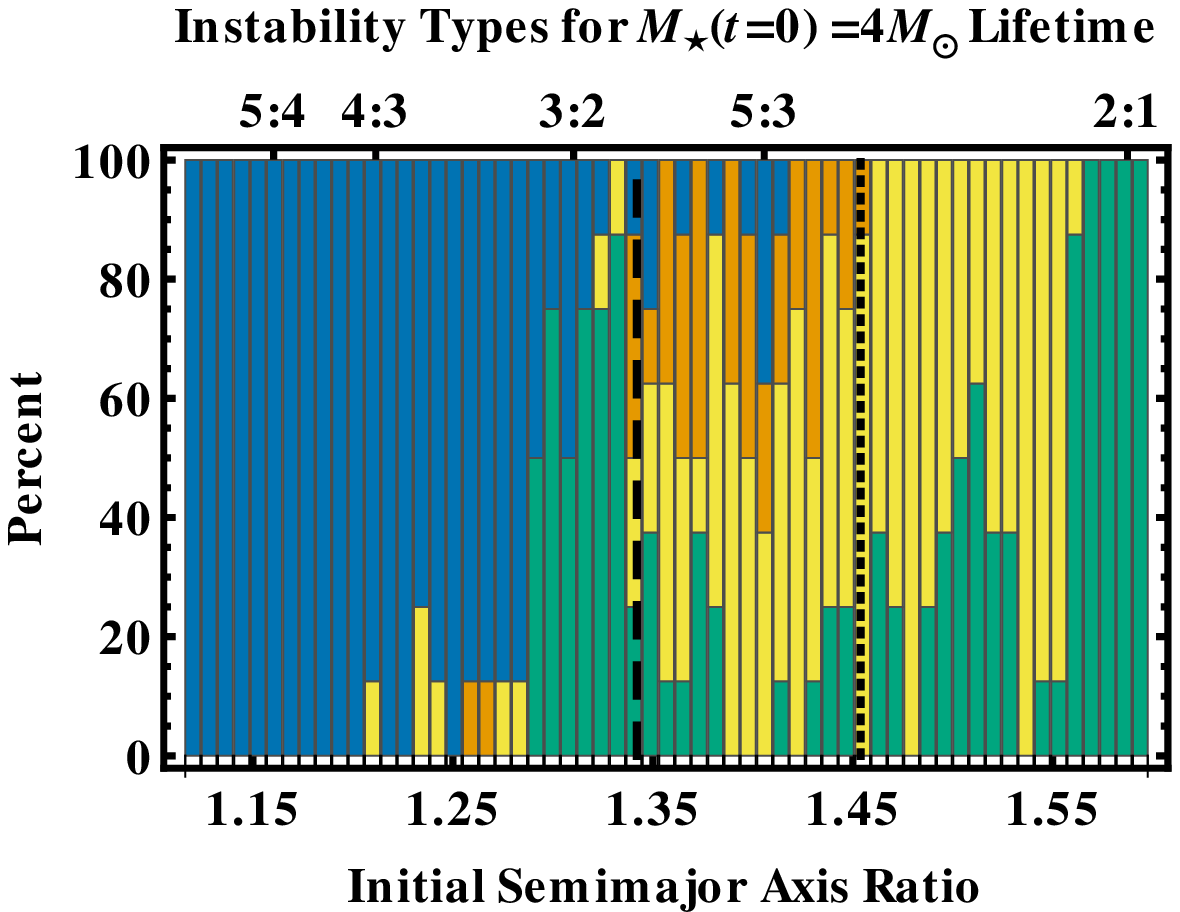,width=8.5cm,height=6.3cm} 
\psfig{figure=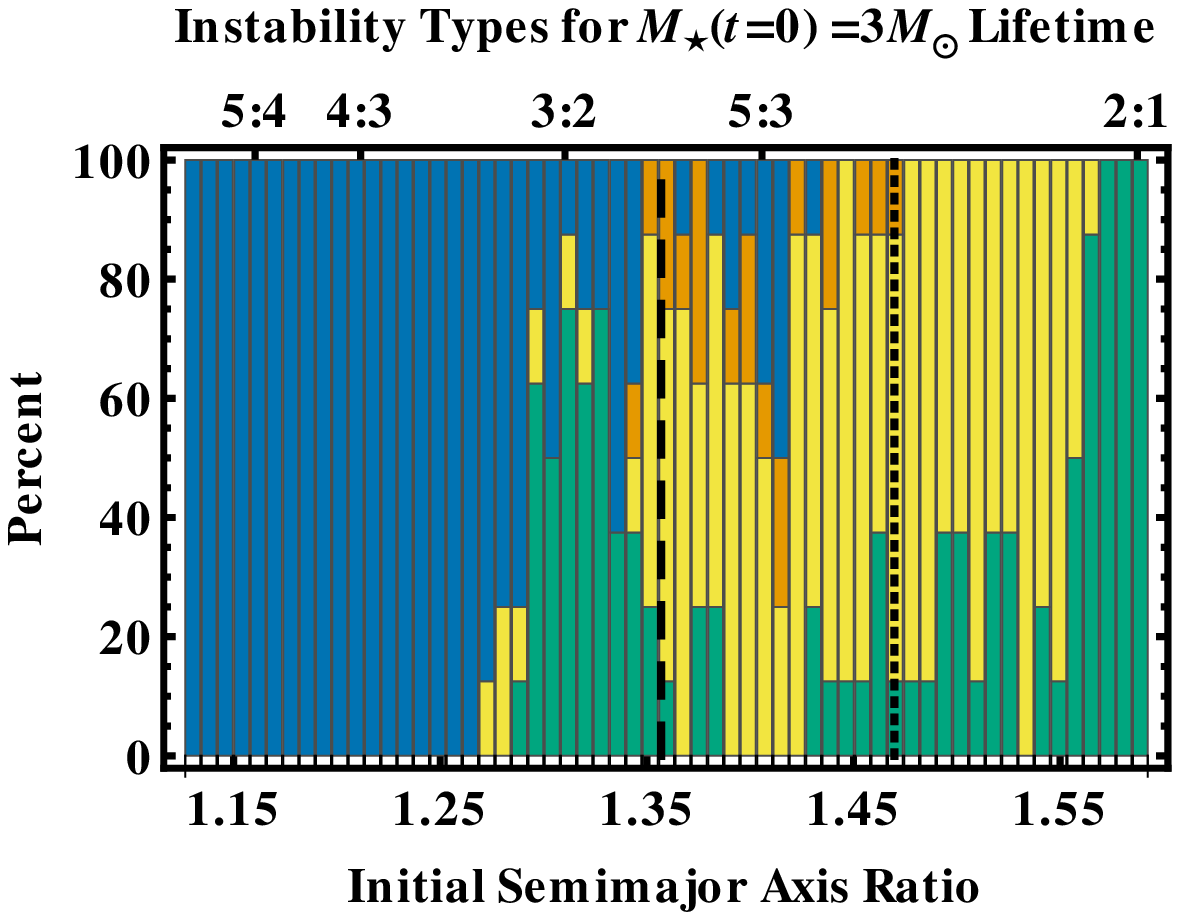,width=8.5cm,height=6.3cm}
}
\caption{
Types of instability.  Blue indicates the fraction of systems 
(out of 8 per bar) that went unstable because of a collision 
between the two planets.
Orange represents instability due to a collision
with the central star.  Yellow indicates any other
type of instability, which predominantly includes ejection.  
Green indicates no instability.
Hence, Hill unstable systems are included in
the blue bars only, and Lagrange unstable systems are included in
the yellow and orange bars only.  The black dashed and dotted
lines are the MS and WD Hill stable boundaries, as in Fig. \ref{Main}.
}
\label{InstType}
\end{figure*}

\begin{figure*}
\centerline{
\psfig{figure=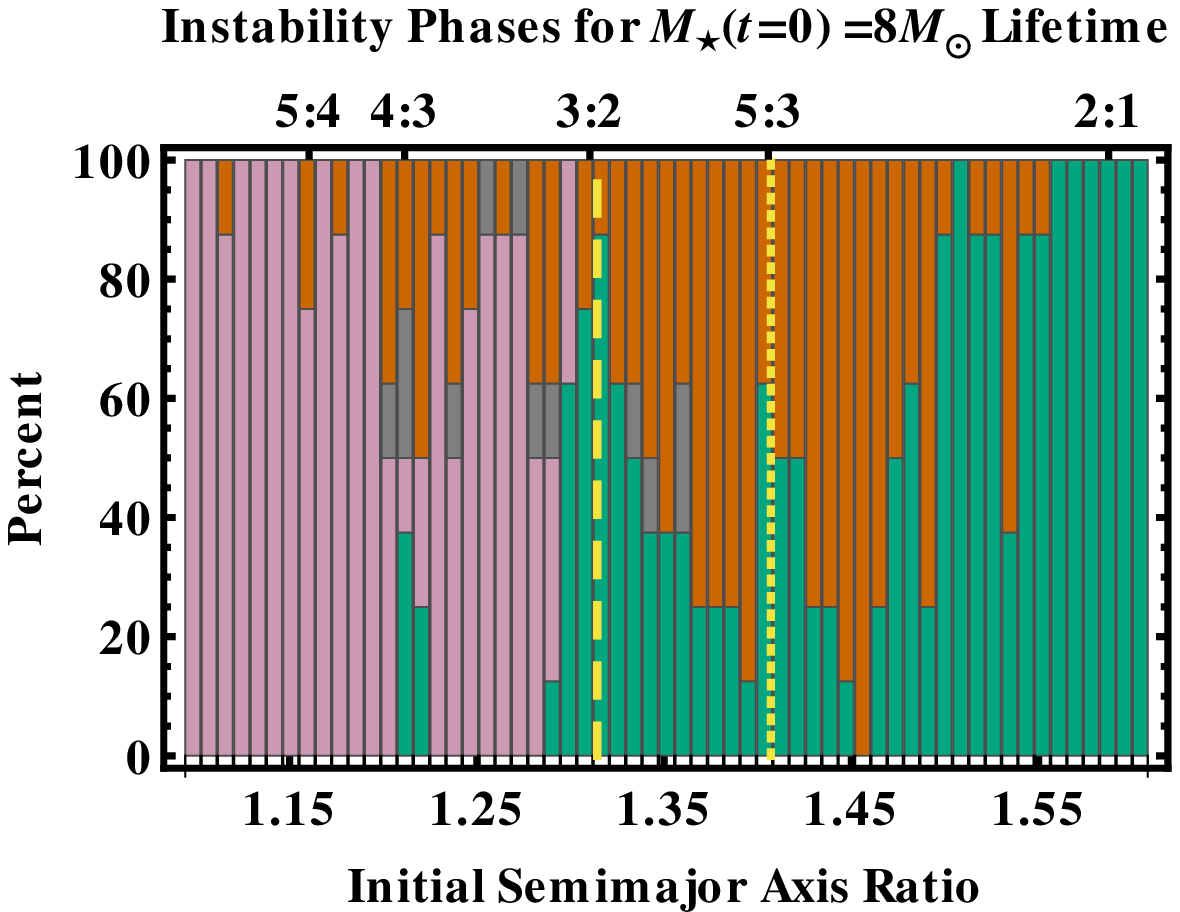,width=8.5cm,height=6.3cm} 
\psfig{figure=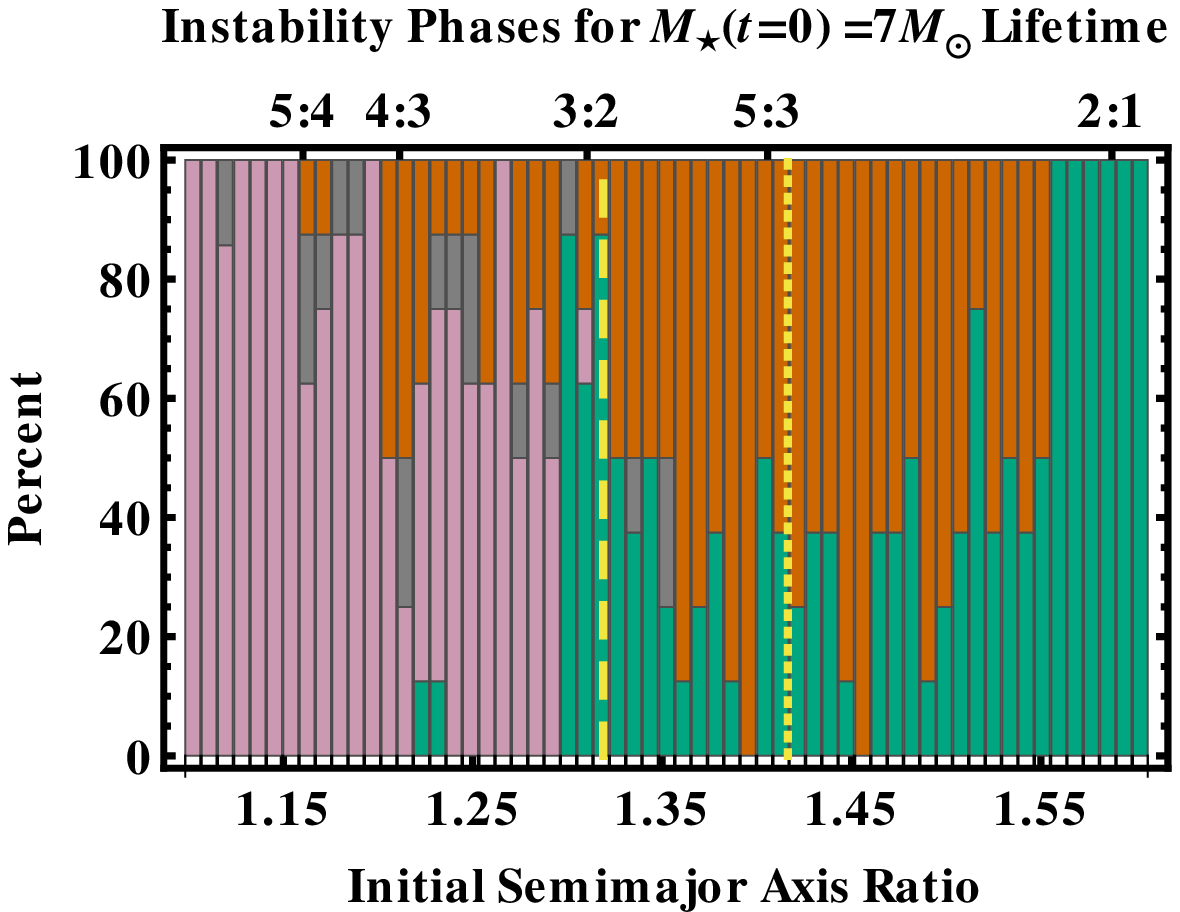,width=8.5cm,height=6.3cm}
}
\centerline{}
\centerline{
\psfig{figure=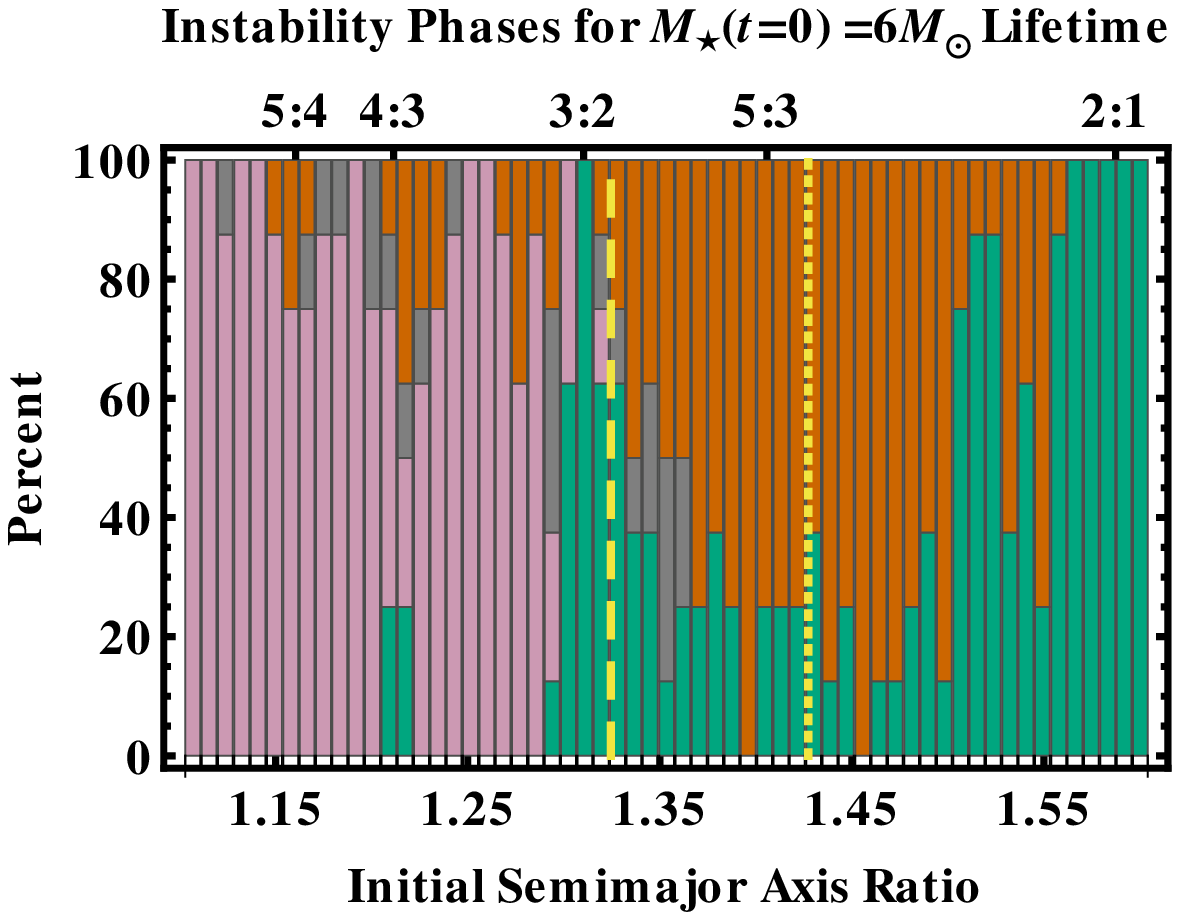,width=8.5cm,height=6.3cm} 
\psfig{figure=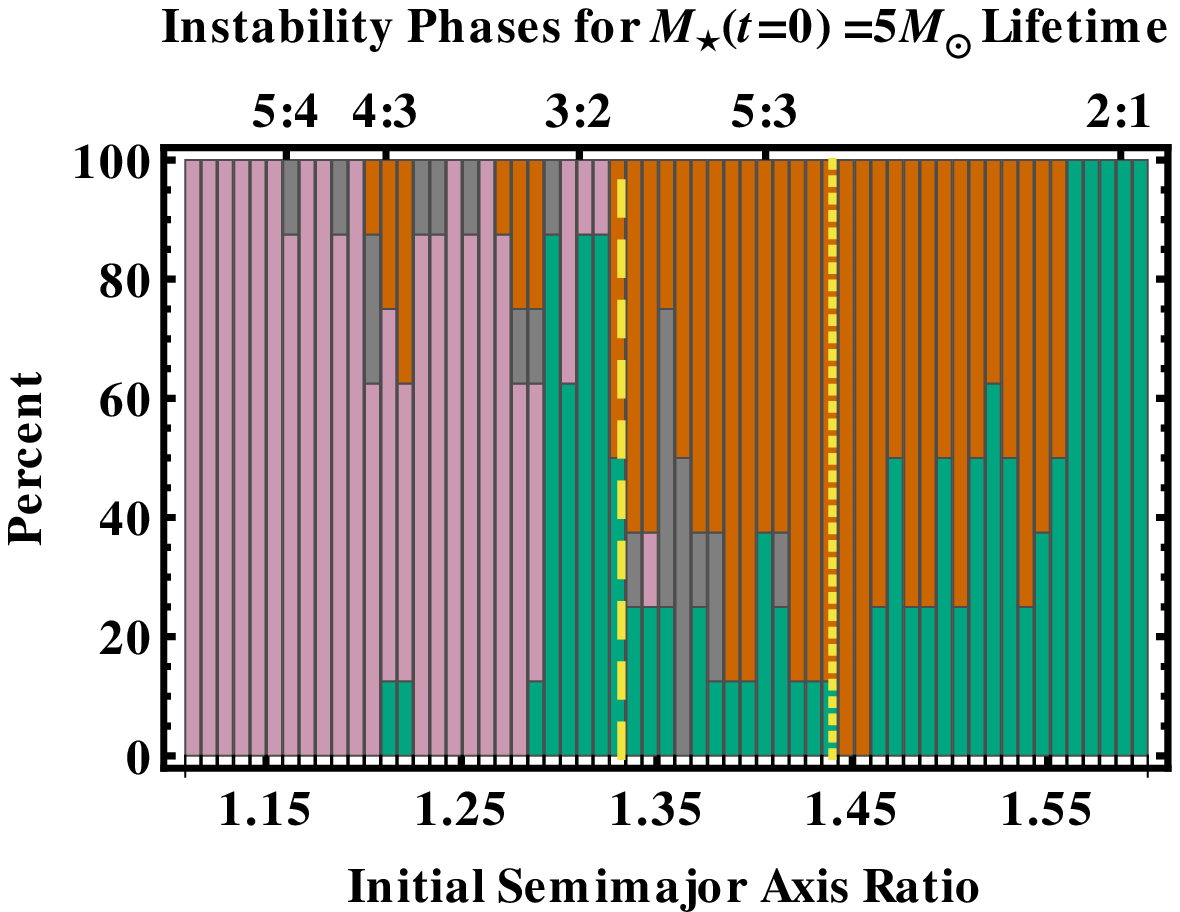,width=8.5cm,height=6.3cm}
}
\centerline{}
\centerline{
\psfig{figure=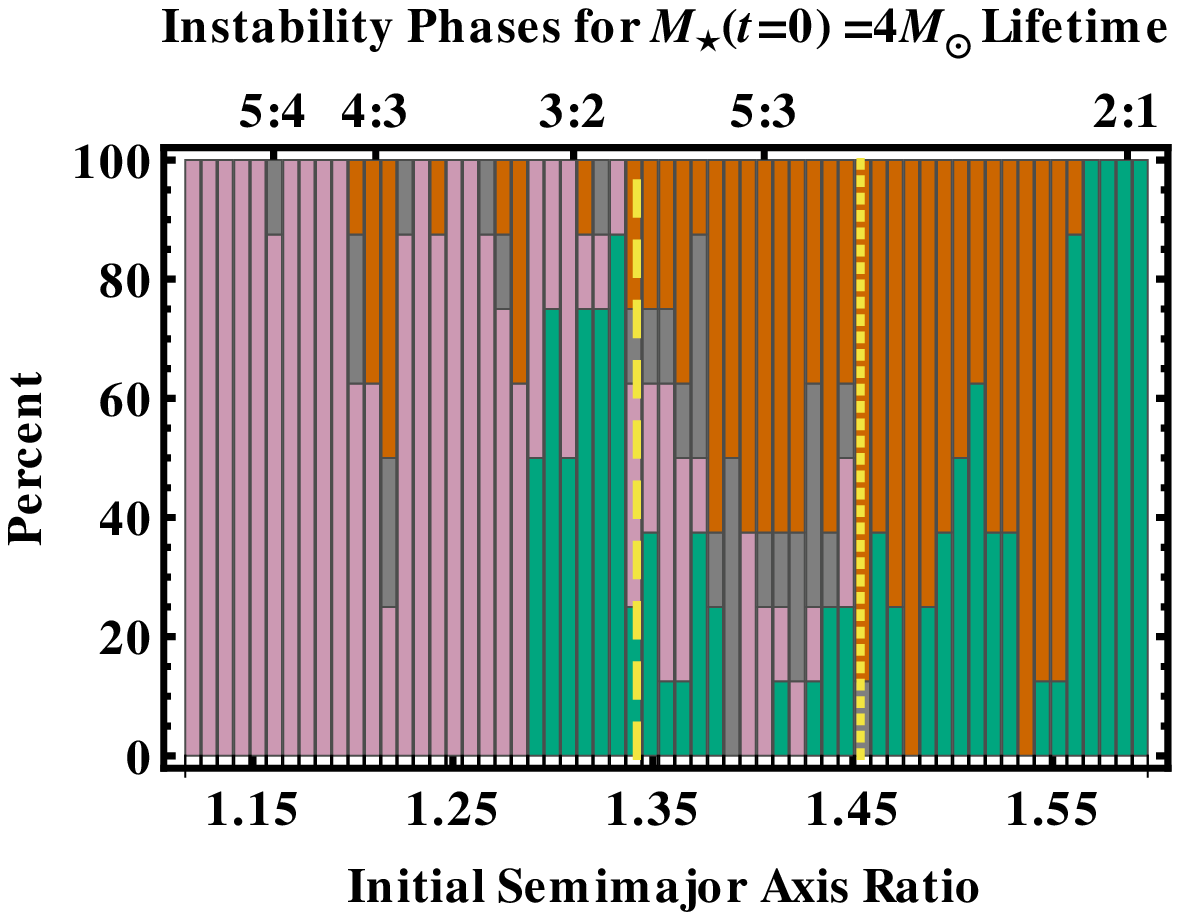,width=8.5cm,height=6.3cm} 
\psfig{figure=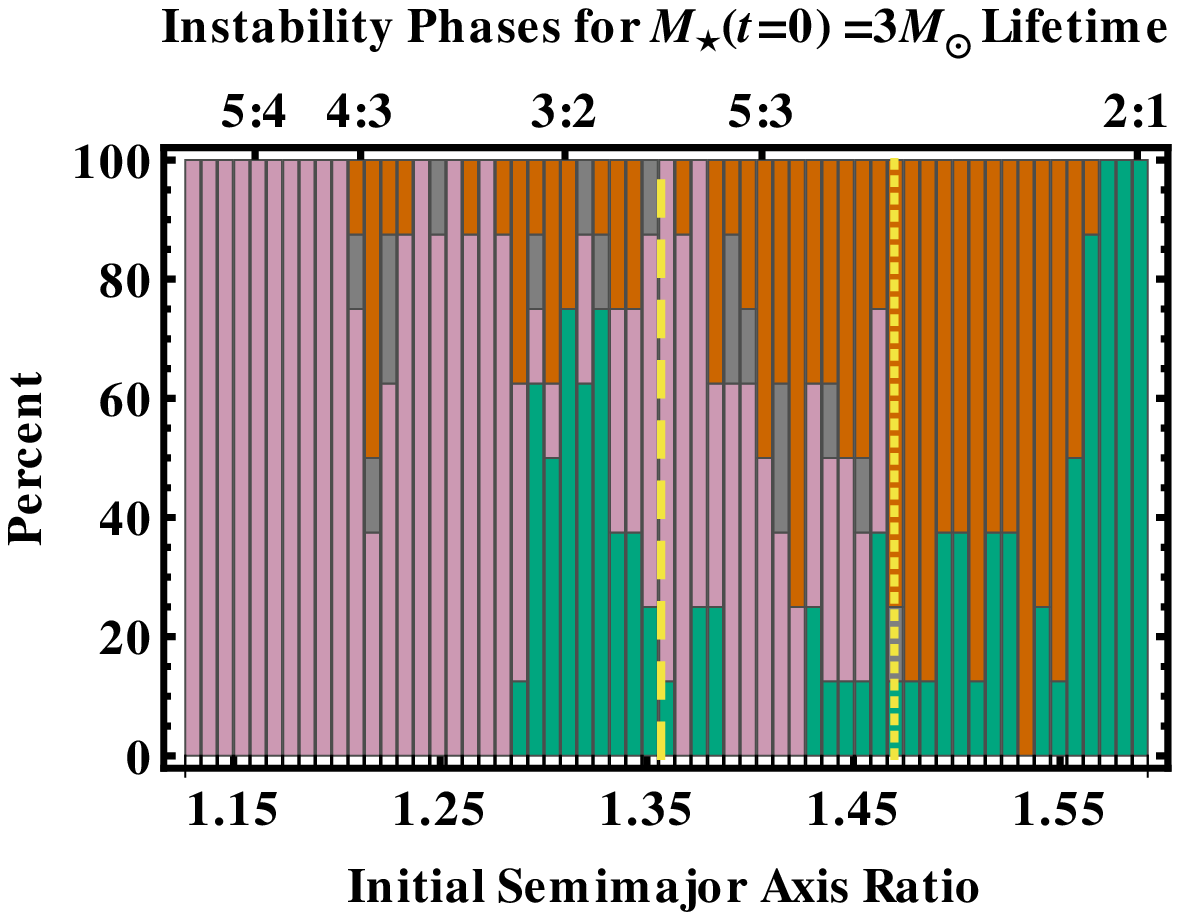,width=8.5cm,height=6.3cm}
}
\caption{
Phases of instability.  Green indicates 
the fraction of systems (out of 8 per bar) that
were stable over 5 Gyr, as in Fig. \ref{InstType}.  
Purple indicates that instability occurred
on the MS.  Gray indicates that instability
occurred between the MS and WD
phases.  Orange indicates that instability occured
during the WD phase.  The plot illustrates that instability
during a giant branch phase does occur, but infrequently.
Also, WD instability can occur for Hill unstable planets
which survive until the WD phase.  The yellow dashed and dotted
lines are the MS and WD Hill stable boundaries, as in Figs. \ref{Main}-\ref{InstType}.
}
\label{FracIns}
\end{figure*}

In support of the above claims, we quantify the types of
instability in Fig. \ref{InstType} for each progenitor mass.  The figure
shows six normalized bar plots. The blue, or topmost, bars represent
the fraction of systems (out of 8) that feature a collision between
both planets; orange bars represent a collision with the central star; yellow
bars represent any other type of instability (including ejection
or periodic instances of a planet attaining a hyperbolic orbit);
green, or bottommost, bars, indicate systems which remained stable for 5 Gyr.
Therefore, Lagrange unstable systems are represented by orange and yellow bars.  
The predominance of the yellow bars towards the right sides of the plots
indicate that any type of collisions become less likely as the initial 
planet separations are increased.  Any blue bars beyond the MS Hill stability
limit indicate planet-planet collisions during the post-MS,
importantly demonstrating that after leaving the MS, planets
are not restricted to (but still predominately experience) Lagrange instability.
There are no blue bars that exceed the WD Hill stability limit, as expected.
The height of the green bars around the $3$:$2$ commensurability demonstrates
how it helps stabilize the simulated systems.

Also of interest is the evolutionary phase at which instability occurs. The 
collision of a planet with a star has been proposed to explain both the 
existence of extreme horizontal branch stars without stellar binary companions -- 
as the envelope of the progenitor giant could be removed by the planet 
\citep{chaetal2011,beasok2012} -- and the enrichment of Lithium seen in a 
few per cent of stars at all parts of the RGB \citep{Lebzelter12}. Of 
particular interest here is the planet candidate proposed orbiting the 
Lithium-rich giant BD+48 740, whose eccentric orbit suggests a past 
strong scattering interaction such as we are considering \citep{adaetal2012}.

Although the fraction of post-MS pre-WD instability 
can be deduced from Figs. \ref{Main} and \ref{Zoom},
we have created a separate figure, Fig. \ref{FracIns}, which better
visualizes the result.  Figure \ref{FracIns} displays the fraction
of systems which are stable (green bars), and those which incur
instability on the MS phase (purple bars), on the WD phase (orange bars)
and in between the MS and WD phases (gray bars).  Instability
during a giant branch phase occurs relatively infrequently:
$[4.0\%,5.4\%,6.9\%,6.7\%,8.7\%,4.6\%]$ of all unstable systems for
each stellar progenitor mass.   WD instability
is not limited to systems with initial separations beyond the Hill
stability limit.

\subsubsection{Potential Resonance Behaviour}

Mean motion commensurabilities, shown on
the upper x-axes of Figs. \ref{Main}-\ref{FracIns},
appear to play an important role in affecting stability.  At these
locations, stability may be either enhanced or additionally disrupted.
A few suggestive instances of these commensurabilities making
a contribution include the $5$:$4$ location in
the $M_{\star}(0) = 8M_{\odot}$ simulations and the $5$:$3$ location in the
$M_{\star}(0) = 3M_{\odot}$ simulations. The $3$:$2$ 
commensurability demonstrably provides a protection mechanism
for planetary systems in each plot.  The $4$:$3$ commensurability
seems to yield all possible outcomes for $M_{\star}(0) \ge 5M_{\odot}$,
but no stable systems for $M_{\star}(0) < 5M_{\odot}$.  

Hence, exploring the resonant character of these systems is 
of potential interest.  However, our 5 Gyr simulations are not 
well-suited to determine if a given
system is locked into mean motion resonance because our output
frequency of 1 Myr is over 4 orders of magnitude greater
than a typical orbital period. The sudden and drastic changes 
in eccentricity and inclination which can arise from purely 
3-body interactions (not including any type of dissipation
nor external forces) may act well within 
1 Myr \citep[e.g.][]{naoetal2011}, and hence disrupt and/or 
create resonances.  Additionally, resonance behaviour may
manifest itself only periodically due to repeated separatrix crossings,
which yield different intervals of libration and circulation
of one or more resonant angles \citep[e.g.][]{fargol2006}.
Recently, \cite{ketetal2012} has classified this behaviour
as ``nodding'' and analytically characterized it.

Despite these caveats, we have considered the evolution from
selected resonant angles from our output.  Identifying
resonant systems requires defining a libration centre, maximum
libration amplitude about this centre, and a duration.
\cite{verfor2009,verfor2010} demonstrated that fully
characterizing potential resonant behaviour for two massive
planets may require the sampling of several libration centres,
as well as computing a mean absolute deviation or root mean
squared deviation about each centre, for each resonant angle.
Here we do not pursue any such analysis, but instead simply
point out that the majority of stable fiducial systems over 5 Gyr 
do not appear to exhibit resonant libration.  These systems
include planets with Hill unstable separations.  Conversely,
a smaller number of systems do appear to exhibit
resonant libration, typically close to the strong first-order
mean-motion resonant commensurabilities.  This result is expected, 
as forming resonances of Jovian-mass planets from uniformly-sampled 
orbital angles should be infrequent (see, e.g., \citealt*{veras2007}).

\begin{figure}
\centerline{
\psfig{figure=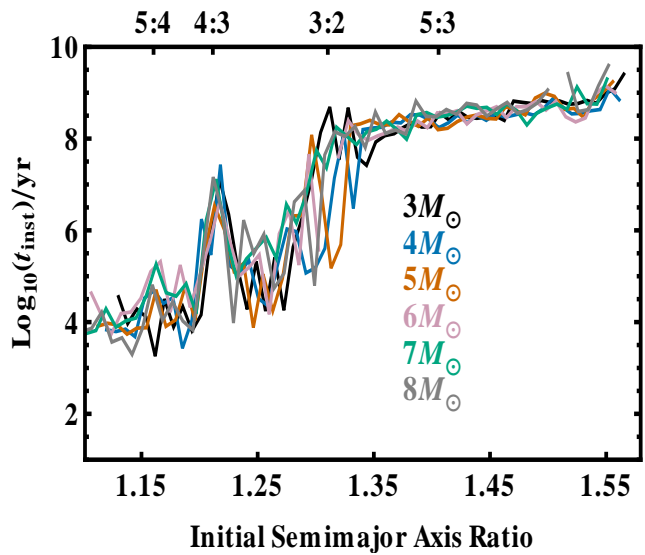,height=8.0cm,width=9.0cm} 
}
\caption{
The geometric mean of instability times when at least
one system out of 8 per semimajor axis ratio goes unstable in 
each fiducial ensemble of simulations.  The plot suggests
that planets near strong mean-motion commensurabilities tend
to survive for longer times before going unstable.
}
\label{GeoMean}
\end{figure}

A perhaps better measure of the effect
of mean motion commensurabilities is by considering the geometric
mean of instability times for each semimajor axis ratio set of 8
simulations that produces at least 1 unstable system.
We plot these mean times in Fig. \ref{GeoMean}, where the $4$:$3$
commensurability is shown to have a clear effect.

\subsubsection{Survivor Orbit Properties}

Of potential interest to WD
pollution investigators and WD planet hunters
is the properties of planets undergoing
Lagrange instability during the WD phase.  
In some cases, the inner planet simply 
collides directly with the WD, creating 
a direct pollutant.  However,  our simulations
suggest that these occurences are rare, occuring
$[1,0,2,2,0,0]$ times
for $M_{\odot}(0) = [8M_{\odot},7M_{\odot},6M_{\odot},5M_{\odot},4M_{\odot},3M_{\odot}]$.


In other cases, the outer planet is ejected
and the inner planet survives on a bounded orbit.
A bound, eccentric inner planet may induce pollution
by scattering comets or asteroids close enough
to the WD to be tidally disrupted and ingested by
the WD.  To gain insight into how an inner planet
survives on an eccentric bounded orbit, we use 
conservation of energy and angular momentum.
Although energy is not conserved in a system
with mass loss -- and certainly not in our integrations -- 
after the parent star has become a WD,
mass loss ceases and then energy is conserved for the future.
Thus, we can compare the states at the beginning of the WD
and at the moment of ejection ($\equiv t_{\rm ins}$).  Further,
because the mass loss is adiabatic, we can relate
the semimajor axes of the planets on the MS and the WD
phases through knowledge of how much mass is lost.
Therefore, we find that the semimajor axis of the
bound planet should be at most:

\begin{equation}
a_1(t_{\rm ins}) \leq 
\frac{
M_{\star}(0)a_1(0)a_2(0)M_1
}{
M_{\star}(t_{WD}) \left[a_2(0)M_1 + a_1(0)M_2 \right]
} 
\label{abound}
\end{equation}

\noindent{where} $M_{\star}(t_{WD})$ is the mass
of the WD.  

Angular momentum is conserved throughout a planetary
system's life, even under the effects of mass loss.
Thus, in principle, one can use conservation of angular momentum
to determine the value of $e_1(t_{\rm ins})$.  However,
doing so requires knowledge of the hyperbolic values of
$a_2(t_{\rm ins})$ and $e_2(t_{\rm ins})$.  These values
are set by the ejection velocity, which is determined
by the strength of the instability in each case.  Our numerical
simulations show that $e_1(t_{\rm ins})$ varies considerably.

\begin{figure*}
\centerline{}
\centerline{
\psfig{figure=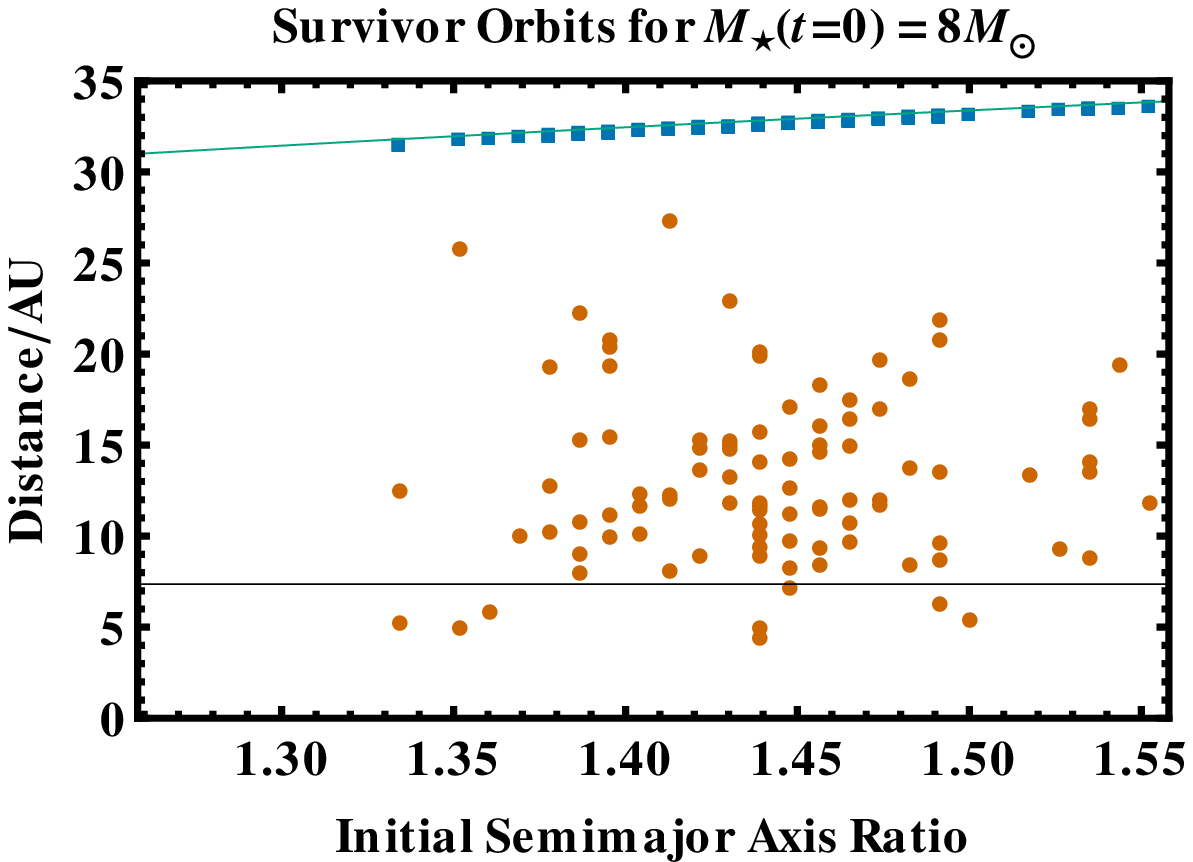,width=8.5cm,height=6.3cm}
\psfig{figure=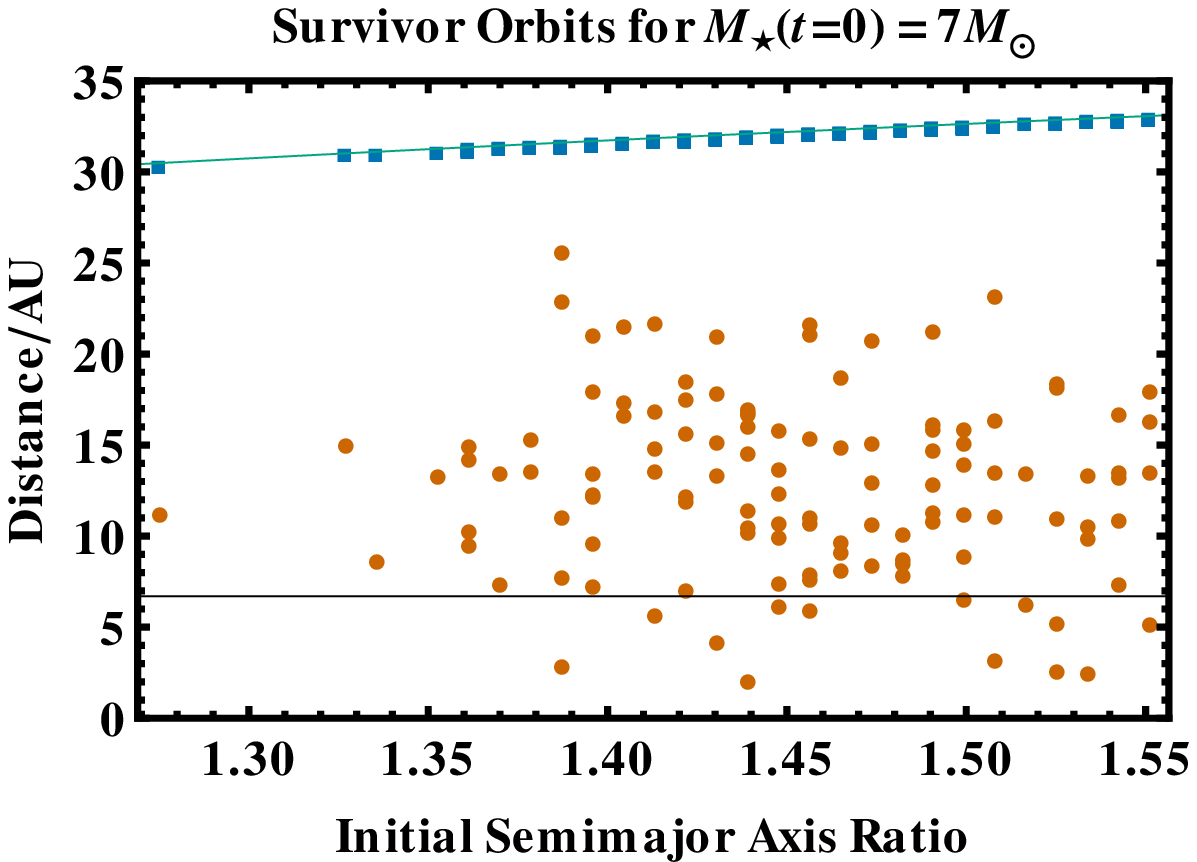,width=8.5cm,height=6.3cm}
}
\centerline{}
\centerline{
\psfig{figure=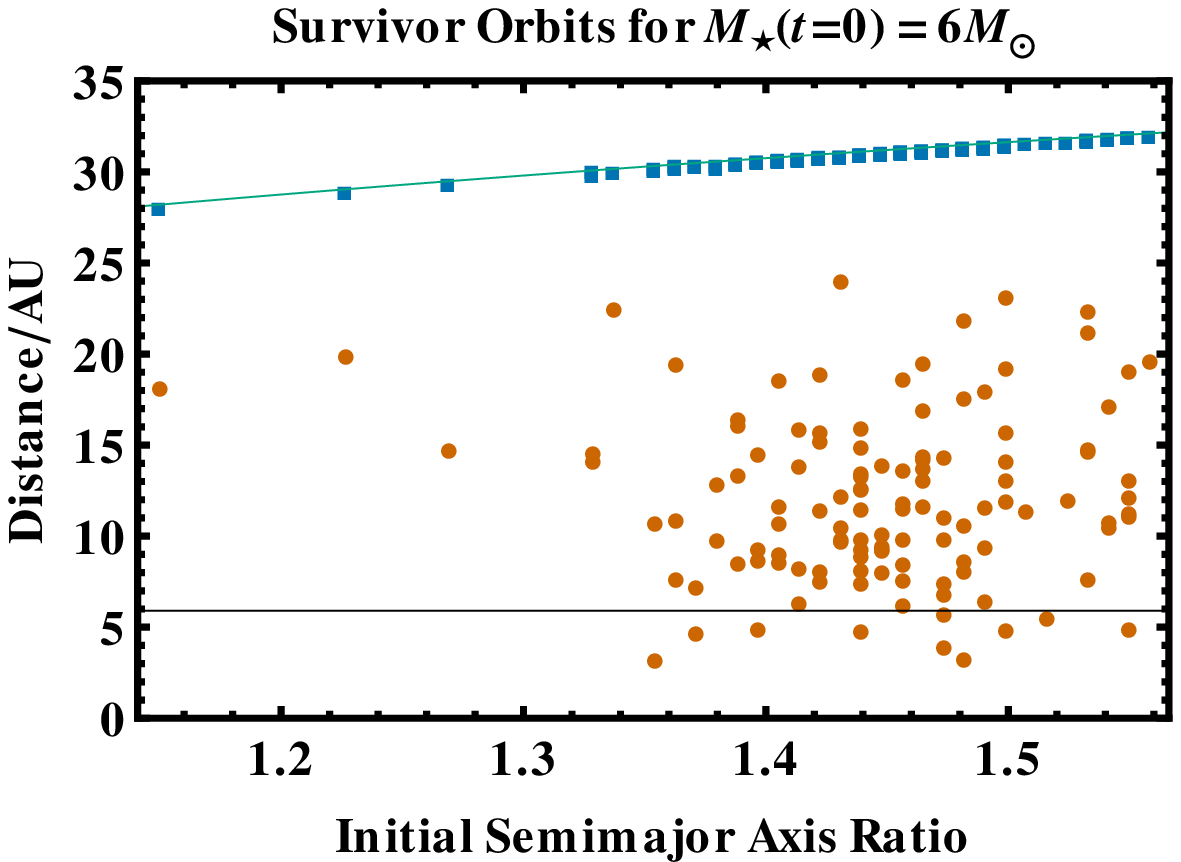,width=8.5cm,height=6.3cm} 
\psfig{figure=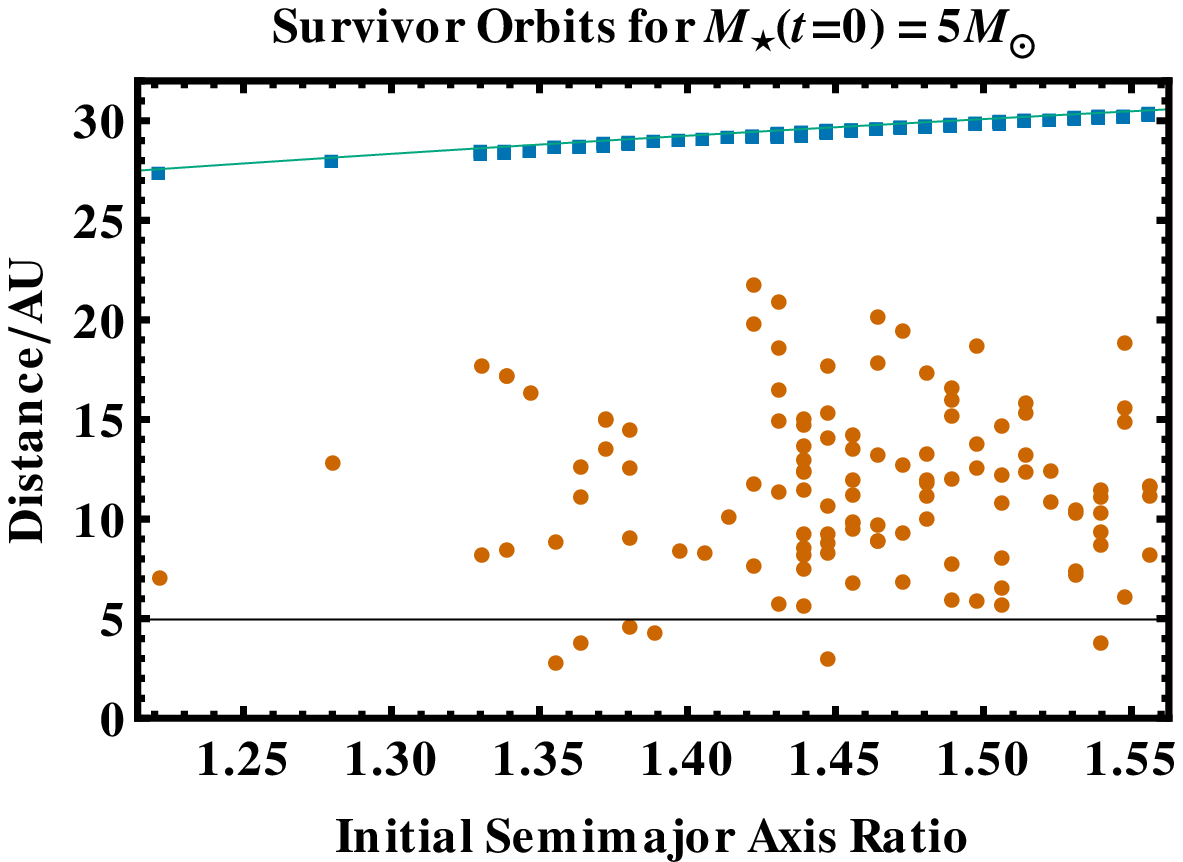,width=8.5cm,height=6.3cm}
}
\centerline{}
\centerline{
\psfig{figure=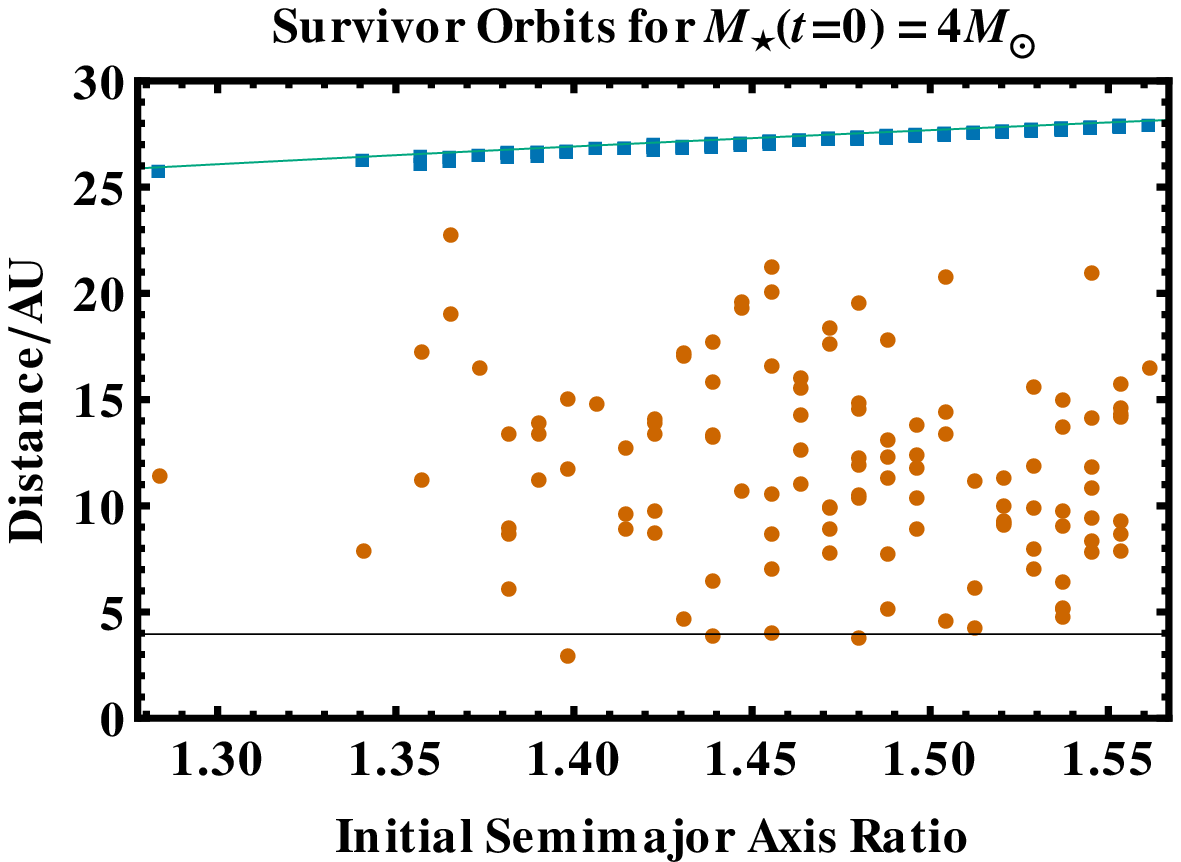,width=8.5cm,height=6.3cm}
\psfig{figure=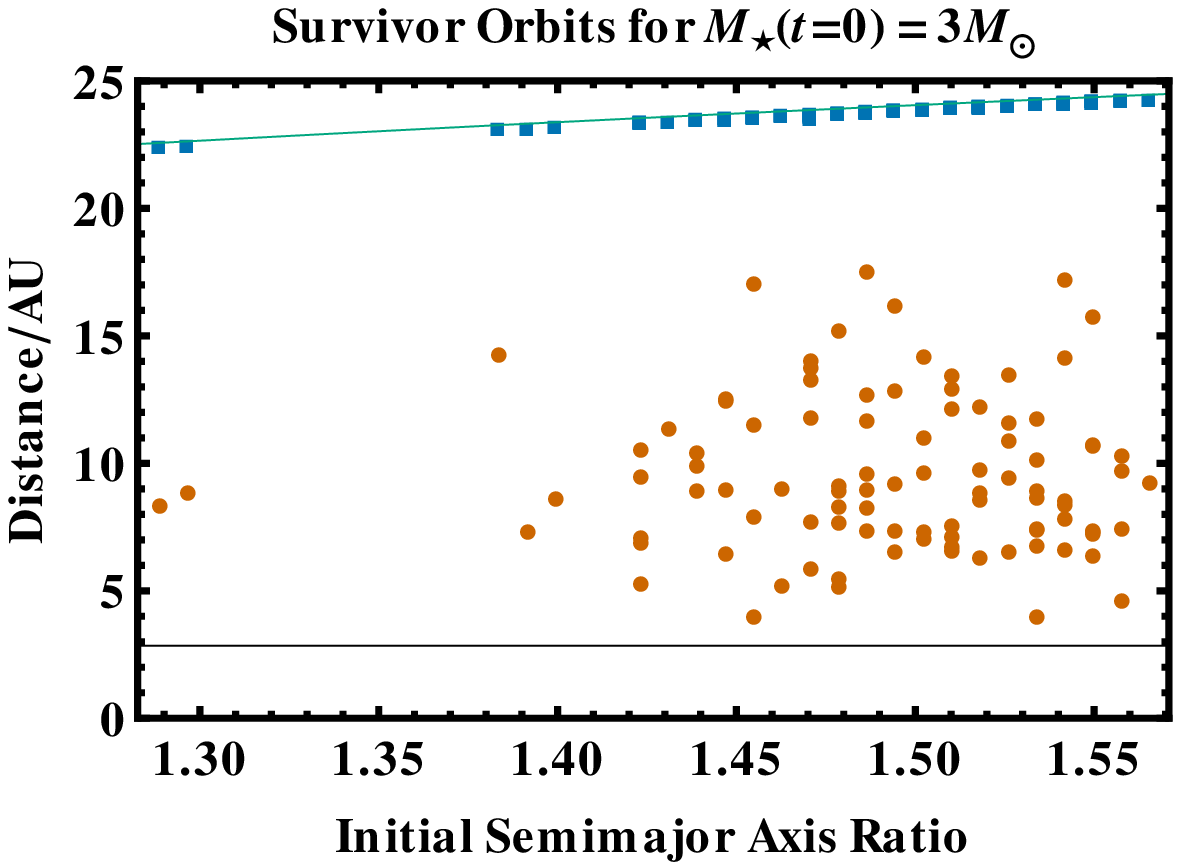,width=8.5cm,height=6.3cm} 
}
\caption{
Orbits of inner planets that survive the ejection of the
outer planet during the WD phase.  Blue squares
represent semimajor axes, and orange dots represent
periastra, all from the simulation outputs.  The aqua 
curve is the analytical estimate 
for where the maximum semimajor axis should be 
(Eq. \ref{abound}).  The black curve indicates the maximum
stellar radius achieved during its evolution.  Dots below
this curve indicate the existence of WD planets orbiting inside
the maximum AGB radius even though the planets were formed elsewhere.
}
\label{Survivor}
\end{figure*}

We plot the semimajor axes (blue squares) and 
periastra (orange dots) of the surviving
planets in systems featuring ejections in the WD phase only,
in Fig. \ref{Survivor}.
Superimposed on the plots through aqua lines are the 
analytically predicted maximum values of $a_1(t_{\rm ins})$ through 
Eq. (\ref{abound}).  Also plotted as a solid black horizontal 
line is the maximum stellar radius (see Figs. \ref{envexp} and \ref{SSERadius})
attained during the star's evolution.
The presence of orange dots below the black line
suggest the presence of a population of highly eccentric planets orbiting 
WDs whose present pericentres take them inside the maximum AGB radius.
These planets cannot have been formed {\it in situ} or anywhere
near their WD locations because otherwise they would have been destroyed or 
suffered radical orbital alterations on the AGB.
The fractions of orange dots below the black
line for each progenitor mass are 8.4\% ($8 M_{\odot}$),
10.9\% ($7 M_{\odot}$), 8.4\% ($6 M_{\odot}$), 5.1\% ($5 M_{\odot}$),
1.7\% ($4 M_{\odot}$) and 0\% ($3 M_{\odot}$).  



\subsection{Simulation Results: Additional Cases}

\subsubsection{Terrestrial Planet Masses}

\begin{figure*}
\centerline{
\psfig{figure=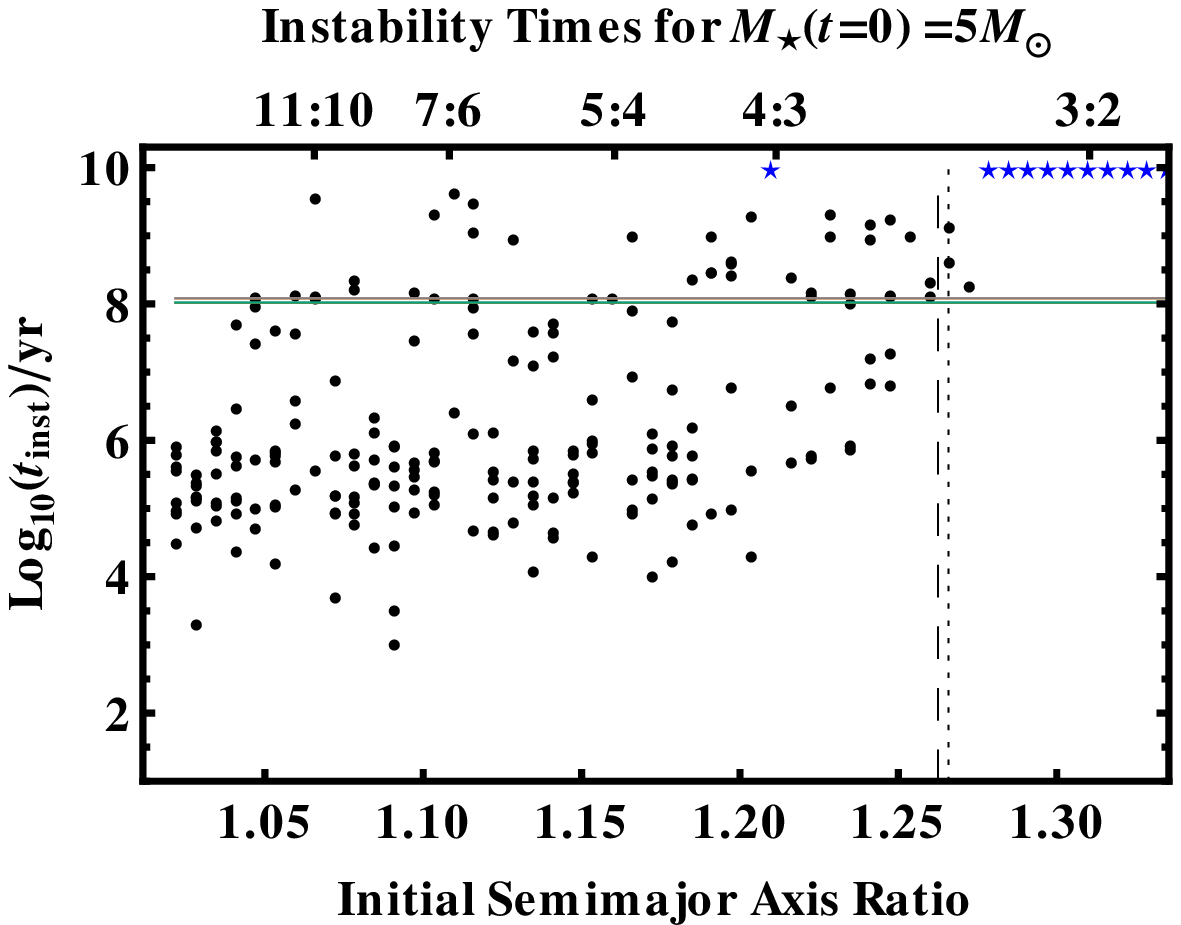,width=8.5cm,height=6.3cm} 
\psfig{figure=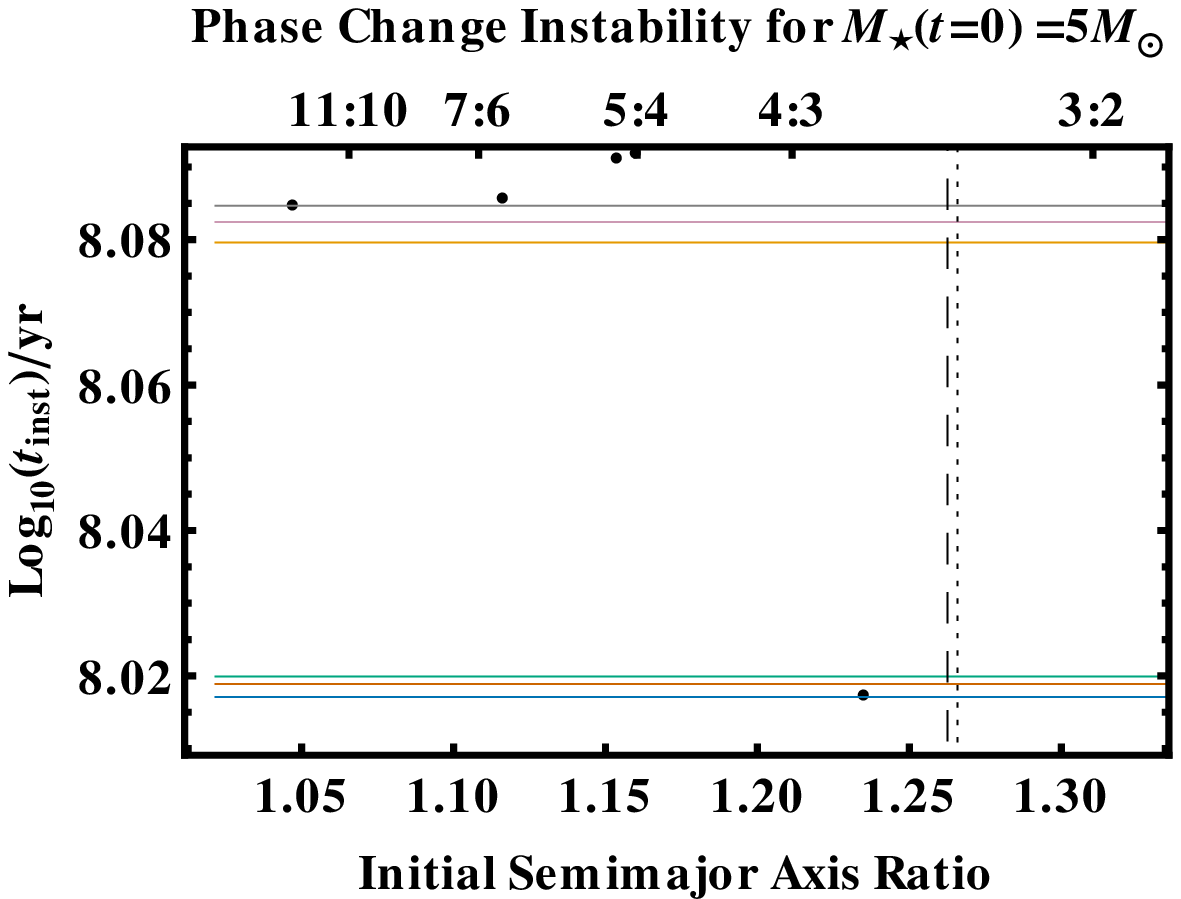,width=8.5cm,height=6.3cm}
}
\centerline{}
\centerline{
\psfig{figure=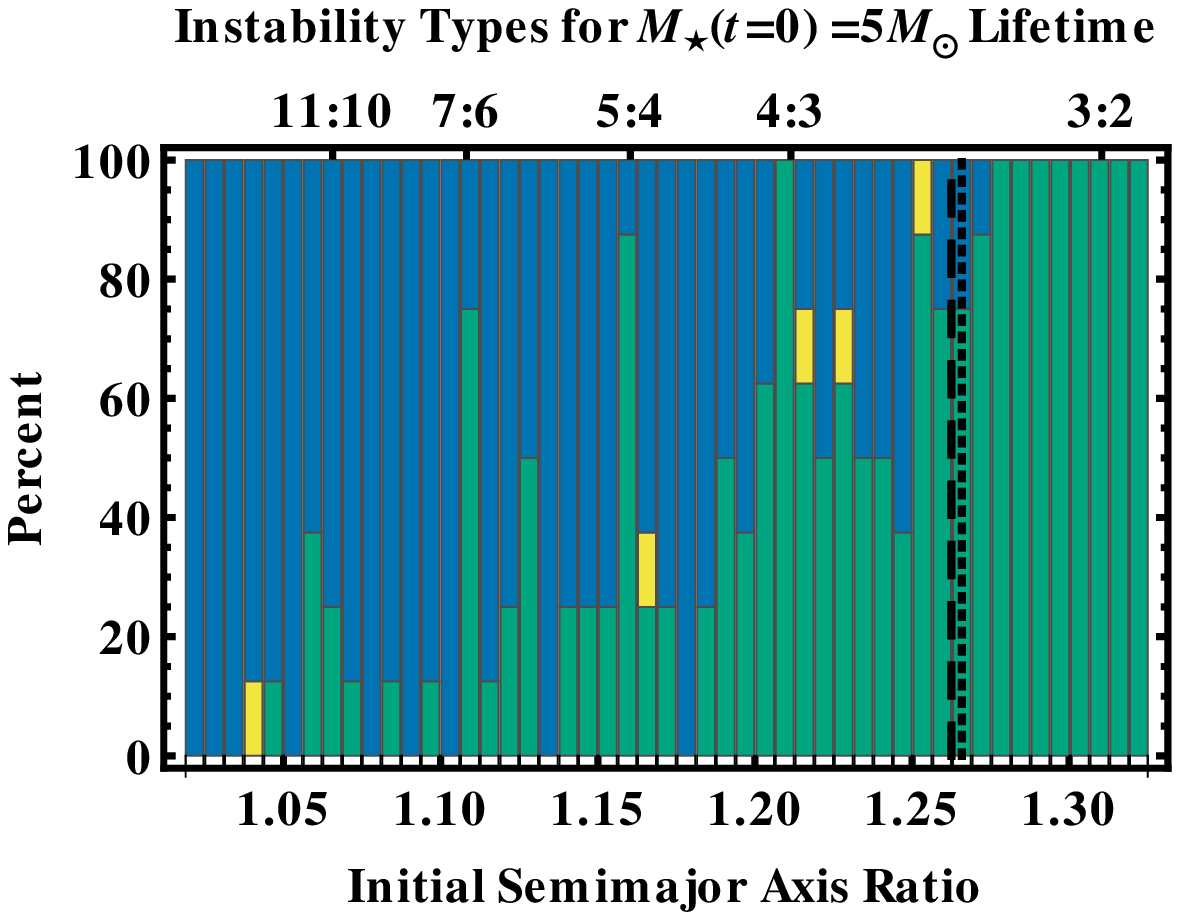,width=8.5cm,height=6.3cm} 
\psfig{figure=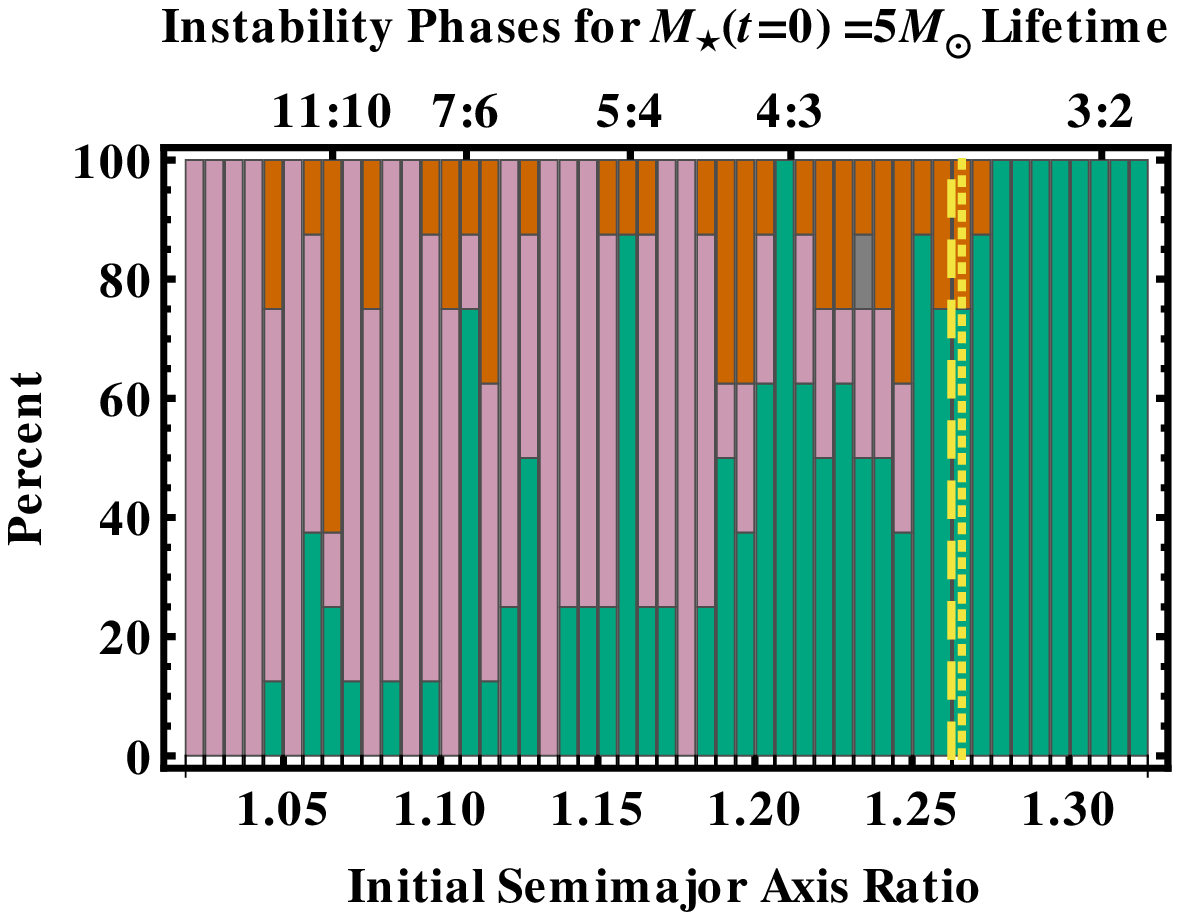,width=8.5cm,height=6.3cm}
}
\centerline{}
\centerline{
\psfig{figure=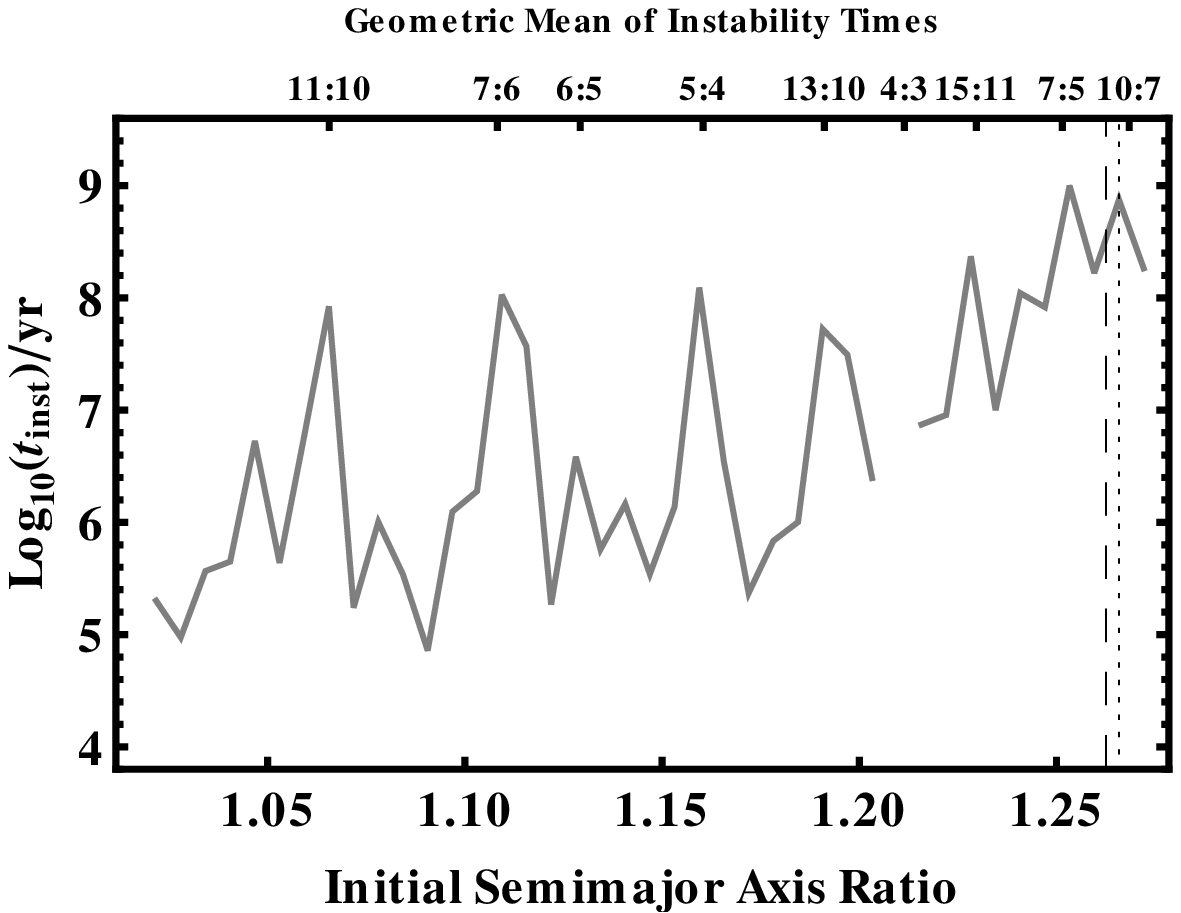,width=8.5cm,height=6.3cm} 
}
\caption{
Here, $M_1=  M_2 = 1 M_{\oplus}$.  The plots
are of the same format as in Figs. \ref{Main}, \ref{Zoom},
\ref{InstType}, \ref{FracIns} and \ref{GeoMean}.  Instability 
during the WD phase occurs almost exclusively at initial separations
inside the MS and WD Hill stability limits.  MS Hill unstable systems protected
by mean motion commensurabilities, many of which feature resonant
behaviour, allow for survival during the MS before becoming
unstable on the WD phase.
}
\label{Earths}
\end{figure*}

Now we consider variations
on the fiducial case.  First, we set
the planet masses such that $M_1 = M_2 = M_{\oplus}$.
The results of those 632 simulations are summarized 
in Fig. \ref{Earths}, which include plots for instability time versus
initial semimajor axis ratio, the types and phases
of instability, and the geometric mean of instability
time.  The smaller planetary masses here shrink
the chaos limit and the Hill stability limit --
as well as the difference between the MS and WD Hill stability 
limits (see bottom panel of Fig. \ref{hille1}) -- and introduces 
a different set of potentially important 
commensurabilities, displayed on the top x-axis of all plots.

Earth-mass planets fail to go unstable beyond the MS Hill
stability limit in all but a few cases; the blue stars
continue in an unbroken chain out to $a_2/a_1 \approx 1.510$.  
Therefore, Hill stability and Lagrange stability appear to have
almost identical boundaries in this case.

No system features a collision of a terrestrial planet with 
the WD.  This result should not imply that this type of collision
cannot occur, but rather that giant planet collisions are likely
to be much more frequent.

Almost all instability involves collision between both planets.  
Interestingly, one planet-planet collision occurs
just beyond the WD Hill stability boundary, meaning that the real
boundary must differ from the line on the plot.  This situation arises
because planets arrive on the WD phase with slightly different
osculating orbital parameters than they harboured at the start of
the MS (primarily due to their mutual interactions, and slightly due to post-MS
mass loss).  Hence, the real WD Hill stability limit for each
individual system is different, and differs from the
line shown in the figure that was computed for $e_1=e_2=0.1$ exactly.

Although only one system undergoes instability in the post-MS
pre-WD phases, instability during the WD phase is common,
and is in fact greatest for separations close to the 
tightly-packed $11$:$10$ commensurability (corresponding
to $a_2/a_1 \approx 1.066$).  Other instances of WD instability
occur around the $7$:$6$ and $5$:$4$ commensurabilities,
and to either side of the $4$:$3$ commensurability, on which
all 8 systems are stable.  The influence meted by these
first-order commensurabilities therefore appears to be extensive. 
However, the bottom plot of the figure might suffer bias due to small
number statistics.  An investigation of individual systems 
indicates that resonant behaviour appears to be more common
for these terrestrial-mass planet systems than for the fiducial 
Jovian-mass systems, perhaps demonstrating the importance of the
planet/star mass ratio in creating resonance behaviour from random
initial conditions.

\begin{figure*}
\centerline{
\psfig{figure=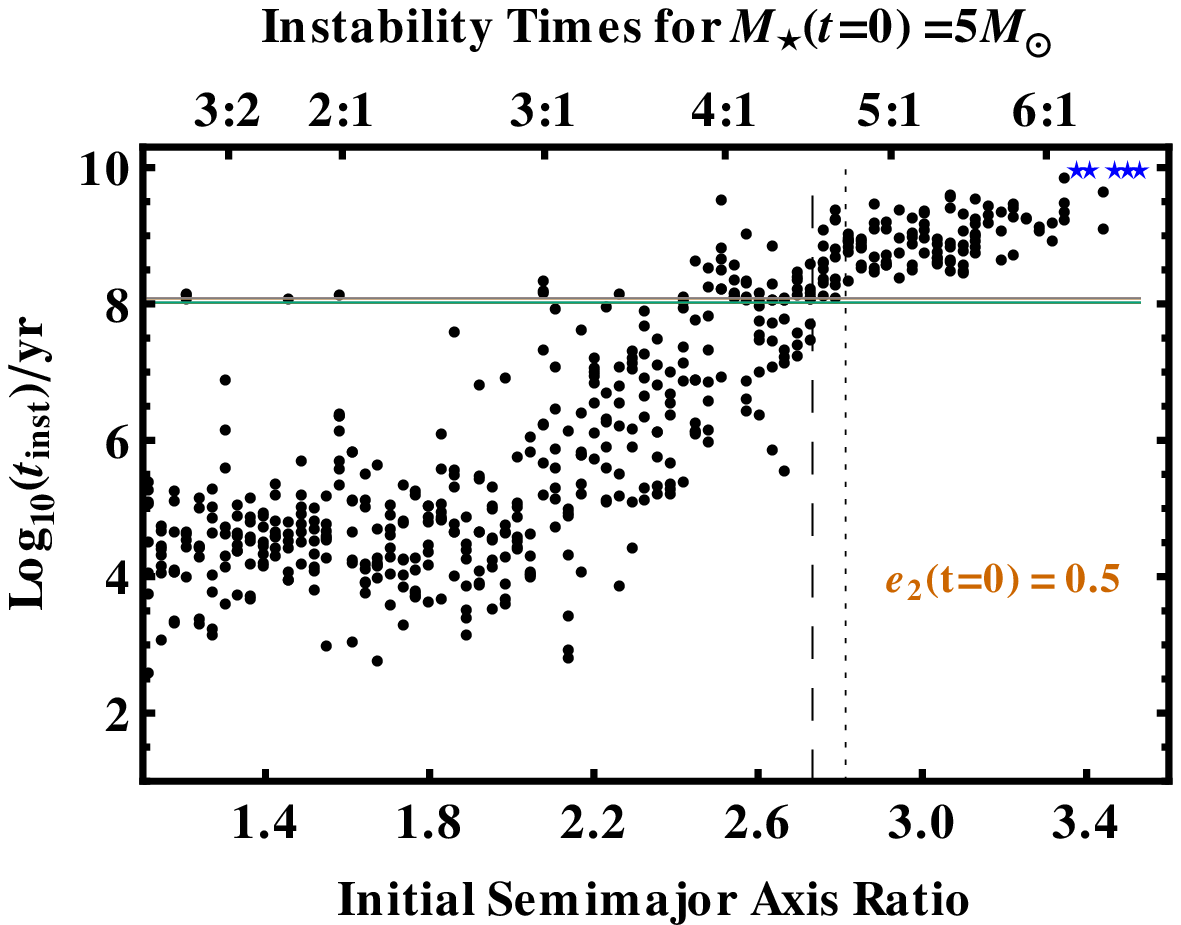,width=8.5cm,height=6.3cm} 
\psfig{figure=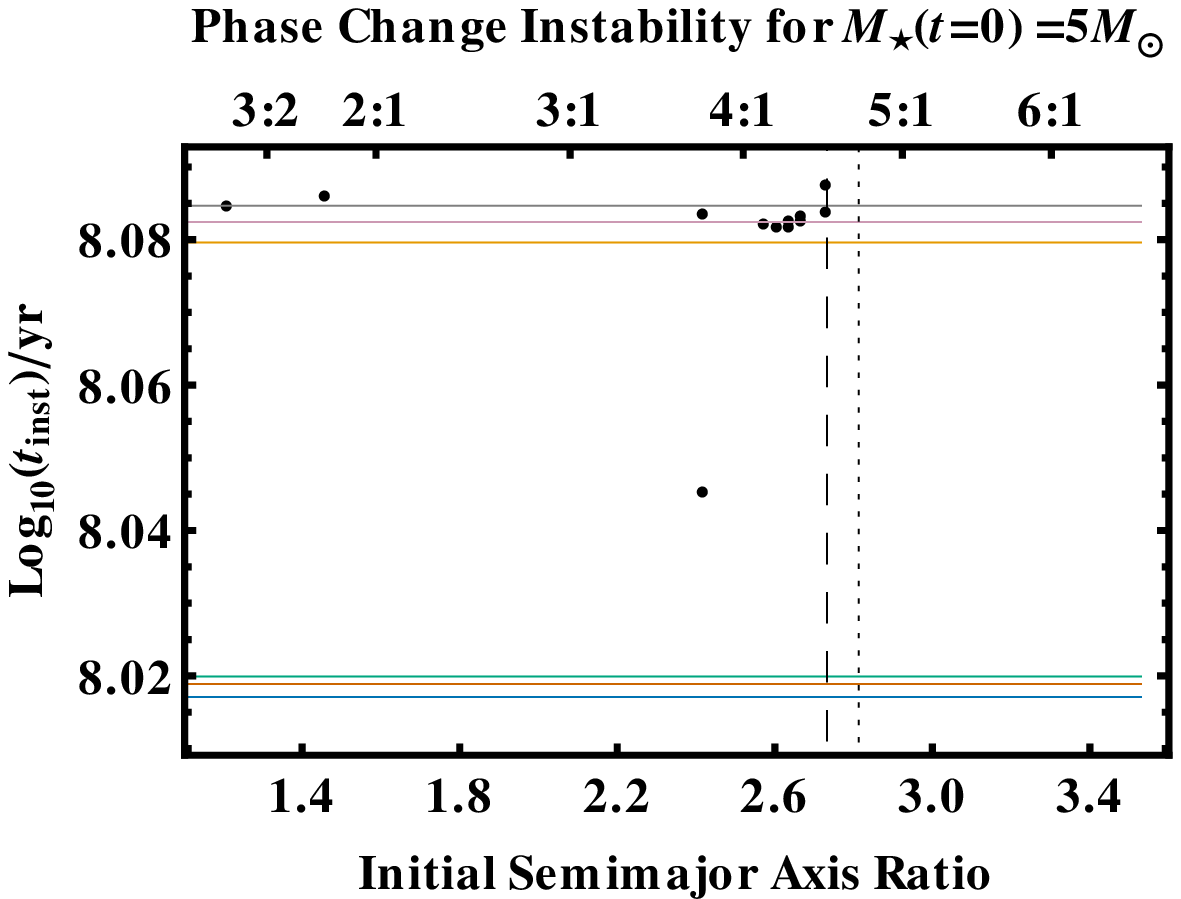,width=8.5cm,height=6.3cm}
}
\centerline{}
\centerline{
\psfig{figure=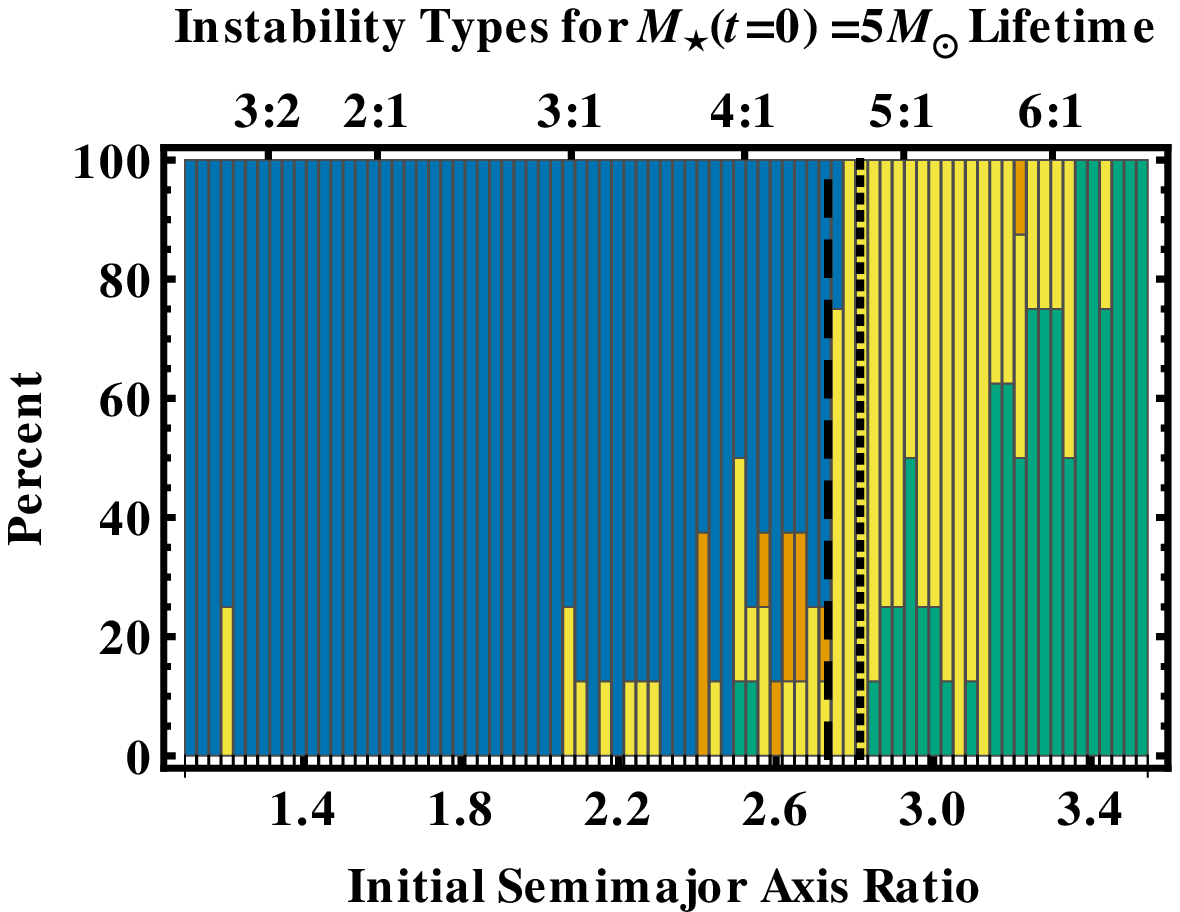,width=8.5cm,height=6.3cm} 
\psfig{figure=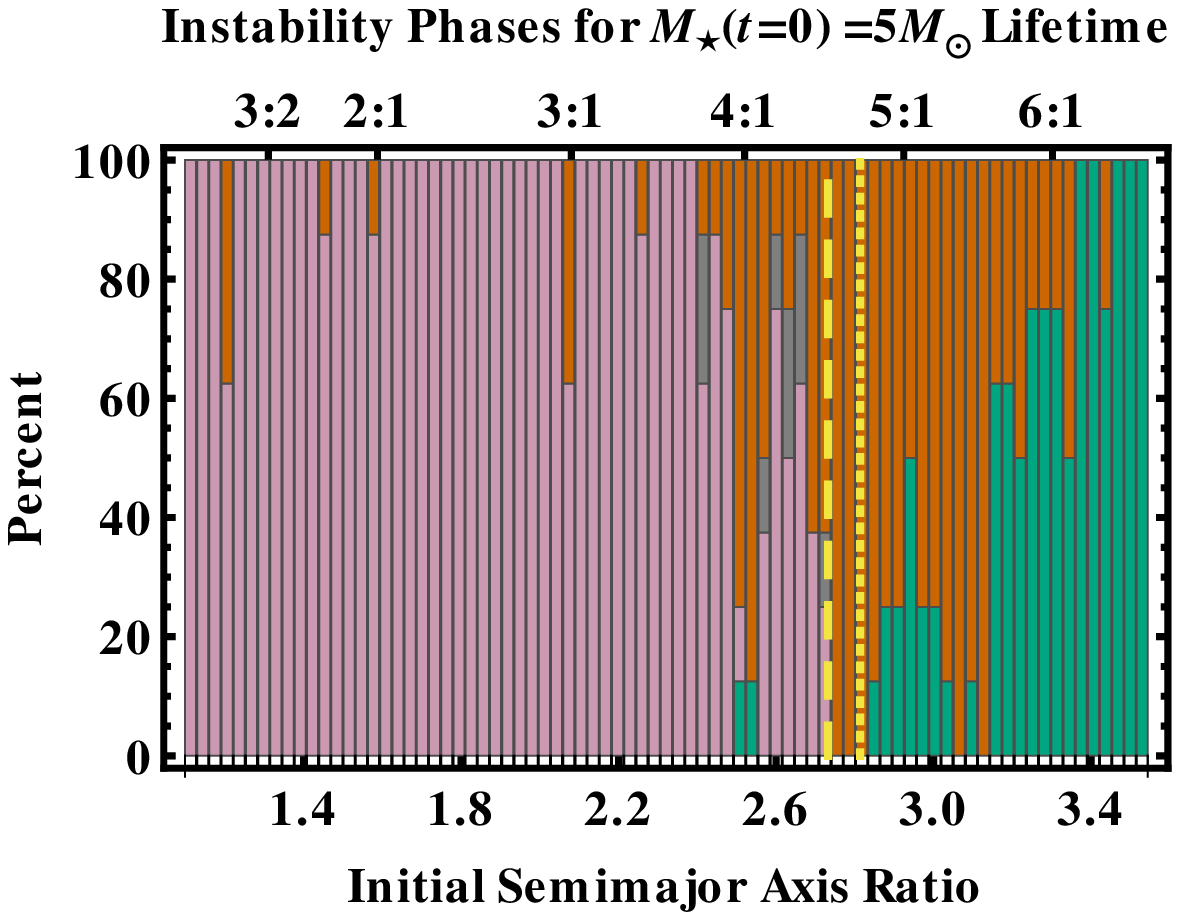,width=8.5cm,height=6.3cm}
}
\centerline{}
\centerline{
\psfig{figure=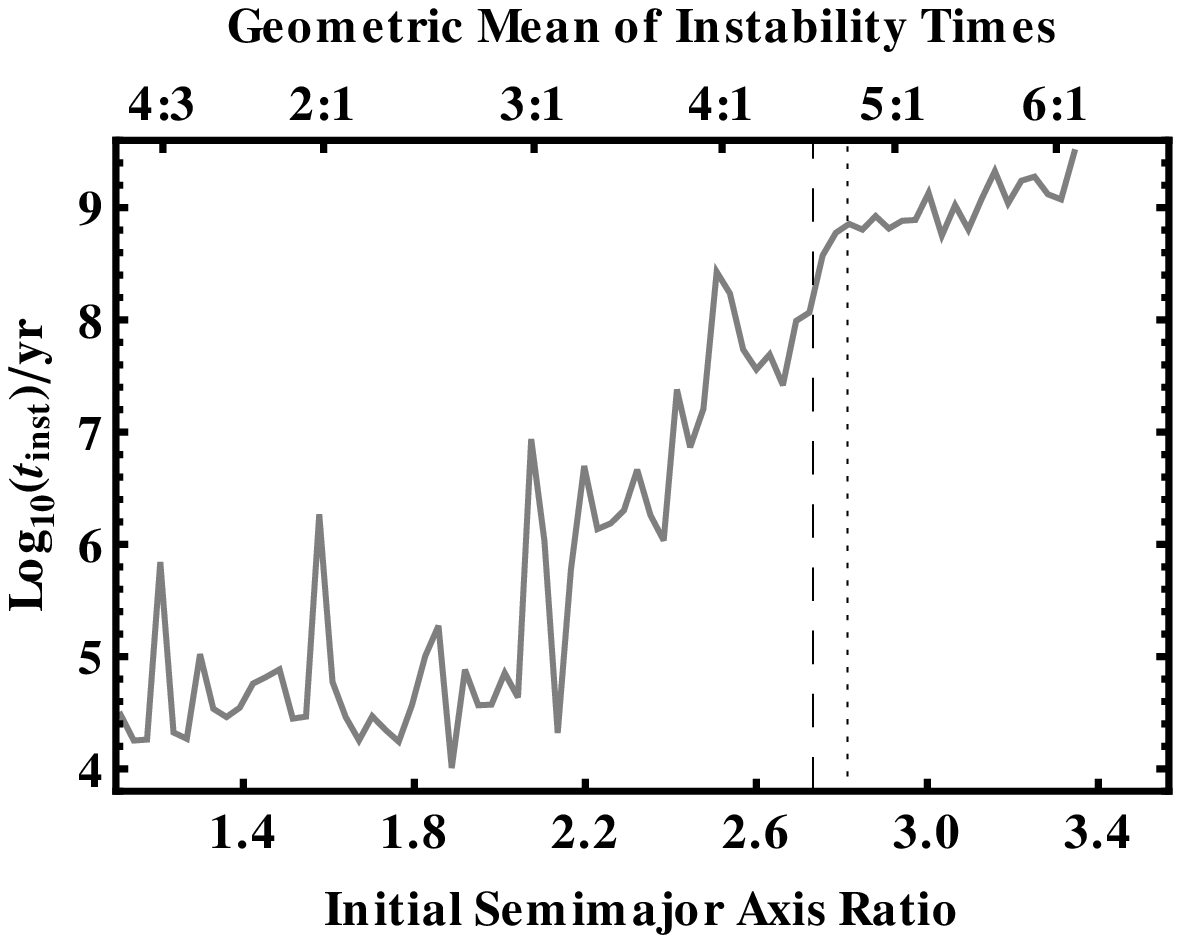,width=8.5cm,height=6.3cm} 
\psfig{figure=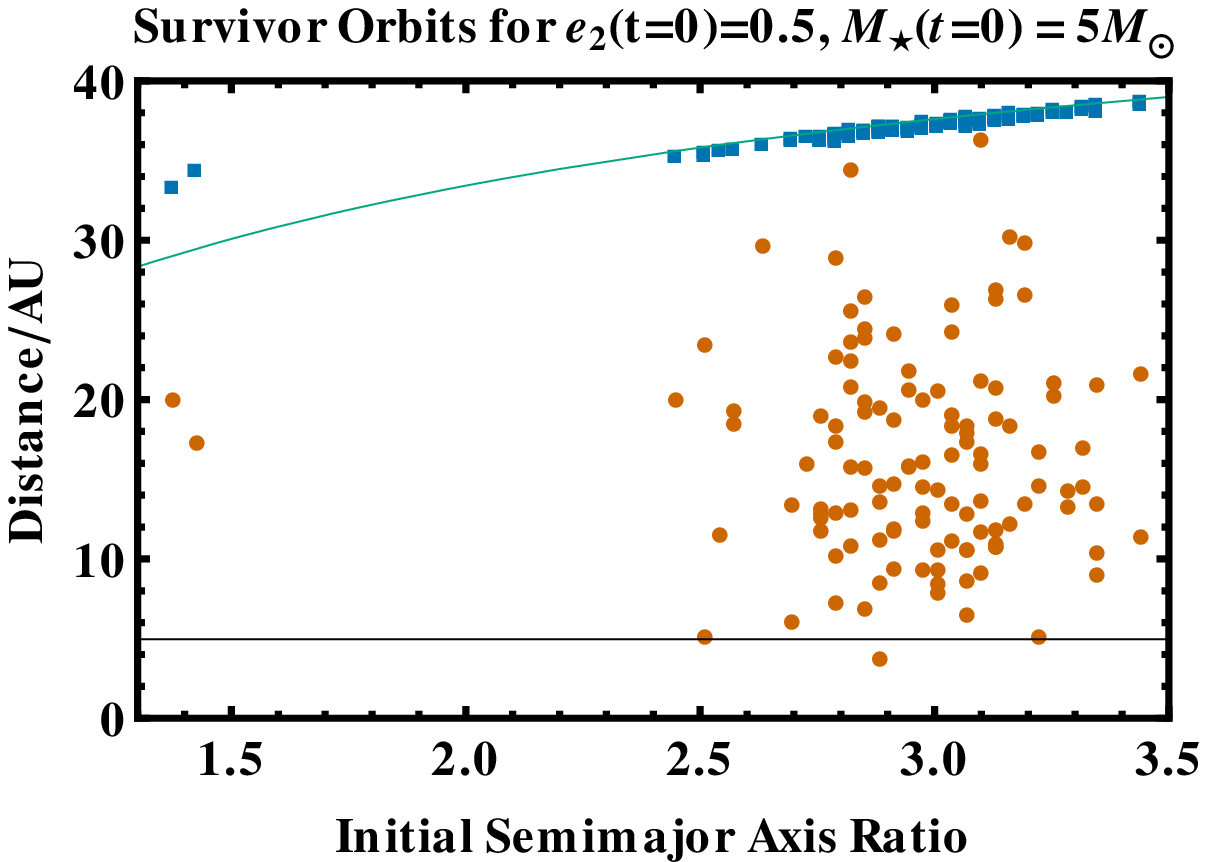,width=8.5cm,height=6.3cm} 
}
\caption{
Here, $e_2(0) = 0.5$.  The plots
are of the same format as in Figs. \ref{Main}, \ref{Zoom},
\ref{InstType}, \ref{FracIns}, \ref{GeoMean} and \ref{Survivor}.  
This figure importantly demonstrates
that Lagrange instability can occur readily in moderately 
eccentric systems.
}
\label{ModEcc}
\end{figure*}

\subsubsection{Different Planetary Eccentricities}

Next, we consider systems with an initially moderately eccentric
outer planet orbit ($e_2 = 0.5$) and the same fiducial inner
planet eccentricity ($e_1 = 0.1$) in Fig. \ref{ModEcc}.  In this case, 
the analytic MS Hill stability limit
is much larger ($a_2/a_1 \approx 2.73$) than in the fiducial case,
which is expected (see Fig. \ref{hille1}).

For these systems, instability beyond the MS and WD Hill limits
is extensive.  In fact, the Lagrange unstable region 
extends out nearly 7 AU from the location of the MS Hill 
stability boundary for $a_1 = 10$ AU.  However, the Lagrange
stability boundary lies at a distance that is $143\%$ and
$135\%$  of the MS and WD Hill stability
limits; the former value is a much smaller factor than in 
the fiducial cases.

Collisions with the star are infrequent and ejections are common, 
perhaps because the inner planet has the much lower eccentricity.
In fact, inner planets which survive outer planet ejections 
largely fail to attain high enough eccentricities to intrude within
the maximum stellar radius achieved during the host star's evolution: 
only 0.85\% of the orange dots are below the black line in the bottom-right plot.  In that
same plot, the two outliers likely represent systems that experienced
pre-WD instability that was missed by our low output frequency. 
Instability during the giant phases is uncommon, and is restricted
to the region around the $4$:$1$ commensurability.

\section{Discussion}

\begin{figure*}
\centerline{
\psfig{figure=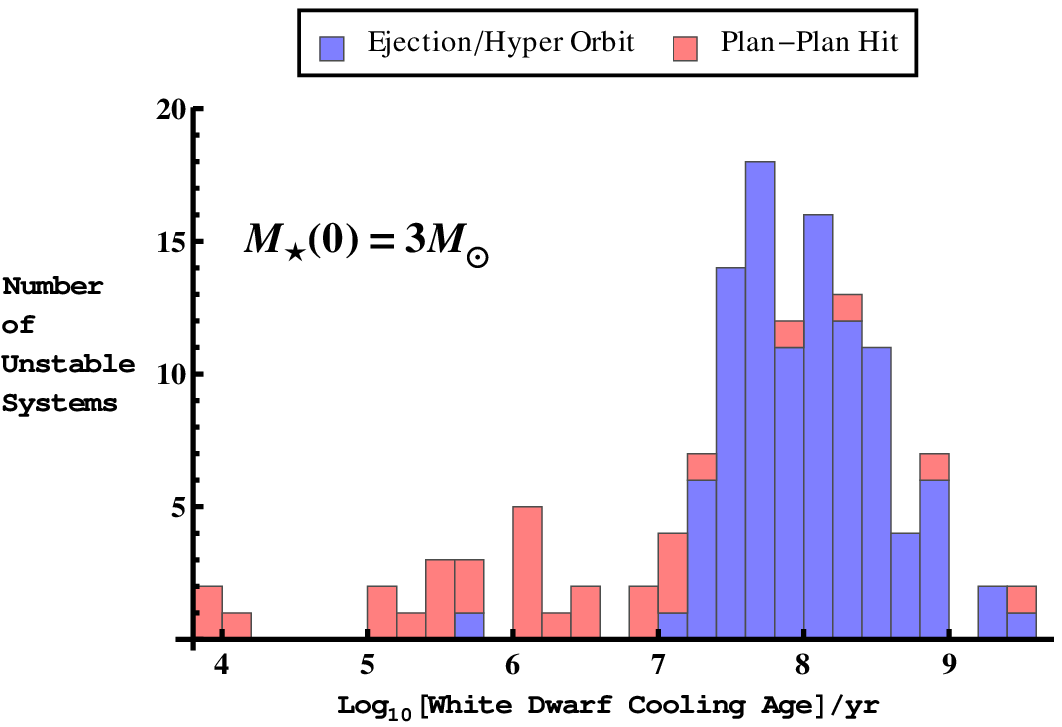,width=8.5cm,height=6.3cm} 
\psfig{figure=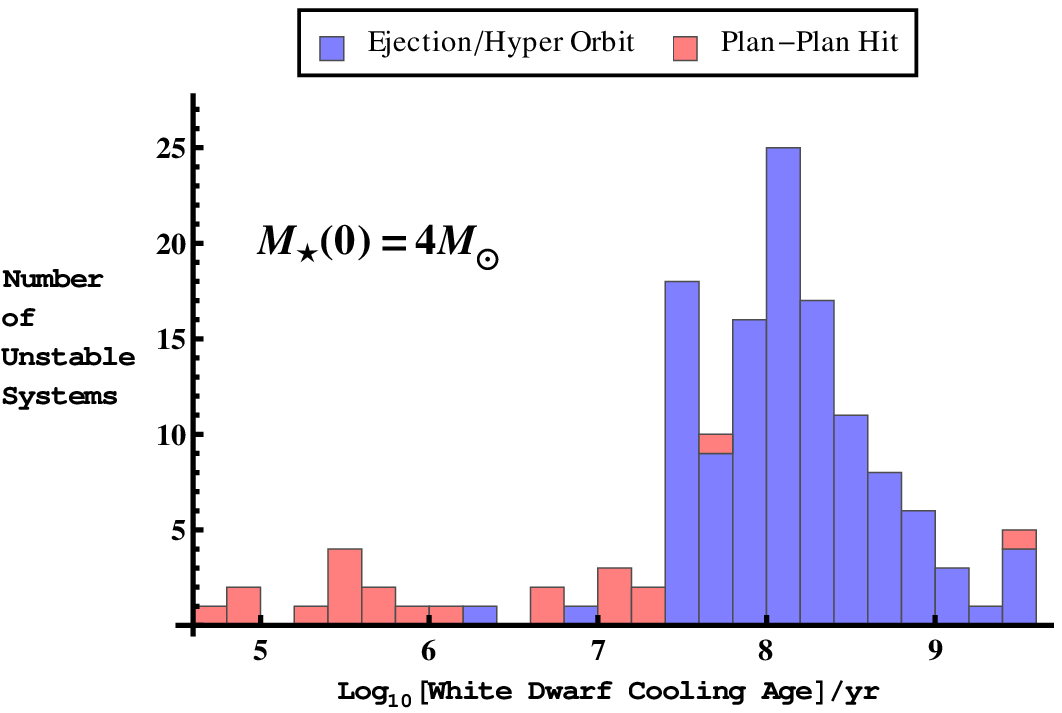,width=8.5cm,height=6.3cm}
}
\centerline{}
\centerline{
\psfig{figure=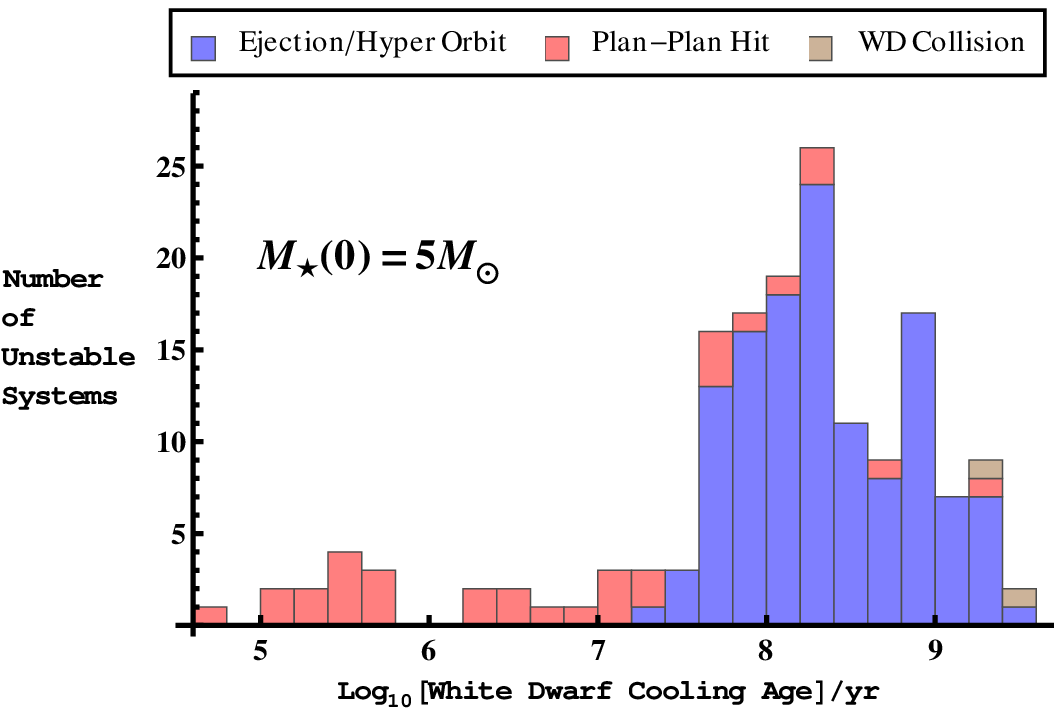,width=8.5cm,height=6.3cm} 
\psfig{figure=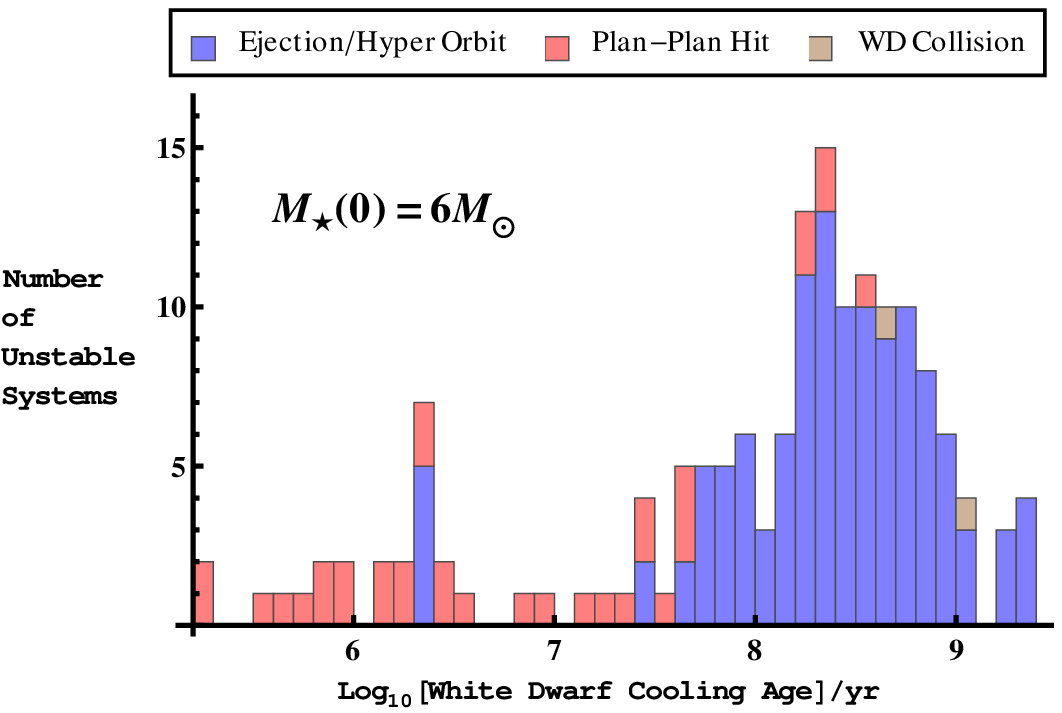,width=8.5cm,height=6.3cm}
}
\centerline{}
\centerline{
\psfig{figure=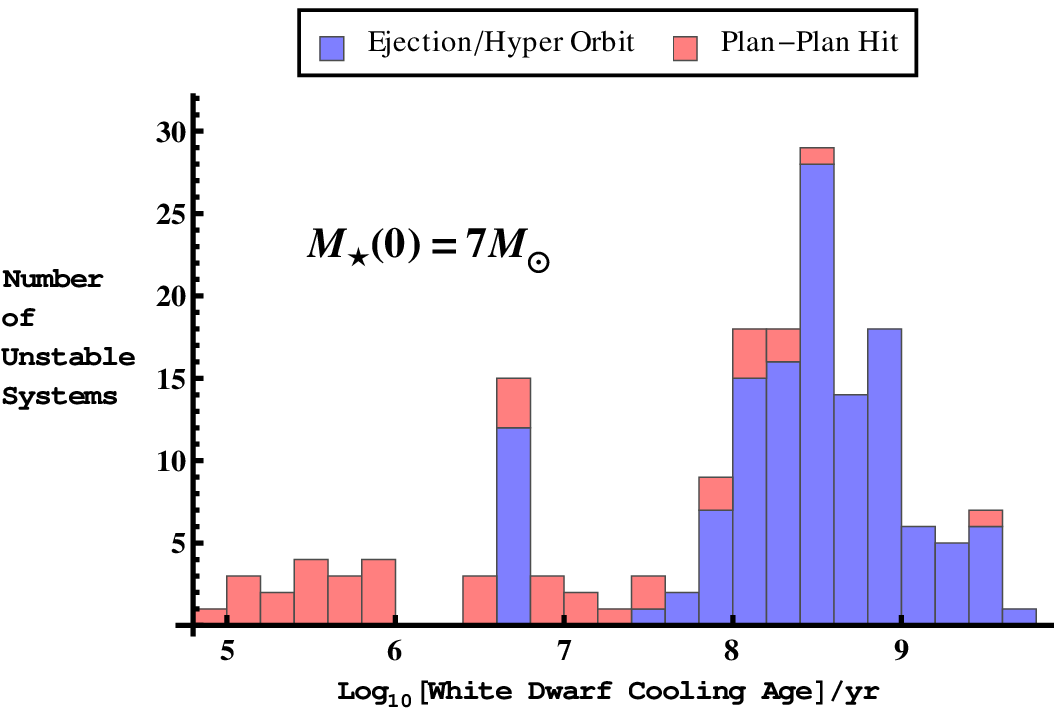,width=8.5cm,height=6.3cm} 
\psfig{figure=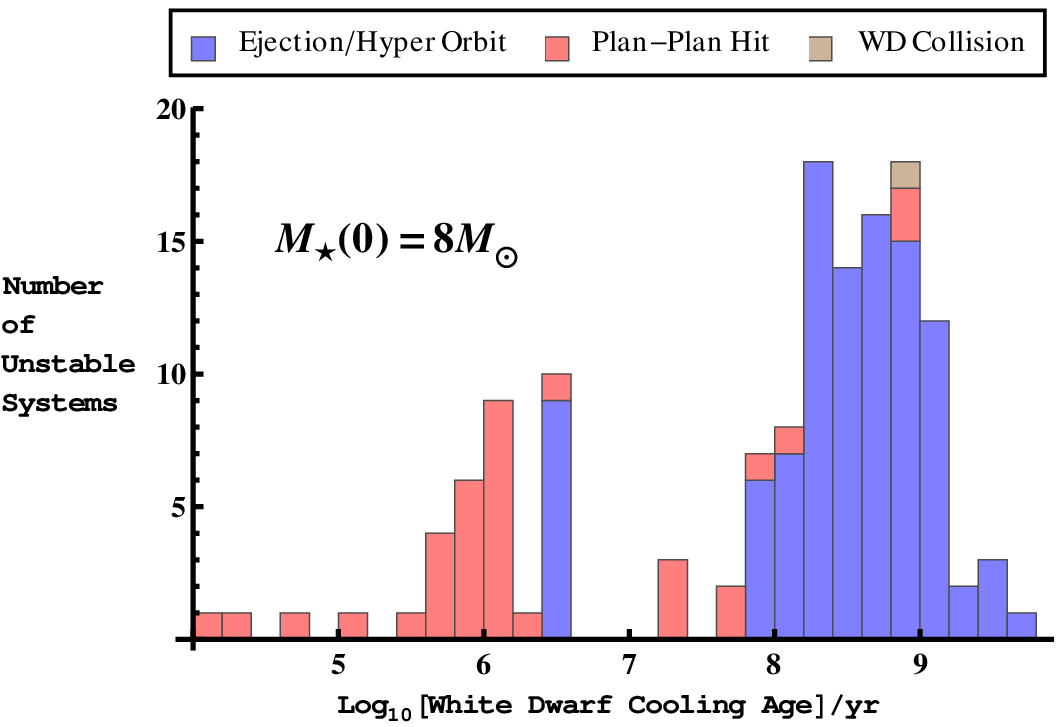,width=8.5cm,height=6.3cm}
}
\caption{
Instability as a function of the time elapsed since
the end of the of the Asymptotic Giant Branch phase, when
the WD was born.  Planet-planet collisions occur 
quickly ($\sim 10^4 - 10^7$ yr),
whereas instances of escape or stellar collisions
typically do not occur until after a few tens of Myr.
}
\label{CoolingTime}
\end{figure*}

\subsection{Consequences for White Dwarf Pollution}

\subsubsection{Background}

One potential consequence of dynamical instabilities induced due to stellar mass loss
is WD pollution.  WDs are surrounded by thin atmospheres of either hydrogen or helium. 
Heavier elements sink rapidly in such thin atmospheres and it is therefore puzzling 
that such a high fraction of WDs have evidence for metal pollutants in their spectra 
(25\% of single DA WDs, \citealt*{zucetal2003}).  Such metal pollution has been associated 
with excess emission in the infrared consistent with a close-in dust disc in more than a 
dozen cases \citep{faretal2009} and gas discs in a handful of 
cases \citep{ganetal2006,ganetal2007,ganetal2008}.  It has been suggested that these 
observations could be explained by planetary material, 
motivated by the similarity of the composition of the accreted material with planetary 
material \citep{kleetal2010,giretal2012}.

In order for a WD to be polluted by material from an outer planetary system, 
comets, asteroids or planets must be scattered at least close enough to the star 
(at about one Solar radius) so that they are tidally disrupted. Changes to the dynamics 
of the planetary system following stellar mass loss has been suggested as a potential 
cause of increased numbers of planetary bodies scattered onto star-grazing orbits 
\citep{debsig2002,jura2008,bonetal2011,debetal2012}. Even in planetary
systems where the planets remain on stable orbits, \cite{bonetal2011} and \cite{debetal2012} show 
that sufficiently many asteroids or comets can be scattered onto star-grazing orbits to 
explain some of the observations of polluted WDs. 
Such mechanisms, however, struggle to explain observations of high accretion rates in old 
polluted WDs \citep{koeetal2011,giretal2012}. 

Instabilities in planetary systems could provide a potential explanation for pollution
in these, and other, WDs. Depending on the exact nature of the instability and structure 
of the individual planetary system, in many cases the number of planetary bodies scattered onto 
star-grazing orbits increases significantly following an instability. This means that 
the planetary systems presented in this work in which instabilities occurred during the 
WD phase have the potential to produce polluted WDs. 

\subsubsection{Our Simulations}

We qualify the following discussion by reminding the reader that we
have not probed progenitor masses between $1 M_{\odot}$ and $2 M_{\odot}$,
where the true WD population appears to peak.  The reason for not 
considering this range is due to computational limitations, as tracking
planetary orbits over such long MS lifetimes is prohibitive.

Figure \ref{Main} showcases the different types of instability which can occur during
the WD phase.  First, many systems with giant planets that were Hill stable on the 
MS become Hill unstable due to mass loss preceding the WD phase.  Second, other systems that 
were technically Hill unstable on the MS, but were protected against instability by commensurabilities, 
become unstable during the WD phase. Third, some of the systems that are 
Hill stable during the WD phase are actually Lagrange unstable, and experience instability 
at a late time. 

We can now relate the WD instability to the cooling age of the WD, which is the time lapsed since
the star became a WD.  The cooling age of real (not simulated) WDs are readily determined from 
observations, and hence provide an opportunity for comparison with and motivation for numerical
simulations.  Figure \ref{CoolingTime} displays the number of planetary systems that 
become unstable during the WD phase as a function of cooling age; each plot corresponds to a 
different progenitor mass.  Any instance of instability could represent a trigger for a potentially
polluting event.  An ejection especially suggests that the inner planetary system will be significantly perturbed, and this perturbation can throw planetesimals or terrestrial planets that were originally in the inner system onto the star.

The figure contains several notable features: 1) planet-planet collisions tend to occur at short 
cooling ages ($\sim 10^4 - 10^7$ yr), 2) escape and stellar collisions becomes prevalent only
after $\sim 10^7$ yr, but tails off after a few Gyr, 3) the cooling age of this type of 
instability increases as the progenitor mass increases, and 4) direct collisions with the WD are rare.
The striking delay of $\sim 10^7$ yr before escape instability becomes dominant
reflects the (largely unexplored) dependencies of Lagrange instability on number of orbits
completed, the initial values of the semimajor axes, and the mass ratios.  In fact, perhaps 
the positive correlation of cooling age instability with progenitor mass is due to more orbits 
being required for smaller mass ratios for Lagrange instability to kick in, a feature also 
apparent on Fig. \ref{Main} on the MS.  Alternatively, the correlation may result from
the much wider orbits of the giant planets around higher-mass WDs.

The few giant planets which hit the WD are particularly interesting as pollution sources, and 
imply that some polluted WDs may result from the disruption of giant planets rather 
than comets or asteroids. Planetary material that is accreted 
onto WDs might be composed of many small asteroids or comets 
\citep{jura2003,bonetal2011,debetal2012}, but could also be the result of a single 
large object. Detailed compositional analysis of some objects concludes that the accreted material 
resembles more closely the bulk Earth in composition than chondritic 
material \citep[e.g.][]{zucetal2007,kleetal2010}; it is feasible that a disrupted 
planet would produce high levels of pollution in a WD. Such disrupted planets could provide the 
explanation for a handful of polluted WDs, in particular old, heavily polluted WDs. However, our simulations indicate that star-grazing giant planet collisions are too rare to alone account for the abundance of observed differentiated material.  This result agrees with the implied low fraction of planetary collisions with WDs from the finding that at most 1-5\% of WDs have high accretion rates due to dust disks \citep{debetal2011,faretal2012c}.

\begin{figure}
\centerline{
\psfig{figure=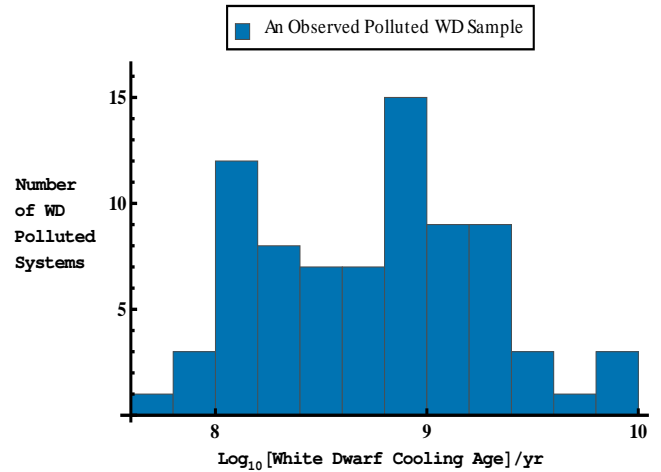,width=8.5cm,height=6.3cm} 
}
\caption{
Cooling ages for a sample of 78 observed WDs from Farihi et al. (2012b) and
Girven et al. (2012).  Although this figure may be compared with 
Fig. \ref{CoolingTime}, the progenitor masses of these WDs are unknown, and
the sample suffers from observational bias.  
}
\label{ObservedWD}
\end{figure}

Observed polluted WDs can provide the inspiration for extensions to our work,
with numerical simulations tailored to particular observational samples or campaigns.  
However, the purpose of this paper is not to make a detailed comparison with the observed polluted WD 
population, which is heavily biased and suffers from uncertainty of the distribution of their 
progenitor masses.  Our computationally-expensive simulations were set up to demonstrate that 
full-lifetime orbital integrations including realistic mass loss are now achievable, and to explore 
dynamical properties of the resulting instability.  Nevertheless, we can make a crude comparison of 
the potentially WD-polluting events in our simulations (Fig. \ref{CoolingTime}) 
and the observed distribution of polluted WDs (in Fig \ref{ObservedWD}).  Data for the observed 
distribution of 78 WD ages was taken from \cite{faretal2012b} and \cite{giretal2012}.  

Both the observations and simulations show a broad consensus.  Observations confidently detect 
pollution at cooling ages from a few tens of Myr \citep[e.g.][]{ganetal2012} to several Gyr 
\citep{koeetal2011}, but cannot detect pollution at earlier times when the WD is too hot 
to rule out a primordial origin for metals\footnote{These young WDs are able to 
radiatively levitate metals, masking what could be external pollution.} nor at later times when the WDs are too intrinsically faint.  The simulations begin 
to demonstrate ejection instability only at times greater than tens of Myr due to a true physical 
effect which may be explained by the timescale for Lagrange instability to act on the WD phase.  
This instability appears to have largely run its course after a few Gyr have passed, so that 
relatively fewer systems are likely to become unstable beyond the 5 Gyr integration time.  
Observationally, comparing the pollution frequency at 5 Gyr with that at tens of Myr is difficult 
due to intrinsic biases, despite the suggestive nature of the distribution in Fig. \ref{ObservedWD}.  Nevertheless, \cite{bonetal2011} found that asteroidal accretion onto WDs from particles scattered due to the overlap of mean motion resonance exterior to the planet follows a similar trend of being present beyond a few tens of Myr after post-MS evolution before eventually tailing off after a few Gyr.  \cite{debetal2012} later found that asteroidal accretion onto WDs from both exterior and interior resonances also follow this trend.

\subsection{Phase Space Extrapolation}

The results of this study raise several questions.  One important question is how robust our 
conclusions are to variations of the initial conditions.  Because of the computational expense of running 
simulation ensembles for 5 Gyr with the Bulirsch-Stoer algorithm, we could not perform a wider phase space 
exploration.  However, we can guess how the outcomes would vary in other circumstances.

Varying $a_1$ would change the number of orbits completed over the main sequence; the effect is similar to 
varying $M_{\star}$.  Hence, Fig. \ref{Main} demonstrates the likely consequence:  if the planets complete 
enough orbits, they will be prone to long-term MS instability.  If the initial planetary separation is beyond 
the Hill stability limit, then this instability must be Lagrange instability.  Otherwise, the type of 
instability is unrestricted.

The consequence of varying planetary eccentricities and mutual inclinations is less clear.  What is clear is 
that planets on eccentric orbits will feature in post-MS systems.  Observations suggest that if giant 
planets are formed at several AU on circular orbits, they are unlikely to retain their primordial eccentricities 
on the MS.  According to the Extrasolar Data Explorer\footnote{at http://exoplanets.org/}, as of 31 Oct, 2012, 
only 20.4\% of exoplanets with $M_p \ge 0.1M_{J}$ and $a \ge 1$ AU have $e < 0.1$.  This percentage shrinks to 
10.4\% for $e < 0.05$\footnote{Admittedly, there is a bias towards fitting higher eccentricity values than 
the true values. The disparity worsens with sparser data, which is often associated with the most distant
radial velocity planets.}.  If one were to include only planets in multi-planet systems, these percentages 
become 18.0\% and 14.0\%, respectively.  Therefore, if the planets we currently observe predominately survive until the 
post-MS, the significant majority will enter those phases with nonzero eccentricities.  Nevertheless, it is
of interest to determine if MS Lagrange instability can occur for planets formed on perfectly circular orbits.
Therefore, we have performed 48 additional simulations with $e_1(0) = e_2(0) = 0.0$ for $M_{\star}(0) = 3M_{\odot}$ at 
different locations beyond the Hill stability limit.  We can confirm that a few of those systems become Lagrange unstable.
A detailed statistical comparison is best left to a more comprehensive phase space study.

We can estimate the dependence on planetary mass by extrapolating from Figs. \ref{Main} and \ref{Earths}.
The trend suggests that as the test particle limit is approached, where there is no mutual interaction between secondaries,
the difference between the Hill and Lagrange boundaries might tend to zero faster than the Hill boundary itself.  In 
the opposite limit, for two brown dwarfs and a more massive evolving primary, we expect the stability boundaries to 
widen even further than in our fiducial case.  In these instances, instabilities might be common, and could 
represent a significant catalyst for WD pollution.

Adding more planets to model real systems such as HR 8799 introduces a significant complication, but is a viable avenue of
future study with our numerical code (Mustill \& Veras, in prep).  If additional planets are of similar masses, then we expect 
them to represent destabilizing influences, particularly in the post-MS stages.

\subsection{Comparison with Radial Velocity Exosystems}

The majority of known exoplanets are likely to be engulfed 
during the post-MS phase, as the population of planets beyond 10 AU is largely unknown.
Nevertheless, scaled versions of many known systems could represent genuine test cases for 
our model, because of the observed close separations of pairs of exoplanets.  Further, our findings of MS instability 
warrant a closer look at these systems.

We have compiled a list of all pairs of planets in the same system with $a > 1$ AU each that were detected by radial 
velocity measurements.  We obtained the data from the Extrasolar Planets Explorer on 6 November, 2012.  Using 
Eqs. (\ref{first})-(\ref{last}), and assuming minimum masses and coplanarity, we determined if the systems are 
currently Hill stable.  

We found that 6 pairs of planets are not (HD 181433 c,d; HD 204313 b,d; 24 Sex b,c; BD +20 2457 b,c; HD 128311 b,c; 
HD 200964 b,c).  Hence, these planets are likely to be protected by their proximity to mean motion commensurabilities 
inside small islands of stability, as in, e.g. \cite{witetal2012}.  This behaviour is reflective of some of our Hill 
unstable systems in Figs. \ref{Main}, \ref{Earths} and \ref{ModEcc}.  BD +20 2457 is notable because its mass is about 
$2.8 M_{\odot}$, and may be related to our $3 M_{\odot}$ simulations.  However, both planets in the system are likely 
instead to be classified as brown dwarfs ($M_1 \approx 22.7 M_{J}$ and $M_2 \approx 13.2 M_{J}$).  The correspondingly 
high mass ratios in the system, coupled with their close separations ($a_2/a_1 \approx 1.39$), perhaps intuitively 
suggest that instability should be imminent.  Hence, the existence 
of such a system is stark confirmation that Hill unstable systems can remain stable for long periods.  The system's 
future prospects for stability on the MS would require detailed modelling.

Three pairs of planets are Hill stable, but have a separation ratio which exceeds the Hill stability limit by
no more than 30\% (HD 37124 c,d; 47 Uma b,c; HD 183263 b,c).  Based on our simulations, the close proximity of these systems to the 
Hill stability limit suggests that they might not be Lagrange stable for the remainder of their MS lifetimes.  Testing 
this suggestion would require a detailed suite of long-term simulations for each system.  If Lagrange instability 
scales strongly with planet/star mass ratio, then the planets in HD 183263 are perhaps in the greatest danger:  the 
mass ratios in those systems are about 9.5 times as great as any mass ratio that we considered in our simulations.

Three pairs of Hill stable planets have separations exceeding the Hill stability limit by between 45\% and 65\% (HD 
108874 b,c; HD 159868 c,b; HD 10180 g,h), and two planet pairs (HD 4732 b, c; mu Ara b, c) have separations that 
are over twice the Hill limit.  Although the Lagrange stability boundary is likely dependent on several variables, 
we have performed additional simulations as proof of concept to show that in at least one case, this boundary can 
extend out to twice the Hill stability boundary.  Hence, HD 108874, HD 159868 and HD 10180 are not guaranteed to be 
Lagrange stable without further detailed analyses.

If the planets in any of these observed systems are not coplanar, then they are more likely to be Hill unstable 
(see the middle panel of Fig. \ref{hille1}).  A mutual inclination of just $12.3^{\circ}$ would render the planets in HD 37124 
Hill unstable.  At the opposite extreme, for the widely separated planets in HD 4732, a mutual inclination of $61.0^{\circ}$ 
would be required.

\subsection{Non-Adiabatic Mass Loss}

Adiabaticity in the two-body problem with mass loss is well-defined 
\citep[e.g.][]{veretal2011}.  If we were to assume the two-body definition 
of adiabaticity for each of the planets in each of our simulations, then
we can claim that at no time did our stable planetary systems 
(with $a_1(0) = 10$ AU) approach a regime that 
featured non-adiabatic mass loss.  However, if 2 
planets were stably orbiting at separations of hundreds of AU on the MS, then the
mutual planet-planet interaction coupled with stellar mass loss could yield 
unpredictable evolution during the post-MS.

\cite{voyetal2013} has recently explored non-adiabatic mass loss in 
the three-body
problem.  By using Lypaunov characteristic numbers to create dynamical 
stability maps, they found that non-adiabatic mass loss can cause a stably
interacting pair of planets to shift to a choatic region of phase space.
The resulting instability (manifested by escape or collision) may be 
latent, sometimes not appearing for a time
that exceeds the duration of the mass loss by a factor of tens.  In cases
where the outer planet escapes, distinguishing whether the instability
was triggered by Lagrange instability, non-adiabatic mass loss, or both might
require detailed follow-up simulations.


\subsection{Sharpness of the Hill Stability Limit} \label{sharpness}

\begin{figure}
\centerline{
\psfig{figure=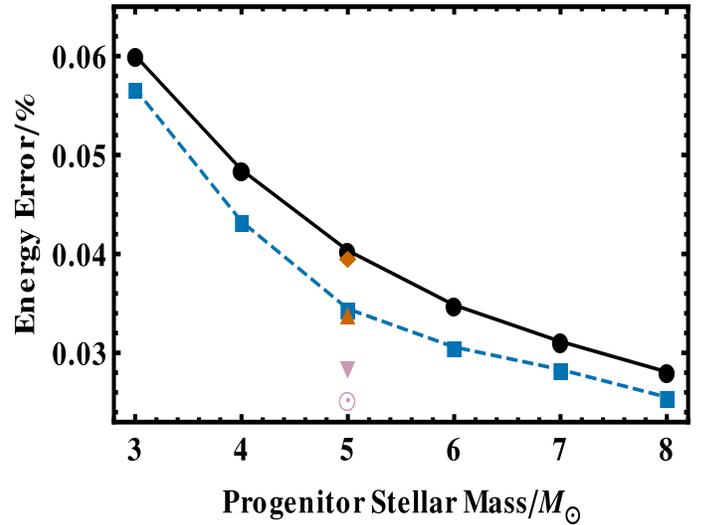,height=7.5cm,width=9.0cm} 
}
\caption{
The error of using the two-body approximation to model
the total energy of the system.  The black curve and
filled dots indicate the mean energy error; the blue
dashed curve and filled squares the median energy error.
These curves are drawn for fiducial initial orbital parameters
described in Subsection \ref{subsec:fiducial}.  The 
orange filled diamond and upwards-pointing
triangle represent the mean and median energy error for 
the $e_2 = 0.5$ simulation.  The purple filled downwards-pointing
triangle and open dotted circle are for the simulations with the pair of 
Earth-mass planets.  These errors correspond to semimajor axis
differences on the order of $10^{-3}$ AU for $a \approx 10$ AU.
}
\label{EError}
\end{figure}

No violations of the Hill stability limit have occurred in our simulations,
despite us using the two-body approximation for the energy in the analytical
formulation (Eqs. \ref{first}-\ref{last}).  Nevertheless, we can estimate the 
error in semimajor axis as a result.  Figure \ref{EError} plots
the mean and median energy error incurred by using the two-body
approximation for the energy of the system.  In all cases, the
energy error is between $0.02\%$-$0.06\%$.  This range corresponds to
semimajor axis differences of several $10^{-3}$ AU for planets
with $a \approx 10$ AU.  Our gradient of semimajor axis values sampled
for our fiducial simulations exceeded this value in each case, helping to confirm
why the analytic limit was not violated and suggesting that the 
unknown integration error was comparably small, if not smaller.

\section{Conclusion}

Architectures of planetary systems during each stellar phase
may represent historical tracers of formation and presage future 
evolutionary instability and death.  We have performed 5 Gyr simulations 
that consistently treat the dynamics of two massive planets and 
every phase of stellar evolution for a wide range of progenitor stellar masses ($3M_{\odot}$ - $8M_{\odot}$).  
These computationally-demanding simulations suggest that stable MS systems are in danger of future 
instability.  The zone of danger is wide, reaching out to 163\%-178\% of the MS Hill stability 
limit and 123\%-137\% of the WD Hill stability limit for our low eccentricity 
($e_1(0) = e_2(0) = 0.1$) simulations.  
The consequences for WD pollution may be significant:
For example, the inner planet can be perturbed onto a highly eccentric 
orbit which takes the planet close to the WD or hits the star directly.

\section*{Acknowledgments}

We thank the referee for a careful read of the manuscript
and astute and helpful suggestions.  We also thank
Jay Farihi for clarifying the observational biases, 
J. Richard Donnison for helping us confirm a typo in the previous 
literature about Hill stability, and Boris T. G\"{a}nsicke and 
Mukremin Kilic for useful discussions.  AJM is funded by the Spanish 
National Plan of R\&D grant AYA2010-20630, 
``Planets and stellar evolution''.  AB acknowledges the support of the 
ANR-2010 BLAN-0505-01 (EXOZODI).  This work made use of facilities funded by the 
European Union through ERC grant number 279973.

\label{lastpage}

\end{document}